\documentclass[a4paper,english,cleveref,autoref,thm-restate,nolineno]{socg-lipics-v2021}

\pdfoutput=1 
\hideLIPIcs  

\usepackage{tikz}
\usetikzlibrary{positioning,arrows.meta}
\usetikzlibrary{decorations.pathmorphing}
\newcommand{\xleadsto}[1][]{
	\mathrel{%
		\tikz[baseline={(lab.base)}]{
			\node[inner sep=1pt] (lab) {$\phantom{\scriptstyle #1}$};
			\draw[->,decorate,
			decoration={snake,amplitude=0.75pt,segment length=6pt}]
			(lab.west) -- (lab.east)
			node[midway,above=0.2ex] {$\scriptstyle #1$};
		}%
	}%
}
\usepackage{tikz-cd}
\usepackage{float}
\usepackage[ruled,vlined,linesnumbered,noend]{algorithm2e}
\usepackage{longtable}

\newcommand{\C}{\mathbb C}
\newcommand{\R}{\mathbb R}
\newcommand{\Z}{\mathbb Z}
\renewcommand{\d} { \mathrm{d} } 
\newcommand{\geomreal}[1] { |#1| } 
\newcommand{\carr}[1] { \mathrm{carr}(#1) } 
\newcommand{\Del}[1] { \mathrm{Del}(#1) } 
\newcommand{\conv}[1] { \mathrm{conv}(#1) } 
\newcommand{\conesimp}[1] { C_\mathrm{simp}(#1) } 
\newcommand{\cone}[1] { C(#1) } 
\newcommand{\sect}[1] { \mathrm{Sect}(#1) } 
\newcommand{\GG}{\mathcal G} 
\newcommand{\VV}{\mathcal V} 

\newif\iffullversion
\fullversiontrue      
\newcommand{\fullonly}[1]{\iffullversion #1\fi}
\newcommand{\shortonly}[1]{\iffullversion\else #1\fi}

\title{Simplicial approximation to CW complexes with spherical Delaunay triangulations}

\author{Raphaël Tinarrage}{Institute of Science and Technology Austria, Klosterneuburg, Austria \and \url{https://raphaeltinarrage.github.io/} }{raphael.tinarrage@ist.ac.at}{https://orcid.org/0000-0002-1404-1095}{}

\authorrunning{R.~Tinarrage}
\Copyright{Raphaël Tinarrage} 



\shortonly{
\relatedversiondetails{Full Version}{https://arxiv.org/abs/2112.07573}
}
	
\supplementdetails{Software}{https://github.com/raphaeltinarrage/cw2simp}

\nolinenumbers 

\EventEditors{Hee-Kap Ahn, Michael Hoffmann, and Amir Nayyeri}
\EventNoEds{3}
\EventLongTitle{42nd International Symposium on Computational Geometry
	(SoCG 2026)}
\EventShortTitle{SoCG 2026}
\EventAcronym{SoCG}
\EventYear{2026}
\EventDate{June 2--5, 2026}
\EventLocation{New Brunswick, NJ, USA}
\EventLogo{socg-logo.pdf}
\SeriesVolume{367}
\ArticleNo{XX}     

\begin{document}

\maketitle

\begin{abstract}
	Simplicial approximation provides a framework for constructing simplicial complexes that are homotopy equivalent to a given manifold, provided a CW structure is explicitly known.
	However, its conventional implementation quickly becomes intractable on a computer: barycentric subdivision produces poorly shaped simplices, and the star condition introduces many vertices. 
	To address these limitations, this article develops a subdivision scheme based on spherical Delaunay triangulations, which attains better refinement properties than barycentric subdivisions.
	Moreover, the star condition is reframed as two independent problems, one geometric and the other combinatorial, respectively tackled in the language of locally equiconnected spaces and the list homomorphism problem, allowing an exponential reduction in the number of vertices.
	Via a prototype implementation, we obtain simplicial complexes homotopy equivalent to Grassmannians and Stiefel manifolds up to dimension~5.
\end{abstract}

\section{Introduction}
\label{sec:intro}

\subsection{Topology software}

In computational topology, a popular way to represent a topological space is via a simplicial complex. 
Once a space is triangulated, it can be explored algorithmically, and a range of homotopy invariants can be evaluated \cite{DBLP:conf/soda/CadekKMSVW12,vcadek2014computing,cadek2014polynomial,Green_2015,vcadek2017algorithmic,10.5555/3174304.3175344,filakovsky2018computing,filakovsky2020two,vokvrinek2017decidability,manin2024algorithmic}.
More generally, triangulations open the door to many further developments: they allow us to test conjectures and discover new properties \cite{benedetti2014random,benedetti2024random}, they serve as a benchmark for comparing software \cite{bell2018computing,ballester2024mantra,telyatnikov2024topobench} and they lay the foundation for new data analysis techniques \cite{sheth2003measuring,stoecker2013interactive,ericok2022quotient,tinarrage2022computing}.

Over the past two decades, software for computations on simplicial complexes has advanced substantially. 
Existing libraries span a wide range of goals, from algebra to geometry and data analysis.
They have underpinned numerous concrete advances: proofs and counterexamples in 3-manifold topology with \texttt{Regina} and \texttt{SnapPy} \cite{regina,SnapPy}; large-scale enumerations that tested conjectures with \texttt{BISTELLAR} and \texttt{Twister} \cite{bjorner2000simplicial,Twister}; new homotopy and cohomology computations in \texttt{Kenzo} and \texttt{HAP} \cite{Kenzo,Ellis2025}; and persistent homology pipelines that surfaced new properties of datasets with \texttt{GUDHI}, \texttt{Ripser}, and \texttt{TTK} \cite{maria2014gudhi,bauer2021ripser,Tierny2017TTK}, among many others.

On the other hand, computational topology lacks \textit{explicit} examples of triangulated manifolds, as noted in \cite{benedetti2014random,filakovsky2020two,ballester2024mantra,telyatnikov2024topobench}.
By explicit, we mean stored on a computer as a list of simplices, or obtainable in reasonable time by an implemented algorithm.
This is especially striking in dimension 4 and above, as summarized in \Cref{tab:known_triangulations}, which collects known triangulations of certain classical manifolds.
Apart from the real and complex projective spaces, most triangulations are ``accidental'', i.e., arising from special homeomorphisms with known spaces.
The aim of this article is to develop an algorithm for triangulating new spaces.

\begin{table}[!htbp]
	\centering
	\begin{tabular}{lll}
		\hline
		\textbf{Space}                    
		& \textbf{Known cases}
		& \textbf{References / Known identifications} \\
		\hline
		Real projective space $\R P^n$                          
		& $n \geq 1$
		& \cite{vonKuhnel1987Kummer,balagopalan2017vertex,adiprasito2022subexponential} 
		\\
		Complex projective space $\C P^n$                          
		& $n \geq 1$
		& \cite{kuhnel19839,sergeraert2010triangulations,bagchi2011icosahedron,bagchi2012triangulation,Sarkar2014CPn,datta2025simplicial} \\
		Special orthogonal group $\mathrm{SO}(n)$                  
		& $n \leq 4$
		& $\mathrm{SO}(3)\cong\R P^3$, $\mathrm{SO}(4)\cong S^3\times\mathrm{SO}(3)$                              
		\\
		Special unitary group $\mathrm{SU}(n)$                  
		& $n\leq 2$
		& $\mathrm{SU}(2)\cong S^3$ 
		\\
		Unitary group $\mathrm{U}(n)$                   
		& $n=1$
		& $\mathrm{U}(1)\cong S^1$ 
		\\
		Real Stiefel manifold $\VV(d,n)$             
		& $d = 1$ or $n \leq 4$
		& 
		$\VV(1,n)\cong S^{n-1}$, 
		$\VV(d-1,d)\cong \mathrm{SO}(d)$ 
		\\
		Real Grassmannian $\GG(d,n)$             
		& $d = 1$ or $d = n-1$
		& $\GG(1,n)\cong \GG(n-1,n) \cong \R P^{n-1}$                                            
		\\
		\hline
	\end{tabular}
	\caption{
		Known explicit triangulations of a selection of manifolds.
		}
	\label{tab:known_triangulations}
\end{table}

\subsection{Related work}

Traditionally, explicit triangulations of manifolds are obtained by two main approaches.
The first is combinatorial and relies on specific descriptions of the manifolds under consideration.
For example, small triangulations of $\R P^n$ have recently been constructed by exploiting its realization from a symmetric polytope \cite{adiprasito2022subexponential}.
Topological properties can also guide the enumeration of combinatorial manifolds, as in the work of Lutz \cite{MR2410566,MR2552676}.
However, combinatorial complexity grows rapidly with the dimension.
Thus, in dimensions 3 and 4, more structured constructions are preferred, based on layered triangulations \cite{jaco2006layeredtriangulations3manifolds,jaco2009minimal,jaco2011coverings}, Dehn fillings \cite{DunfieldExceptionalDehnFillings}, Heegaard diagrams \cite{ennes_et_al:LIPIcs.ESA.2025.37,He_Morgan_Thompson_2025,ennes2025compresseddatastructuresheegaard}, and Kirby diagrams \cite{Burke_2025,burke_et_al:LIPIcs.SoCG.2025.28}.
The present work follows this perspective by exploiting the \textit{CW structure} of the spaces involved, which is well understood.

Sampling-based methods offer a different viewpoint.
A large body of work studies how to reconstruct an embedded manifold $\mathcal{M} \subset\R^n$ from a finite sample $X \subset\R^n$.
The goal is usually not to recover a triangulation of $\mathcal{M}$ itself, but rather a simplicial complex with the same homotopy type.
A standard construction is the \v{C}ech complex.
Recovery is guaranteed when the sampling density is sufficiently fine relative to an appropriate condition number, typically the reach of $\mathcal{M}$ \cite{niyogi2008finding}.
Quantitative refinements use less restrictive geometric quantities, such as the $\mu$-reach \cite{chazal2006sampling,kim_et_al:LIPIcs.SoCG.2020.54}, weak feature size \cite{chazal2007stability}, local feature size \cite{chazal2008smooth}, convexity defects \cite{attali2011vietoris}, and convexity radius \cite{fasy2022reconstruction}.
In higher dimensions, however, the \v{C}ech complex may be prohibitively expensive to compute, and one may instead use the Vietoris--Rips complex, which depends only on pairwise distances.
Following the foundational results of Hausmann and Latschev \cite{hausmann1995vietoris,latschev2001vietoris}, quantitative guarantees have been established in terms of the reach \cite{majhi:LIPIcs.SoCG.2024.73,majhi2025demystifying}, the $\mu$-reach \cite{kim_et_al:LIPIcs.SoCG.2020.54}, and convexity defects \cite{attali2011vietoris}.
Other approaches rely on Delaunay complexes \cite{10.1145/1810959.1811013,boissonnat2010manifold,boissonnat2018delaunay,aamari2018stability}, witness complexes \cite{10.2312:SPBG/SPBG04/157-166,guibas2008reconstruction,boissonnat2007manifold,Boissonnat2009}, and metric thickenings~\cite{adamaszek2018metric,adams2019metric}.

These methods nevertheless remain computationally demanding. 
Vietoris--Rips complexes are the cheapest to compute, but their guarantees are weaker and they typically require many vertices.
\v{C}ech complexes present the same difficulty, as they may contain a large number of simplices, potentially of high dimension. 
In contrast, more refined constructions such as Delaunay complexes require substantial computation time and scale poorly with dimension.

Grassmannians $\GG(d,n)$ are a notable example, regarded as difficult to triangulate.
Theoretical results indicate that the minimal number of facets increases exponentially with $n$ \cite{GovcMarzantowiczPavesic2020}.
Knudson considered triangulating $\GG(2,4)$ by embedding it into $\R^{16}$ and constructing a Vietoris--Rips complex \cite{Knudson2020}.
In practice, the approach amounted to increasing the scale until obtaining a complex with the expected homology.
This proved too costly in memory, and a witness complex was used instead.
Although a complex with the correct homology was found, this construction does not satisfy the hypotheses of the available theorems. 

\subsection{Contributions}

In \Cref{alg:algorithm_full} we present an implementation of \emph{simplicial approximation to CW complexes}, a framework well established in theory yet overlooked in practice. 
Given a CW structure on a topological space, the algorithm outputs a \textit{homotopy equivalent} simplicial complex.
Although not homeomorphic, such a complex still suffices for many of the applications mentioned above.
Crucially, our complexes are equipped with a \emph{point location} routine, making them practical surrogates for the manifold in data-driven applications.

Turning the textbook construction into a practical implementation requires several adjustments; in particular, we aim to keep the resulting complexes as small as possible.
To this end, in \Cref{sec:delaunay}, we adopt Delaunay refinement instead of barycentric subdivision; in \Cref{sec:simplicial_mapping_cones}, we build a simplicial mapping cone that avoids subdividing the complex; and in \Cref{sec:approximation}, we substitute the star condition for a more efficient constraint satisfaction problem.

We implemented two versions of the algorithm, using either global or local refinement. 
Only the first is currently proven to terminate, but the second often produces smaller complexes.
In both cases, when the algorithm terminates, the output is correct (see \Cref{th:algorithmcorrect}).

The software is provided as supplementary material.\shortonly{\footnote{Fully implemented prototype in Python: \url{https://github.com/raphaeltinarrage/cw2simp}}}
The repository includes complexes homotopy equivalent to $\GG(2,4)$ and $\VV(2,4)$ (of dimensions $4$ and $5$), which were not available prior to this work.
We intend to address higher-dimensional examples in future work by increasing computational resources; the current results were obtained on a personal~laptop.

\fullonly{
Although the focus here is theoretical, our work is motivated by \textit{Topological Data Analysis}, a theory that uses topological invariants to study real data. Our two main outputs---triangulations of new manifolds and a framework for simplicial approximation---naturally set the stage for new methods in data science; we suggest two such applications in \Cref{sec:applications}.
}

\fullonly{
The algorithm is sketched in the section below.
In \Cref{sec:delaunay,sec:simplicial_mapping_cones,sec:approximation} we develop the theoretical results required for its implementation; all proofs are deferred to \Cref{sec:proofs}.
Notation is collected in \Cref{sec:background}, and \Cref{sec:execution} showcases a full execution of the algorithm.
Finally, \Cref{sec:background_grassmannian} reviews the basic properties of Grassmannians, one of our main examples.
}

\shortonly{
The algorithm is sketched in the next section.
In \Cref{sec:delaunay,sec:simplicial_mapping_cones,sec:approximation} we develop the theoretical results required for its implementation.
Proofs appear in the full version of the~article.
}

\subsection{Overview and discussion of the construction}
\label{subsec:overview}

\subparagraph{CW complexes}

The theoretical background can be found in Hatcher's book \cite[Section~2.C]{hatcher_algebraic_2002}.
A \textit{CW complex} of dimension $n$ is a topological space $X$ endowed with a decomposition $X = X^n \supset X^{n-1} \supset \cdots \supset X^0$ where $X^0$ is finite and each $X^{d}$ is equal to a disjoint union
\[
	X^d = X^{d-1} \coprod_{1\leq i\leq m(d)} e_i^d,
\]
where each $e_i^d$, called a \textit{$d$-cell}, is homeomorphic to an open ball of dimension $d$, and $m(d)$ is their number.
It is required that the homeomorphism extends to the closed ball $B^d$, yielding a map $\Phi_i^d\colon B^d\to \bar{e}_i^d\subset X^{d}$, called the \textit{characteristic map}.
Its restriction to the boundary of $B^d$---i.e., the sphere $S^{d-1}$---is called the \textit{gluing map} and is denoted $\phi_i^d\colon S^{d-1} \to X^{d-1}$.
In other words, $X^d$ is homeomorphic to the gluing of $d$-balls along their boundary on $X^{d-1}$:
\[
	X^d \cong X^{d-1} \cup_{\phi_1^d} B^d \cup_{\phi_2^d} B^d \cup\dots\cup_{\phi_{m(d)}^d} B^d.
\]

\begin{figure}[!htbp]
	\centering
	\subcaptionbox{
		$S^2$
	\label{fig:examples_cw_complexes:sphere}
	}[0.55\textwidth]{%
		\includegraphics[width=.85\linewidth]{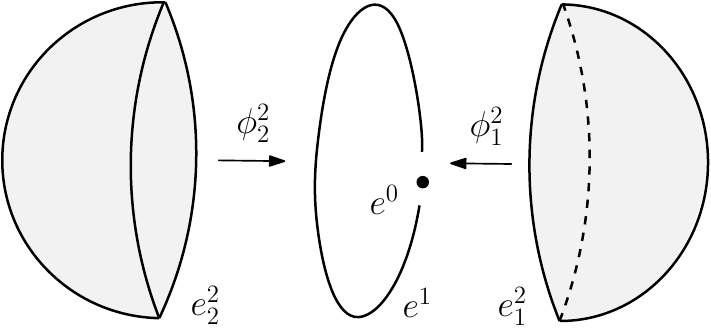}
		}
	\hfill
	\subcaptionbox{
		$\R P^2$
	\label{fig:examples_cw_complexes:rp2}
	}[0.38\textwidth]{%
		\includegraphics[width=.85\linewidth]{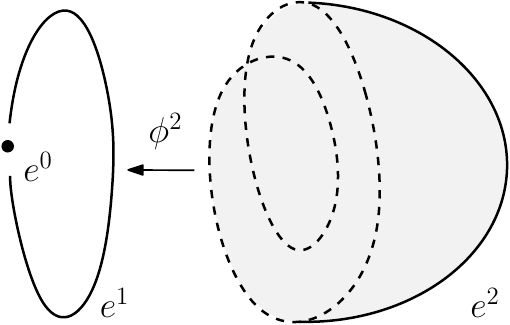}}
	\caption{
		Examples of CW structures on the sphere $S^2$ (four cells in total) and the projective plane $\R P^2$ (one cell per dimension). 
		If only one $d$-cell is attached, we omit the subscript in $e_i^{d}$.}
	\label{fig:examples_cw_complexes}
\end{figure}

\Cref{fig:examples_cw_complexes:sphere} shows a CW structure on $S^2$, made of one $0$- and $1$-cell and two $2$-cells.
\Cref{fig:examples_cw_complexes:rp2} depicts the usual structure on $\R P^2$: one cell per dimension and a gluing map $\phi^2\colon S^1\to S^1$ of degree~2.
Similarly, all the manifolds in \Cref{tab:known_triangulations} admit a well-known CW structure.
We refer the interested reader to \cite[Section~6]{Milnor_1974} for the classical structure on Grassmannians, \cite[Section~3.D]{hatcher_algebraic_2002} for Stiefel manifolds and orthogonal groups, and \cite{Yokota1956CellularUnitary} for unitary groups.
\fullonly{
For reference, the example of the Grassmannian $\GG(2,4)$ is presented in detail in \Cref{sec:background_grassmannian}.
}

\subparagraph{Sketch of the approach}

\begin{figure}[!htbp]
	\centering
	
	\begin{subfigure}[t]{1\textwidth}
		\centering
		\begin{tikzcd}[column sep=.3cm]
			e^0 \arrow[rr, hook]  
			&& e^0 \cup e^1 \arrow[rr, hook]  
			&& e^0 \cup e^1 \cup e^2_1 \arrow[rr, hook]  
			&& e^0 \cup e^1 \cup e^2_1 \cup e^2_2 \\
			& S^0 \subset B^1 \arrow[ul, "\phi^1"] &
			& S^1 \subset B^2 \arrow[ul, "\phi^2_1"] &
			& S^1 \subset B^2 \arrow[ul, "\phi^2_2"] &
		\end{tikzcd}
		\subcaption{%
			Diagram induced by the CW structure on $S^2$ in \Cref{fig:examples_cw_complexes:sphere}.
		}
	\end{subfigure}
	\hfill

	\begin{subfigure}[t]{1\textwidth}
		\centering
		\includegraphics[width=.91\linewidth]{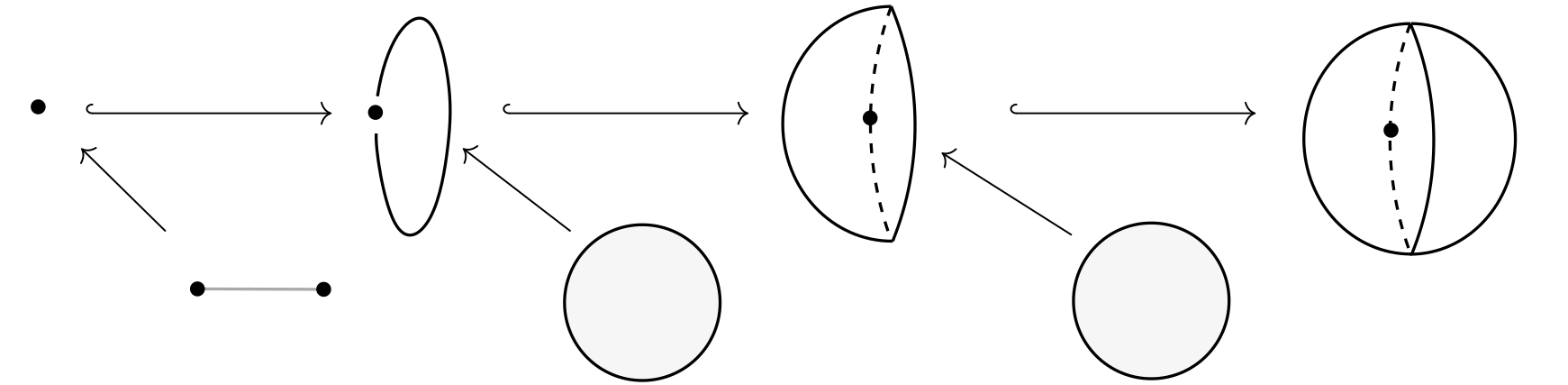}
		\subcaption{%
			Geometric visualization of the diagram above.
		}
	\end{subfigure}
	\hfill

	\begin{subfigure}[t]{1\textwidth}
		\centering
		\includegraphics[width=.91\linewidth]{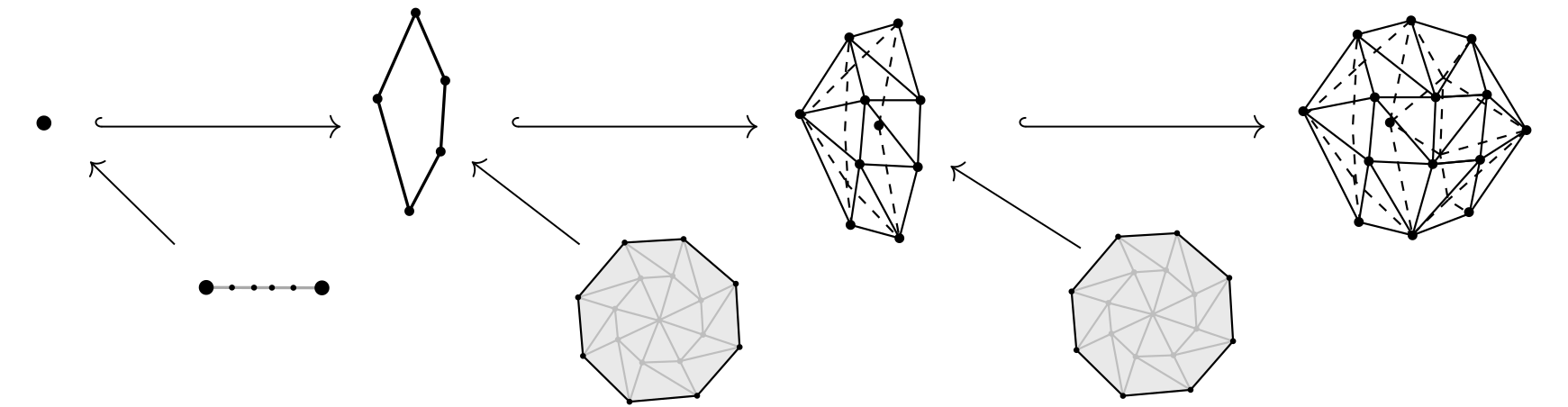}
		\subcaption{%
			Simplicial approximation of the diagram.
			}
	\end{subfigure}
	
	\caption{%
		Schematic view of the simplicial approximation procedure for CW complexes.
	}
	\label{fig:pipeline_triangulation_cw}
\end{figure}

One can ``convert'' a CW complex into a simplicial complex by gluing \textit{simplicial balls} instead of cells; the idea is sketched in \Cref{fig:pipeline_triangulation_cw}.
The construction proceeds inductively on the dimension.
Suppose we have already built a simplicial complex $L^{d}$ homotopy equivalent to the $d$-skeleton $X^{d}$; we denote by $\geomreal{L^{d}}$ its geometric realization.
For each $(d+1)$-cell $e^{d+1}_i$, we perform the following steps:
\begin{itemize}
	\item 
	We apply \textit{simplicial approximation} (see below) to the gluing map $\phi_i^{d+1}\colon S^{d} \to \geomreal{L^{d}}$.
	This yields a triangulation $K_i^{d}$ of $S^{d}$ and a simplicial map $g^{d+1}_i\colon K^{d}_i\to L^{d}$ homotopic to $\phi_i^{d+1}$.
	\item 
	From $K^{d}_i$, we construct a simplicial ball $B(K^{d}_i)$ with boundary $K^d_i$ and glue it to $L^{d}$ along this boundary via $g^{d+1}_i$.
	This amounts to the simplicial mapping cone of $g^{d+1}_i$.
\end{itemize}
After all $(d+1)$-cells have been attached, the resulting complex $L^{d+1}$ is homotopy equivalent to the skeleton $X^{d+1}$, and the procedure can be iterated in the next dimension.

\subparagraph{Computational obstacles}

Conventionally, the simplicial approximation of $\phi_i^{d+1}\colon S^{d} \to \geomreal{L^{d}}$ is obtained as follows: one starts from any triangulation $K$ of $S^{d}$ and repeatedly applies barycentric subdivision until the triangulation is fine enough.
More precisely, one stops when the map satisfies the \textit{star condition}: each closed star of a vertex $v\in K$ is mapped by $\phi_i^{d+1}$ into the open star of some vertex $w\in L^{d}$.
The simplicial approximation theorem ensures that this procedure terminates \cite[Theorem~2C.1]{hatcher_algebraic_2002}.
The assignment of vertices $v\mapsto w$ defines a simplicial map $g^{d+1}_i$ homotopic to $\phi_i^{d+1}$ (through a homotopy that is linear on each simplex).

In practice, this approach quickly becomes intractable for two reasons.
First, barycentric subdivision introduces a large number of vertices: a $d$-simplex is converted into a complex with $2^{d+1}-1$ vertices.
In \Cref{sec:delaunay}, we propose to use instead the \textit{Delaunay complexes} and their refinements.
We show in \Cref{th:shrinkingrefinements} that Delaunay refinement enjoys better approximation properties than barycentric subdivision: the simplices shrink more rapidly.

Secondly, the star condition itself is too coarse: each facet of $L^d$ requires $d+1$ vertices of $K$ that are mapped into it.
This contributes further to the exponential growth in the number of vertices.
In \Cref{sec:approximation}, we avoid the star condition altogether by decomposing the problem into two parts: a geometric step (constructing a homotopy equivalence) and a combinatorial step (constructing a simplicial map).
\Cref{prop:approximationroutine} ensures that this procedure is correct.

Last, the standard construction of a simplicial mapping cone, going back to Cohen \cite{cohen1967simplicial}, relies on the barycentric subdivision of $K$, which again leads to large complexes.
In \Cref{sec:simplicial_mapping_cones}, we consider a lighter triangulation, based on \textit{staircase triangulation} of products.
\Cref{prop:equivalencemappingcones} shows that this object is homotopy equivalent to the standard mapping cone.

\section{Successive refinements of spherical Delaunay triangulations}
\label{sec:delaunay}

Our construction begins with a refinement scheme for spherical Delaunay complexes, which generates arbitrarily fine triangulations and is well suited to concrete implementation.

\subsection{Spherical Delaunay triangulations}
\label{subsec:delaunay_triangulations}

\paragraph*{Definition}

Let $X \subset \R^d$ be finite. 
A subset of $d+1$ points has the \emph{empty circle property} if its circumscribing (open) ball is empty of points of $X$.
These subsets form the facets of an abstract simplicial complex $\Del{X}$, called the \emph{Delaunay complex}.
Under the genericity assumption that no subset of $d+2$ points lies on the same sphere, $\Del{X}$ is naturally embedded in $\R^d$ \cite{boissonnat2018geometric}.

These definitions adapt to the spherical case: given a subset $X$ of the unit sphere $S^d\subset\R^{d+1}$, the empty circle property is understood with respect to the geodesic distance on $S^d$; the resulting complex is called the \emph{spherical Delaunay complex} and is still denoted $\Del{X}$.
If no subset of $d+2$ points lies on the same geodesic sphere, then it is naturally embedded in $\R^{d+1}$.
We shall implicitly make this assumption throughout the article.

It is well-known that the spherical Delaunay complex coincides with the boundary of the convex hull of its vertices, thus reducing the computation of $\Del{X}$ to $\conv{X}$ \cite{b-gtfga-80}.
In our implementation, we used \texttt{Qhull}, a popular software package for computing convex hulls \cite{barber1996quickhull}.

\paragraph*{Point location}

A natural map $\geomreal{\Del{X}}\to S^d$ is given by the scaling $x\mapsto x/\|x\|$, which is a homeomorphism provided that the origin is included in the interior of the convex hull of $X$.
A set $X$ satisfying this assumption will be referred to as \emph{admissible}, and $\Del{X}$ as an \emph{admissible triangulation}.
We call the inverse homeomorphism $r\colon S^d\to\geomreal{\Del{X}}$ the \emph{radial projection}.
Computationally speaking, the radial projection of $x\in S^d$ is found by identifying a facet $\sigma\in\Del{X}$ to which $r(x)$ belongs, then computing a simple line/hyperplane intersection.

We perform point location via a conventional \emph{Jump-and-Walk} strategy \cite{mucke1996fast}: it consists of choosing a first candidate $\sigma\in\Del{X}$, hopefully close to $r(x)$, then walking through its neighbor facets until reaching one that contains $r(x)$.
Unlike popular software packages such as \texttt{STRIPACK} or \texttt{CGAL} \cite{cgal:eb-25b,renka1997algorithm}, our implementation does not make use of spherical geometric predicates.
Instead, we project $\Del{X}$ into the tangent space of $S^d$ at $x$ via the stereographic projection $p\colon S^d\setminus\{-x\}\to T_x S^d$ and work in the induced Euclidean triangulation (the simplices are replaced with the convex hull of their vertices).
This approach lets us reuse Euclidean routines already present in our code.
In this step, we exclude from $\Del{X}$ all simplices incident to $-x$, since their images under $p$ are not defined.

A caveat is that stereographic projection distorts geodesics: in general, the image geodesic simplex $p(\geomreal{\sigma})$ need not coincide with the linear simplex $\conv{\{p(v_0),\dots,p(v_d)\}}$ on the image of its vertices. 
Consequently, a point might map to the ``wrong'' linear simplex.
The next lemma shows that this cannot happen at the pole $x$, which justifies our procedure.

\fullonly{
\begin{lemma}[%
	name={proof p.~\pageref{proof:originstereographicprojection}},%
	restate=originstereographicprojection%
	]
	\label{lem:originstereographicprojection}
	Let $\sigma=[v_0,\dots, v_d]\in\Del{X}$ and $x\in S^d$ such that $r(x)\in\geomreal{\sigma}$.
	Then $p(x)\in\conv{\{p(v_0),\dots,p(v_d)\}}$, where $p$ is the stereographic projection at $x$.
\end{lemma}
}

\shortonly{
\begin{lemma}
	\label{lem:originstereographicprojection}
	Let $\sigma=[v_0,\dots, v_d]\in\Del{X}$ and $x\in S^d$ such that $r(x)\in\geomreal{\sigma}$.
	Then $p(x)\in\conv{\{p(v_0),\dots,p(v_d)\}}$, where $p$ is the stereographic projection at $x$.
\end{lemma}
}

\subsection{Global Delaunay refinements}
\label{subsec:global_delaunay_refinements}

\fullonly{
\paragraph*{Definition and shrinking property}
}

Given a spherical Delaunay triangulation $\Del{X_0}$ with vertex set $X_0\subset S^d$, one obtains a finer triangulation by choosing additional points $Y_0\subset S^d$ and building the Delaunay complex on $X_1 = X_0 \cup Y_0$.
The new points are called \textit{Steiner points}, and this procedure is known as a \textit{Delaunay refinement}.
This construction can be iterated, yielding a sequence of Steiner points $Y_0$, $Y_1$, $Y_2$, $\dots$ and complexes $\Del{X_0}$, $\Del{X_1}$, $\Del{X_2}$, $\dots$

In computational geometry, Delaunay refinement lies at the core of popular algorithms for generating and refining Euclidean meshes; see \cite{cheng2013delaunay} for a modern presentation. 
For instance, Chew's and Ruppert's algorithms \cite{chew1989guaranteed,chew1993guaranteed,ruppert1995delaunay} take a set of input constraints (e.g., boundary segments that must appear as edges) and iteratively improve the mesh by eliminating poor-quality triangles. 
The refinement step consists of marking a bad triangle and inserting its circumcenter as a new vertex, after which the Delaunay triangulation is recomputed.

\begin{figure}[!htbp]
	\begin{subfigure}[t]{1\linewidth}
		\centering
		\includegraphics[width=\linewidth]{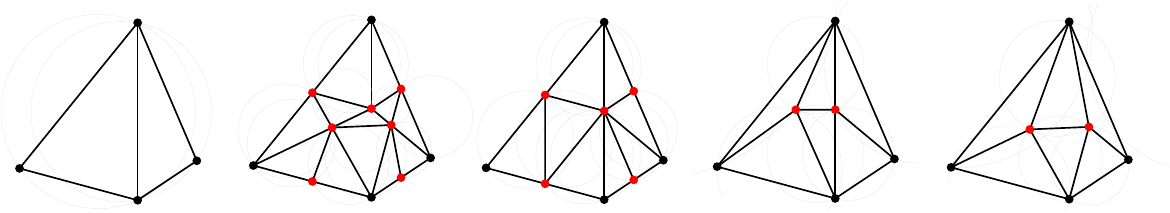}
		\begin{minipage}{0.19\linewidth}
			\centering
			Initial complex
		\end{minipage}
		\begin{minipage}{0.19\linewidth}
			\centering
			Barycentric
		\end{minipage}
		\begin{minipage}{0.19\linewidth}
			\centering
			Edgewise
		\end{minipage}
		\begin{minipage}{0.19\linewidth}
			\centering
			Minicenter
		\end{minipage}
		\begin{minipage}{0.19\linewidth}
			\centering
			Centroid
		\end{minipage}
		\subcaption{%
			Local picture: A Delaunay complex on a set of four points in $\R^2$ and the four refinements proposed.
		}
		\label{subfig:delaunay_refinements:local}
	\end{subfigure}
	\hfill	
	\begin{subfigure}[t]{1\textwidth}
		\centering
		\begin{minipage}{0.19\linewidth}
			\centering
			\includegraphics[width=.99\linewidth]{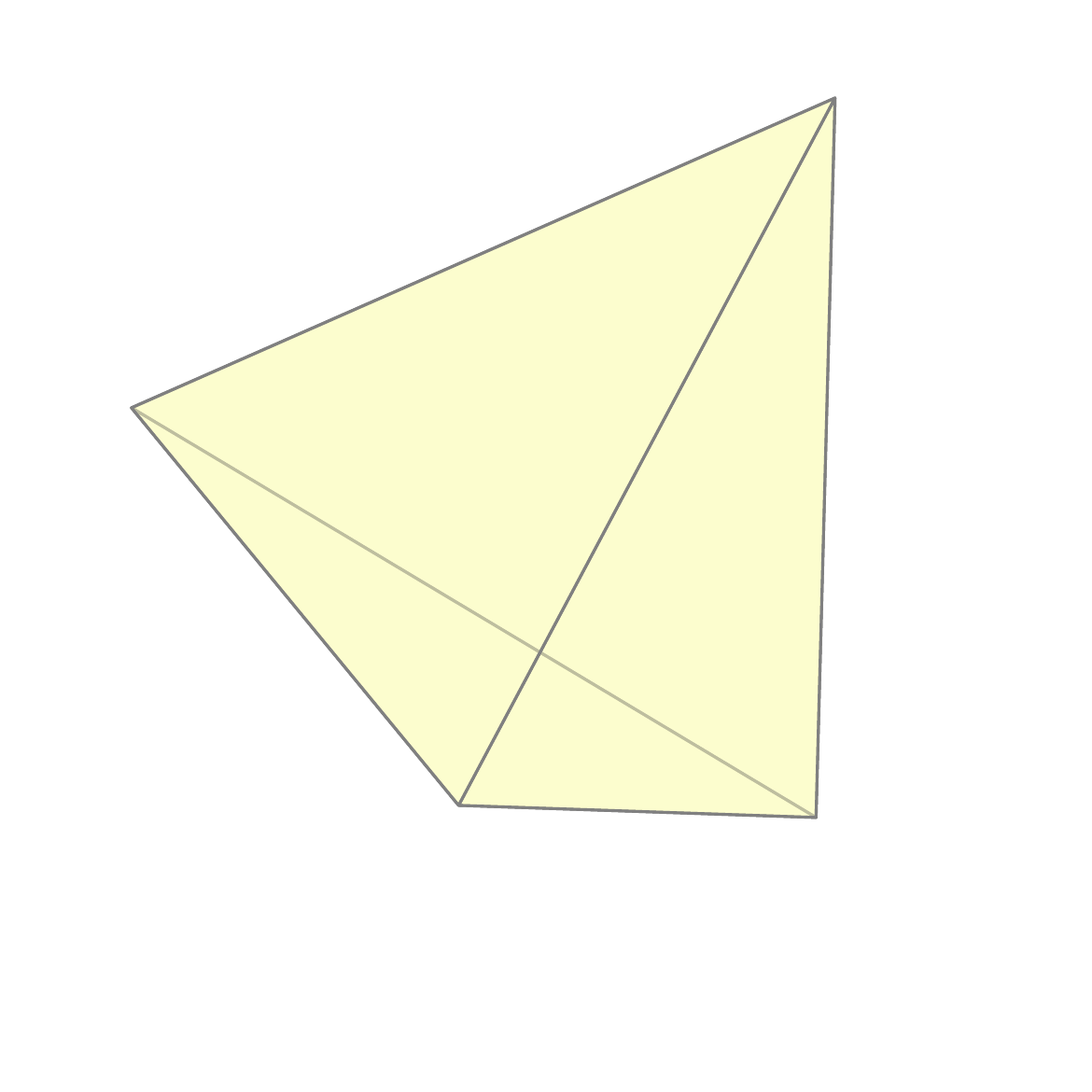}
			Initial complex
		\end{minipage}
		\begin{minipage}{0.19\linewidth}
			\centering
			\includegraphics[width=.99\linewidth]{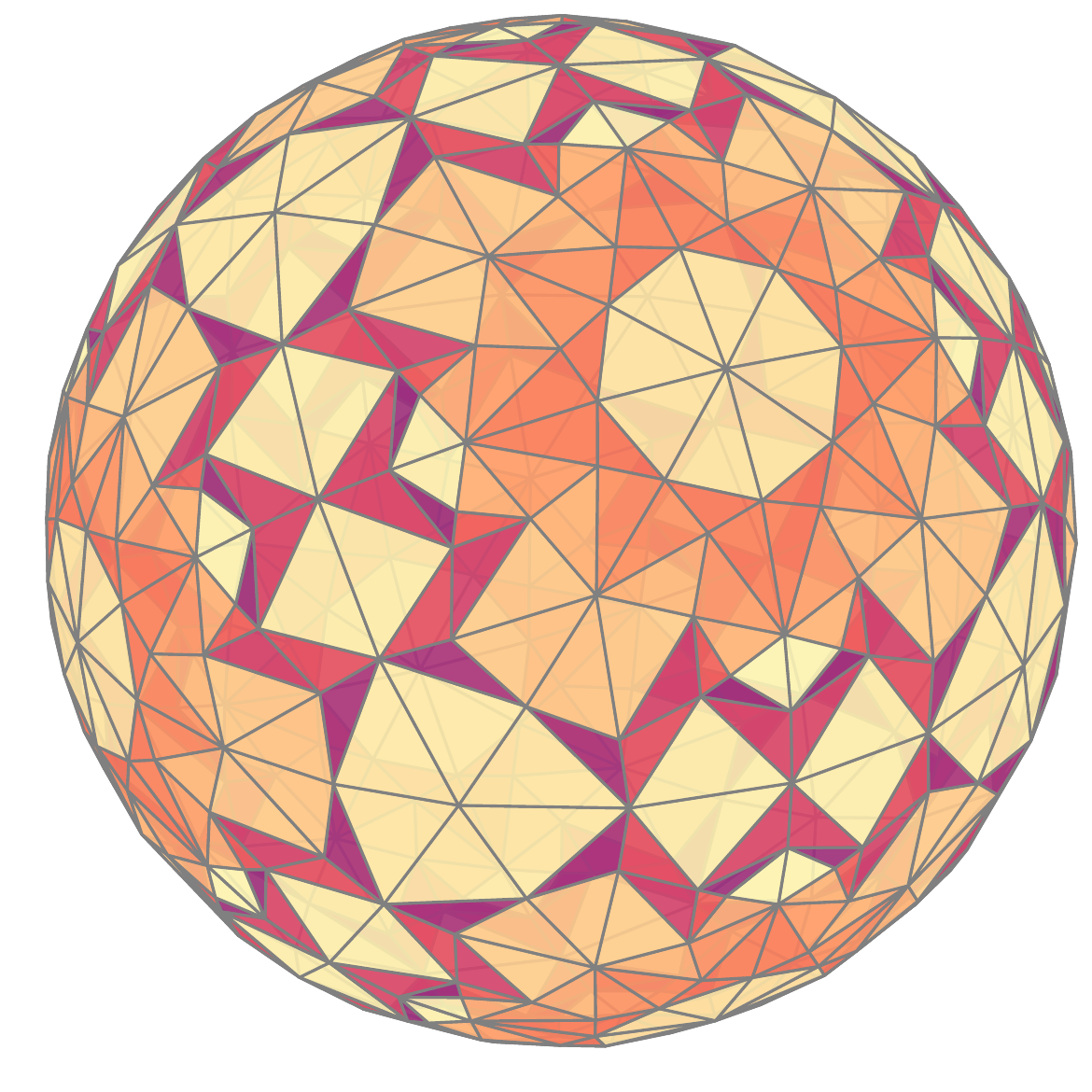}
			Barycentric
		\end{minipage}
		\begin{minipage}{0.19\linewidth}
			\centering
			\includegraphics[width=.99\linewidth]{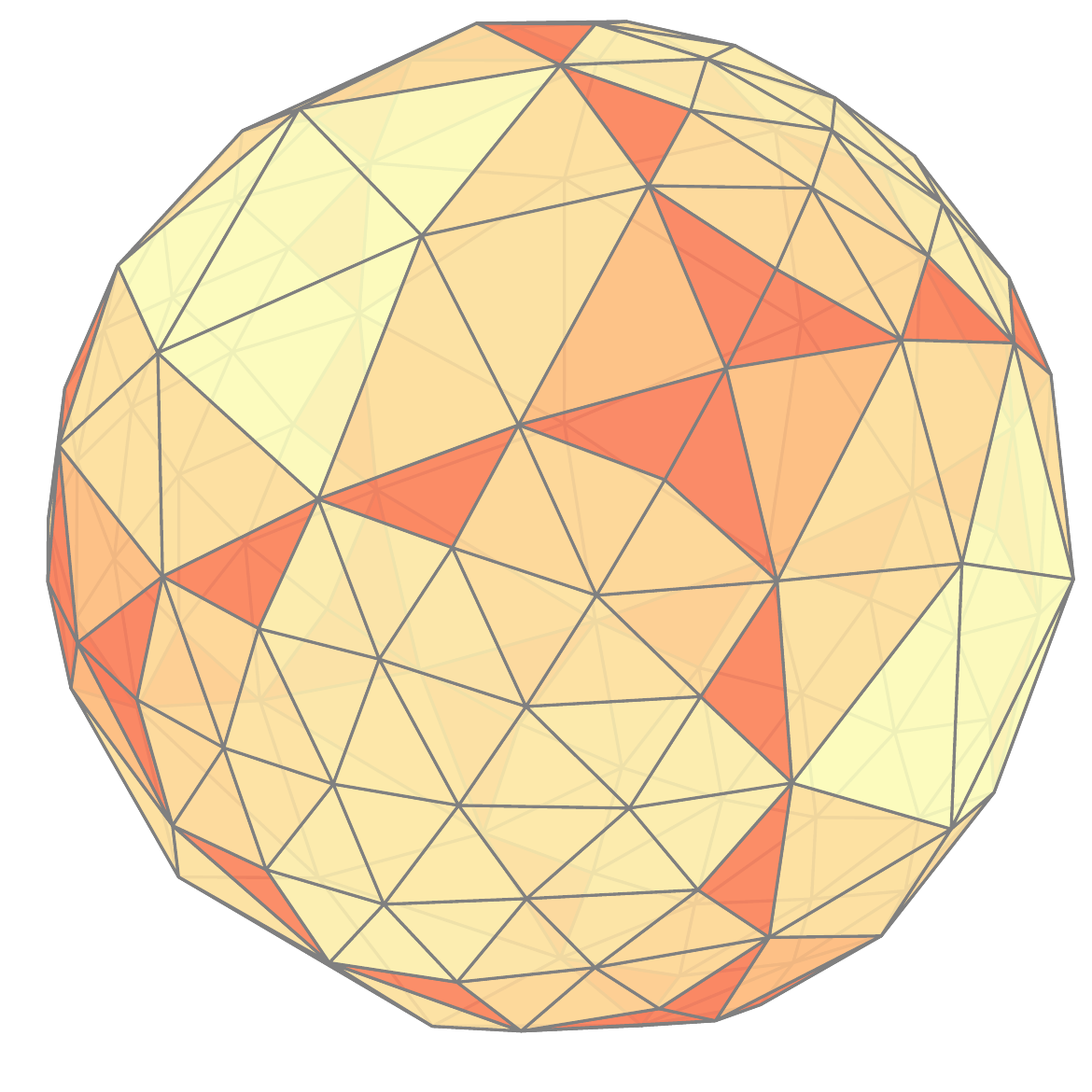}
			Edgewise
		\end{minipage}
		\begin{minipage}{0.19\linewidth}
			\centering
			\includegraphics[width=.99\linewidth]{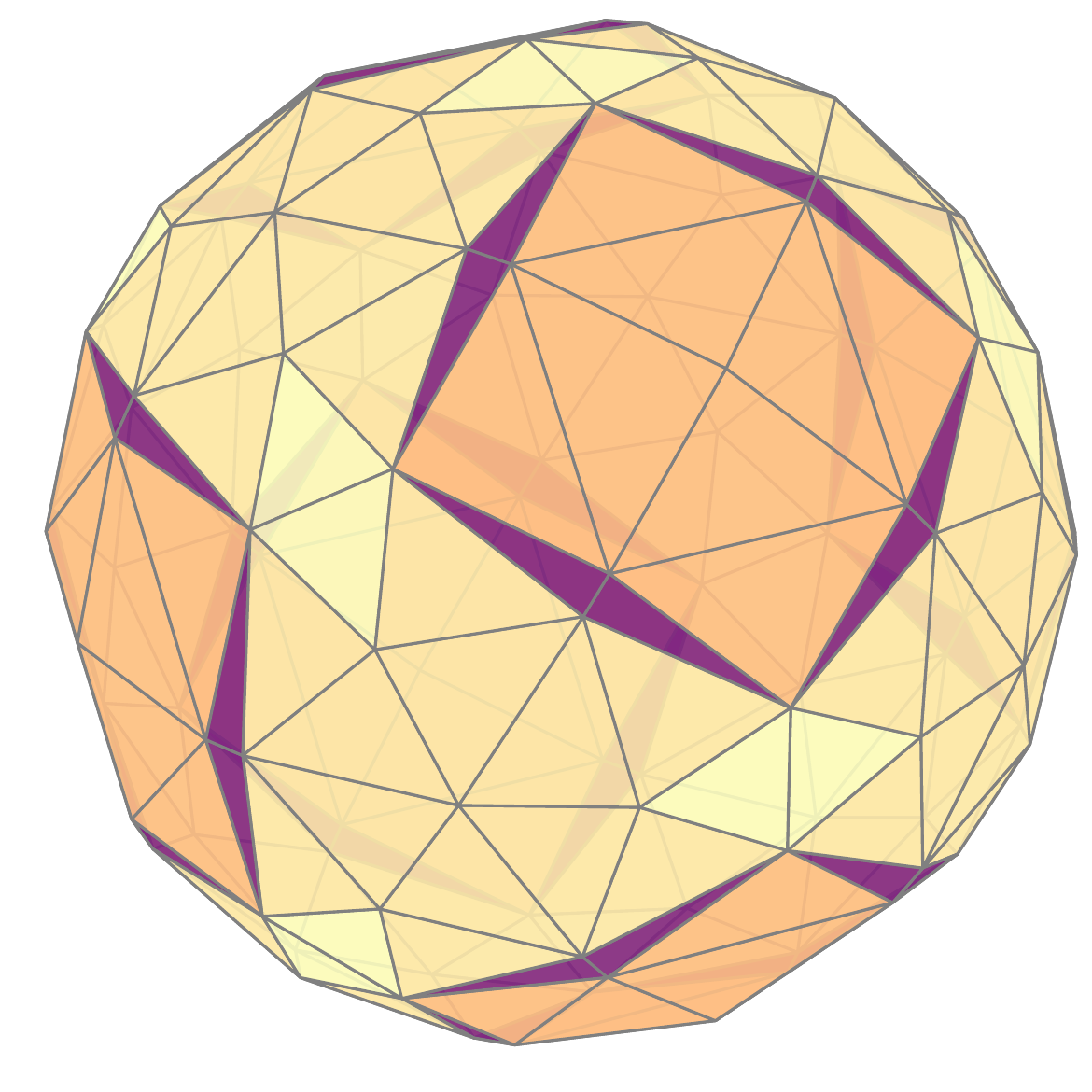}
			Minicenter
		\end{minipage}
		\begin{minipage}{0.19\linewidth}
			\centering
			\includegraphics[width=.99\linewidth]{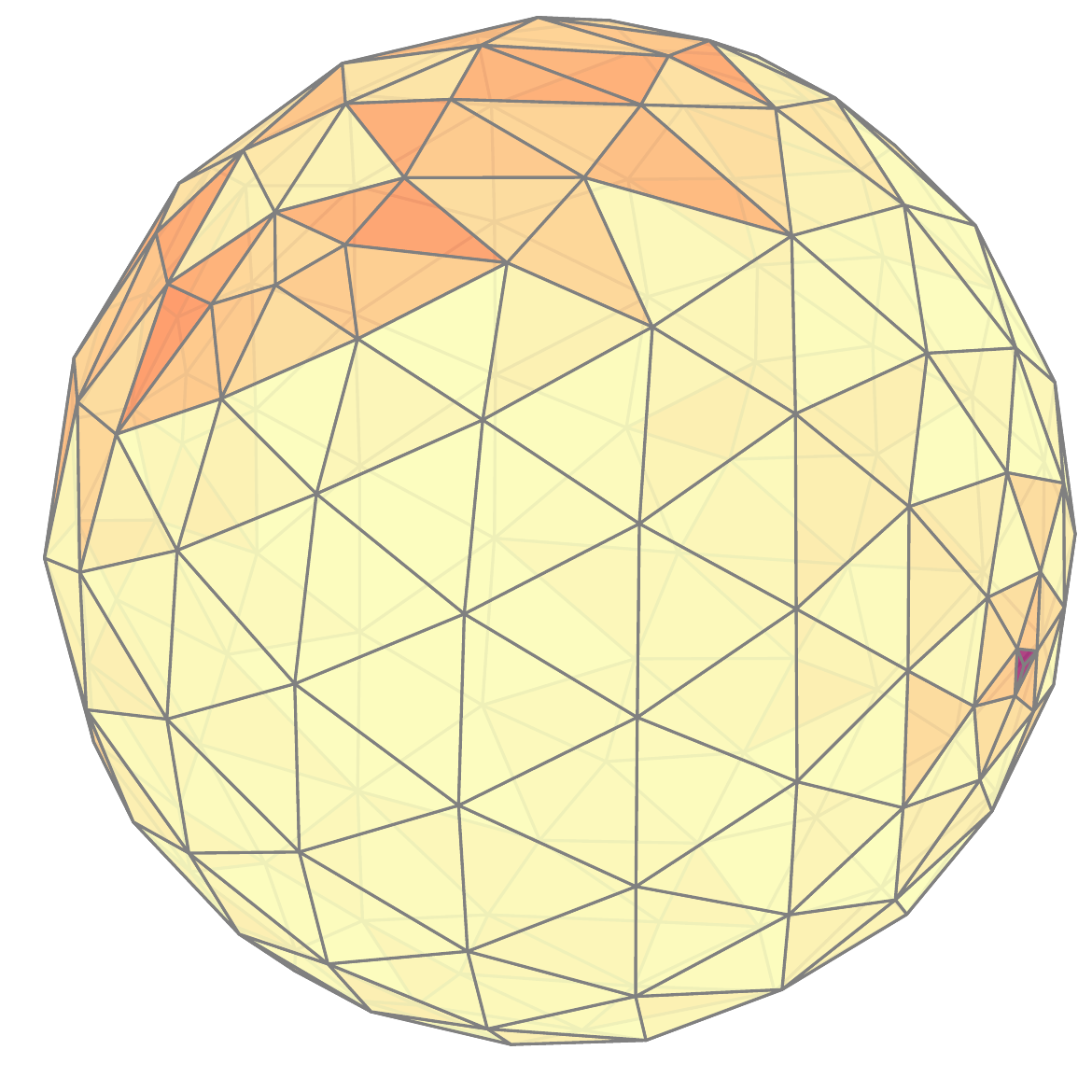}
			Centroid
		\end{minipage}
		\subcaption{%
			Global picture: Spherical Delaunay complexes after three or four refinements of an initial triangulation of $S^2$ (the boundary of the standard simplex).
			The colors indicate the simplex ratio inradius/circumradius.
		}
	\end{subfigure}
	\hfill
	\caption{
		We employ Delaunay refinement as a subdivision scheme for triangulations of $S^d$.
	}
	\label{fig:delaunay_refinements}
\end{figure}

Our objective here is different: we seek triangulations whose simplices can be made arbitrarily small, enabling the simplicial approximation of maps described in \Cref{subsec:overview}.
To this end, we introduce several families of Delaunay-based refinements, visualized in \Cref{fig:delaunay_refinements}. 
\begin{description}
	\item[Barycentric refinement:]
	Inspired by barycentric subdivision, we let the Steiner points $Y_i$ be the barycenters of all simplices of $\Del{X_i}$, except its vertices.
	
	\item[Edgewise refinement:]
	We may instead insert the midpoints of all edges of $\Del{X_i}$, in analogy with the popular Coxeter-Freudenthal-Kuhn subdivisions of simplices, also called edgewise subdivisions \cite{bey2000simplicial,edelsbrunner2000edgewise,gonccalves2007H2,plaza2007eight,kroger2008stability,korotov2014red,ChoudharyKachanovichWintraecken2020,BoissonnatKachanovichWintraeckenCFKCompact,brunck2023iterated,BoissonnatEtAlSimplicialSubdivisionCurved,boissonnat_et_al:LIPIcs.SoCG.2021.17,boissonnat2023tracing}. 
	
	\item[Minicenter refinement:]
	Closer in spirit to conventional Delaunay refinement, we also consider inserting the minicenters of the facets (centers of minimal enclosing balls).
	We favor minicenters over circumcenters because, unlike circumcenters, they always lie inside the simplex that defines them, which better reflects the local nature of classical subdivision.

	\item[Centroid refinement:]
	Finally, we may insert the barycenters of only the \textit{facets} of $\Del{X_i}$ (the maximal simplices), standing as a cheaper alternative to barycentric refinement.
	
\end{description}
In all cases, the \textit{spherical} Steiner points are considered (i.e., spherical barycenters or minicenters). 
This is equivalent to computing their Euclidean counterpart in the ambient space $\R^{d+1}$ and projecting them onto the sphere; see, for instance, Fiedler's book \cite[Section 5.4]{fiedler2011matrices}.

\begin{figure}[!ht]
	\centering
	\includegraphics[width=.55\linewidth]{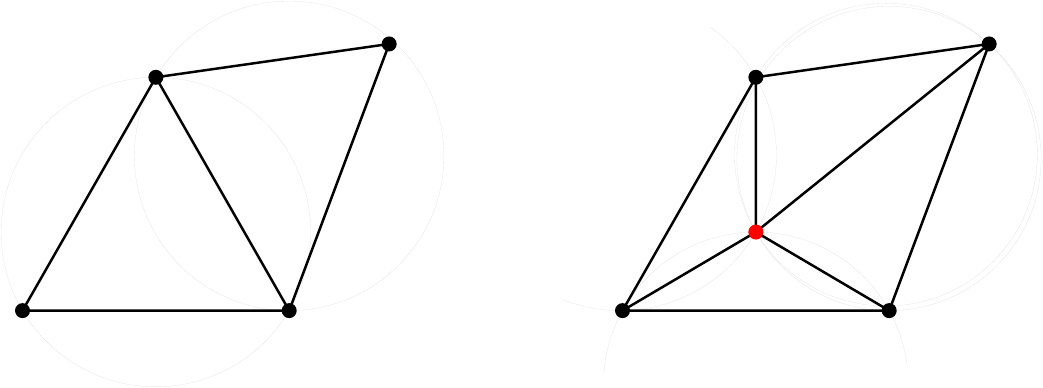}
	\caption{Adding a point to a Delaunay complex may increase the maximal edge length.
	}
	\label{fig:delaunay_length_increase}
\end{figure}

Note that, strictly speaking, $\Del{X_{i+1}}$ is not a subdivision of $\Del{X_i}$: when realized on the sphere, a simplex of $\Del{X_{i+1}}$ need not be contained in a single simplex of $\Del{X_{i}}$.
This can be seen in \Cref{subfig:delaunay_refinements:local}, where only the edgewise and minicenter refinements are subdivisions.
As a consequence, it is not immediate that simplices become smaller under refinement.
In particular, adding a point to a Delaunay triangulation can increase the maximal edge length, as illustrated in \Cref{fig:delaunay_length_increase} (shown in the plane for clarity).
Instead, we show that the \textit{maximal (spherical) circumradius} of $\Del{X_i}$, denoted $\rho_\mathrm{circ}(\Del{X_i})$, decreases monotonically to zero.
Since the maximal diameter of the simplices of $\Del{X_i}$ is at most twice its maximal circumradius, it follows that the maximal diameter also tends to zero.

A convenient quantity to study a Delaunay complex on a subset $X\subset S^d$ is its \textit{covering radius} (also known as sampling radius) \cite{boissonnat2014delaunay,boissonnat2018geometric}.
On the sphere, it is defined by
\[
\rho_\mathrm{cov}(X) = \sup_{y \in S^d} \inf_{x \in X \vphantom{S^d}} \d(x,y),
\]
where $\d(x,y)$ is the geodesic (great-circle) distance on the sphere.
We note that the set $X$ is admissible as long as $\rho_\mathrm{cov}(X)<\pi/2$ (i.e., $X$ is not contained in a closed hemisphere).

The maximal circumradius and covering radius are related by the following standard~result.

\begin{lemma}
	\label{lem:equality_covering_circumradius}
	For every admissible finite subset $X\subset S^d$, it holds that $\rho_\mathrm{circ}(\Del{X}) = \rho_\mathrm{cov}(X)$.
\end{lemma}

We take advantage of this shift to the covering radius to prove the following lemma.

\fullonly{
\begin{lemma}[%
	name={proof p.~\pageref{proof:decreasecoveringradius}},%
	restate=decreasecoveringradius%
	]
	\label{lem:decreasecoveringradius}
	Consider a finite subset $X\subset S^d$ and let $Y$ denote the Steiner points associated with $\Del{X}$.
	Assume that $\rho_\mathrm{cov}(X)\leq \pi/3$. Then
	\[
	\rho_\mathrm{cov}(X\cup Y)\leq \alpha \rho_\mathrm{cov}(X),
	\]
	where $\alpha = \alpha'/\cos(\rho_\mathrm{cov}(X))$, and $\alpha'$ depends on the chosen refinement, as given in the table
	\begin{table}[H]
		\centering
		\begin{tabular}{l|lll}
			Refinement & Edgewise     & Minicenter   & Centroid  \\ \hline
			$\alpha'$  & $1/\sqrt{2}$ & $1/\sqrt{2}$ & $d/(d+1)$
		\end{tabular}
	\end{table}
\end{lemma}
}

\shortonly{
\begin{lemma}
	\label{lem:decreasecoveringradius}
	Consider a finite subset $X\subset S^d$ and let $Y$ denote the Steiner points associated with $\Del{X}$.
	Assume that $\rho_\mathrm{cov}(X)\leq \pi/3$. Then
	\[
	\rho_\mathrm{cov}(X\cup Y)\leq \alpha\rho_\mathrm{cov}(X),
	\]
	where $\alpha = \alpha'/\cos(\rho_\mathrm{cov}(X))$, and $\alpha'$ depends on the chosen refinement, as given in the table
	\begin{table}[H]
		\centering
		\begin{tabular}{l|lll}
			Refinement & Edgewise     & Minicenter   & Centroid  \\ \hline
			$\alpha'$  & $1/\sqrt{2}$ & $1/\sqrt{2}$ & $d/(d+1)$
		\end{tabular}
	\end{table}
\end{lemma}
}

The quantity $\cos(\rho_\mathrm{cov}(X))$ reflects the spherical distortion of lengths; it goes to 1 as the covering radius goes to zero.
A similar lemma can be obtained in the Euclidean case, without this factor.
We point out that we have not been able to obtain a satisfactory bound for barycentric refinement; for lack of a better estimate, we use the bound for edgewise~refinement.

By iterating this lemma, we obtain our main result on Delaunay refinement.

\begin{theorem}
	\label{th:shrinkingrefinements}
	Assume the initial sample $X_0\subset S^d$ is sufficiently dense so that $\alpha < 1$, where $\alpha=\alpha(X_0)$ is defined in \Cref{lem:decreasecoveringradius}.
	Then the $n^\mathrm{th}$ iteration of Delaunay refinement satisfies
	\[
	\rho_\mathrm{cov}(X_n) \leq \alpha^n \rho_\mathrm{cov}(X_0).
	\]
	In particular, the maximal diameter of simplices of $\Del{X_n}$ goes to zero.
\end{theorem}

\begin{remark}
	This result highlights a notable distinction between Delaunay refinement and standard subdivision. 
	While barycentric refinement reduces diameters asymptotically by $1/\sqrt{2}$ (at most), the best bound for barycentric subdivision is $d/(d+1)$.	
	Likewise, centroid refinement reduces them by $d/(d+1)$, even though it introduces only one vertex per facet.
\end{remark}

\fullonly{
	\paragraph*{Simplex quality}
	
	A central goal in mesh generation is to produce triangulations with well-shaped (non-flat) simplices, since this typically improves the performance of subsequent algorithms. 
	This is also our aim, which is why we work with Delaunay complexes. Indeed, Delaunay triangulations enjoy several optimality properties: in $\R^2$ they maximize the minimum angle over all triangulations of a fixed point cloud \cite{sibson1978locally}; in $\R^n$ they minimize the maximal miniradius of the simplices \cite{rajan1991optimality,Rajan1994}; and they also minimize a certain weighted sum of edge lengths \cite{musin1997properties}.
	
	To encourage large, well-shaped simplices, we start from ``evenly distributed'' samples $X\subset S^d$.
	A convenient seed is given by the standard simplex $\Delta^{d+1}$ embedded in the unit ball of $\R^{d+1}$.
	Its vertices $v_0, \dots, v_{d+1}$ are regularly spaced on $S^d$, with pairwise Euclidean distance $\sqrt{2(d+2)/(d+1)}$.
	If more points are required, we rely on approximate evenly spaced configurations.
	On $S^2$, we draw points on the Fibonacci lattice \cite{gonzalez2010measurement}. 
	In higher dimensions, we use configurations obtained by solving the Thomson problem---placing points so as to minimize their total Coulomb energy---which yields well-distributed samples \cite{saff1997distributing}.
		
	However, in dimension three and higher, evenly spaced samples do not suffice to ensure high-quality simplices \cite{dey1991good,ruppert1995delaunay}.
	Famously, near-cocircular 4-tuples of points may have a large circumradius but form a tetrahedron with almost zero volume; they are known as \textit{slivers} \cite{cheng2000sliver,alliez2005variational}.
	We have observed a remarkably simple example of degeneracy: starting from the boundary of the standard simplex $\partial \Delta^{5}$ embedded in $S^4$, adding the midpoints of its edges already yields slivers.
	In our implementation, each construction of a Delaunay complex is followed by a quality check to ensure the simplex volumes are not too small; if such a configuration appears, the algorithm is relaunched with a different refinement method.
	
	We stress that the relationship between simplex quality and the output of our main algorithm, presented in \Cref{subsec:full_algorithm}, is somewhat indirect. 
	Indeed, the algorithm yields an abstract simplicial complex, not embedded in Euclidean space.
	A geometric realization arises only locally, through the charts provided by the cells of the underlying CW structure. 
	Accordingly, simplex quality should be understood within these local realizations.
}

\section{Simplicial mapping cones with staircase triangulations}
\label{sec:simplicial_mapping_cones}

In this section, we assume that a simplicial approximation $g\colon K\to L$ of $f\colon S^{d}\to\geomreal{L}$ is given.
We build a simplicial mapping cone $\conesimp{g}$ and show it is homotopy equivalent to the usual mapping cone $\cone{f}$.
The construction proceeds by building a simplicial ball $B(K)$ through staircase triangulation, which we first recall.
Although this construction already appears in the literature, notably in \cite[Exercise~E, p.~151]{Spanier_1981} and \cite[Proposition~2]{Bell_2004}, we include a detailed account here in order to collect the properties that will be used in the sequel.

\subsection{Triangulation of the ball}\label{subsec:triangulation_ball}

\paragraph*{Staircase triangulation of the prism}

To triangulate the Cartesian product of a $d$-simplex $\sigma$ with an interval $I = [0,1]$, the \textit{staircase triangulation} consists of first ordering the vertices $\{v_0,\dots,v_d\}$ of $\sigma$, taking a copy $\{v_0',\dots,v_d'\}$, and inserting the simplices $\sigma_k = [v_k,\dots,v_d,v_0',\dots,v_k']$ for all $k\in \{ 0,\dots,d \}$ \cite{lee2017subdivisions}.
We shall refer to the product $\geomreal{\sigma}\times I$ as a \emph{prism}, and its face $[v_k,\dots,v_d]$ (resp.\ $[v_0',\dots,v_k']$) as the \emph{inner face} (resp.\ the \emph{outer face}). 
Geometrically, they correspond to $\geomreal{\sigma}\times\{0\}$ and $\geomreal{\sigma}\times\{1\}\subset\geomreal{\sigma}\times I$.

\begin{figure}[!htbp]
	\centering
	\subcaptionbox{
		The square $\geomreal{\Delta^1}\times I$  is triangulated with two triangles
		\label{fig:triangulation_prism_1}
	}[0.32\textwidth]{%
		\includegraphics[width=.65\linewidth]{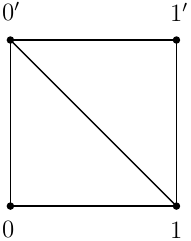}}
	\hfill
	\subcaptionbox{
		The prism $\geomreal{\Delta^2}\times I$  is triangulated with three tetrahedra.
		\label{fig:triangulation_prism_2}
	}[0.32\textwidth]{%
		\includegraphics[width=.71\linewidth]{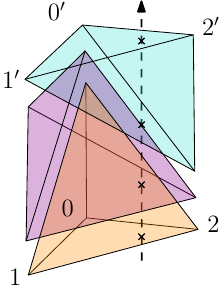}}
	\hfill
	\subcaptionbox{
		Horizontal sections inherit a subdivision into polygonal cells.
		\label{fig:triangulation_prism_3}
	}[0.32\textwidth]{%
		\includegraphics[width=.7\linewidth]{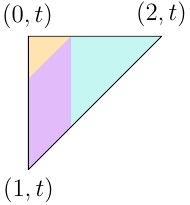}}
	\caption{
		A triangulation of the prism $\geomreal{\sigma}\times I$ is obtained by ordering the vertices $\{v_0,\dots,v_d\}$ of $\sigma$, taking a copy $\{v_0',\dots,v_d'\}$, and inserting the simplices $\sigma_k=[v_k,\dots,v_d,v_0',\dots,v_k']$ for $0\leq k \leq d$.
	}
	\label{fig:triangulation_prism}
\end{figure}

The relation $\sigma_k < \sigma_{k+1}$ defines an order on the prism's $d+1$ facets.
In particular, the following observation will be useful later: \textit{vertical} straight lines in the prism---i.e., of the form $t\mapsto (x,t)$ for a certain $x \in \geomreal{\sigma}$---cross each facet consecutively.
Indeed, for $k\in \{ 1,\dots,d-1\}$, the only neighboring facets of $\sigma_k$ are $\sigma_{k-1}$ and $\sigma_{k+1}$; see \Cref{fig:triangulation_prism_1,fig:triangulation_prism_2}.
On the other hand, \textit{horizontal} sections of the prism---i.e., sets $\geomreal{\sigma}\times\{t\}$ for $t \in [0,1]$---inherit subdivisions in polyhedral cells; see \Cref{fig:triangulation_prism_3}.
These are closely related to \textit{mixed subdivisions} \cite{lee2017subdivisions,MR2134766}.

\paragraph*{Filling the sphere}

Let $K$ be an admissible triangulation of the sphere $S^d\subset \mathbb{R}^{d+1}$ (a convex hull of unit vectors that contains the origin).
We obtain a triangulation of the unit ball $B^{d+1}$ in three steps.
\begin{itemize}
	\item 
	The polyhedron $\geomreal{K}$ is embedded in $\R^{d+1}$ and called the \emph{outer layer}.
	We prime its vertex labels ($0'$, $1'$, \textit{etc.}).
	Besides, a copy of $\geomreal{K}$ is taken and scaled by a factor $\rho_\mathrm{inner}\in (0,1)$; it is seen as a triangulation of the sphere of radius $\rho_\mathrm{inner}$, referred to as the \emph{inner layer}.
	\item 
	Each simplex of the outer layer is connected to the corresponding inner-layer simplex via staircase triangulation, forming a prism.
	Together, these prisms yield a triangulation of the spherical shell of radii $(\rho_\mathrm{inner},1)$, which we call the \emph{outer shell}.
	\item 
	Last, the origin is added to the triangulation as a new vertex, over which the inner layer is coned, forming the \emph{inner ball}.
\end{itemize}
The resulting geometric simplicial complex, denoted $B(K)$, is a triangulation of the unit ball.

\begin{figure}[!htbp]
	\centering
	\subcaptionbox{
		$0\to1\to2\to3\to4\to5\to0$
	}[0.32\textwidth]{%
		\includegraphics[width=.75\linewidth]{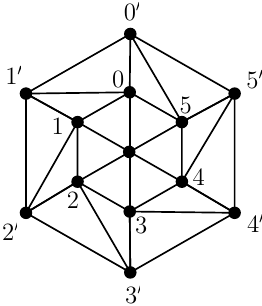}}
	\hfill
	\subcaptionbox{
		$5\to4\to3\to2\to1\to0\to5$
	}[0.32\textwidth]{%
		\includegraphics[width=.75\linewidth]{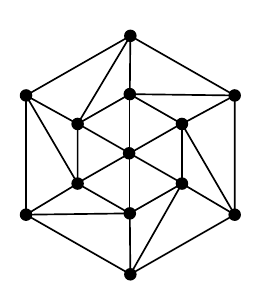}}
	\hfill
		\subcaptionbox{
		$0\to1\leftarrow2\to3\leftarrow4\to5\leftarrow0$
	}[0.32\textwidth]{%
		\includegraphics[width=.75\linewidth]{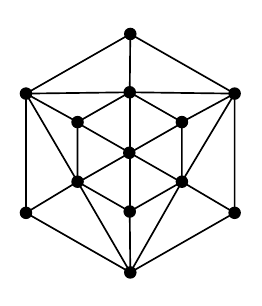}}
	\hfill
	\caption{
		The simplicial ball $B(K)$ built from $K$ depends on an orientation of the edges of $K$.
	}
	\label{fig:triangulation_ball}
\end{figure}

In our implementation, we chose $\rho_\mathrm{inner}=1/2$.
Besides, because the construction uses staircase triangulations, it depends on a choice of ordering of the vertices in each facet of $K$.
These orderings must be compatible across neighboring facets.
Equivalently, the construction requires an ``orientation'' on $K$, by which we mean an orientation of its edges such that no facet contains a directed cycle.
Different choices may lead to different $B(K)$; see \Cref{fig:triangulation_ball}.

Each facet $\sigma \in K$ generates a \textit{sector}, defined as the subcomplex $\sect{\sigma}\subset B(K)$ containing the origin, its cone with $\sigma$ seen in the inner layer, and the prism built on it.
Most of the constructions to follow will be carried out sector by sector.

\paragraph*{Radial normalization}

A number of natural homeomorphisms $\geomreal{B(K)}\to B^{d+1}$ exist.
For instance, one could lift the vertices of $B(K)$ to the upper $(d+1)$-hemisphere of $\R^{d+2}\supset\R^{d+1}\times\{0\}$ via orthographic or stereographic projection, build geodesic simplices, and take them back to $B^{d+1}\subset\R^{d+1}$.
However, we found that the idea of \emph{radial normalization} was more appropriate for our~problem.

As for any convex domain, the \emph{gauge function} of $\geomreal{B(K)}$ is defined for all $x\in \R^{d+1}$ as 
\[
J(x) = \inf\{t>0\mid t^{-1} x \in \geomreal{B(K)}\}.
\]
It is a convex, positively homogeneous function.
The reciprocal of the gauge is known as the \emph{radial function}, used in the study of star-shaped sets \cite{hansen2020starshaped}.
It can be written as
\[
J(x)^{-1} = \sup\{t>0\mid t x \in \geomreal{B(K)}\}.
\]
If $x$ is a unit vector, then $J(x)^{-1}$ is the length of the part of $[0,x]$ contained in $\geomreal{B(K)}$.
In particular, the minimum of $J(x)^{-1}$ over the unit sphere is equal to the polyhedron's inradius.

\begin{figure}[!htbp]
	\centering
	\subcaptionbox{
		The polyhedron $B(K)$ seen in $\R^{d+1}$.
		\label{fig:apothem_normalization_1}
		}[0.49\textwidth]{%
			\includegraphics[width=.7\linewidth]{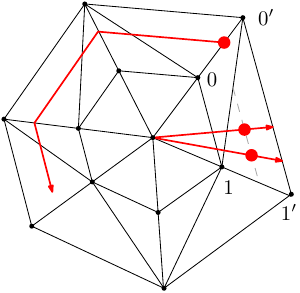}
		}
	\hfill
	\subcaptionbox{
		Its image, the Euclidean ball $B^{d+1}\subset\R^{d+1}$.
		}[0.49\textwidth]{%
			\includegraphics[width=.7\linewidth]{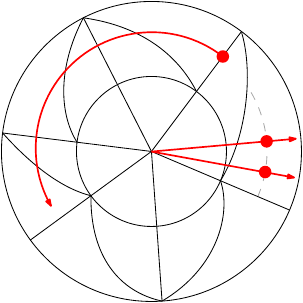}
		}
	\caption{
		Radial normalization yields a homeomorphism $\nu\colon\geomreal{B(K)}\to B^{d+1}$.
		}
	\label{fig:apothem_normalization}
\end{figure}

Our preferred homeomorphism $\nu\colon\geomreal{B(K)}\to B^{d+1}$ is the \textit{radial normalization}, defined as
\[
\nu(x) = J\bigg(\frac{x}{\|x\|}\bigg)x,
\]
with inverse $\nu^{-1}(x) = J(x/\|x\|)^{-1}x$.
Although not explicitly written, $\nu$ depends on $\geomreal{B(K)}$.
One of its main advantages is its simple geometric behavior, illustrated in \Cref{fig:apothem_normalization}.

\fullonly{
\begin{lemma}[%
	name={proof p.~\pageref{proof:apothemnormalization}},%
	restate=apothemnormalization%
	]
	\label{lem:apothemnormalization}
	Under inverse radial normalization $\nu^{-1}\colon B^{d+1}\to \geomreal{B(K)}$,
	\begin{itemize}
		\item 
		rays through the origin are mapped to rays through the origin;
		\item 
		circular arcs---i.e., intersections of planes with spheres centered at the origin---are mapped to paths which are linear in each sector $\geomreal{\sect{\sigma}}\subset\geomreal{B(K)}$ and parallel to $\geomreal{\sigma}$.
	\end{itemize}
\end{lemma}
}

\shortonly{
\begin{lemma}
	\label{lem:apothemnormalization}
	Under inverse radial normalization $\nu^{-1}\colon B^{d+1}\to \geomreal{B(K)}$,
	\begin{itemize}
		\item 
		rays through the origin are mapped to rays through the origin;
		\item 
		circular arcs---i.e., intersections of linear planes with spheres centered at the origin---are mapped to paths which are linear in each sector $\geomreal{\sect{\sigma}}\subset\geomreal{B(K)}$ and parallel to $\geomreal{\sigma}$.
	\end{itemize}
\end{lemma}
}

\fullonly{
	From an algorithmic viewpoint, the gauge function $J(x)$ is obtained as a byproduct of our implementation, requiring no further computations.
	Indeed, to locate a point $x\in B^{d+1}$ in $B(K)$, we first locate $x/\|x\|$ on $K$, thus identifying the sector containing $x$ and its normalization $\nu^{-1}(x)$, then find the facet of $B(K)$ to which it belongs by parsing the sector.
}

\subsection{Homotopy equivalence between the mapping cones}
\label{subsec:equivalence_mapping_cones}

\begin{figure}[!htbp]
	\centering
	\subcaptionbox{		
		$g\colon K\to L$
		}[0.24\textwidth]{%
		\includegraphics[width=.99\linewidth]{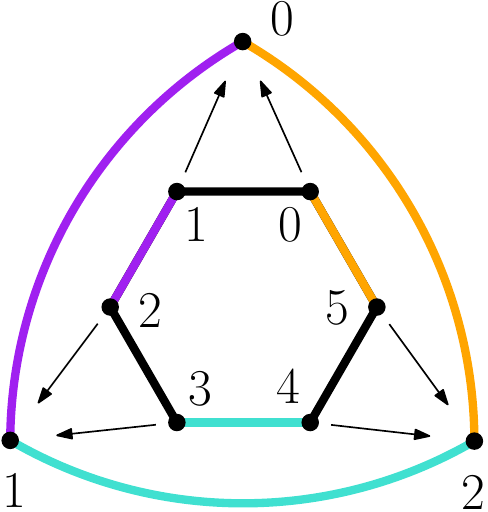}
	}
	\hfill
	\subcaptionbox{
		$\geomreal{L}\cup_f B^{d+1}$
		}[0.24\textwidth]{%
		\includegraphics[width=.99\linewidth]{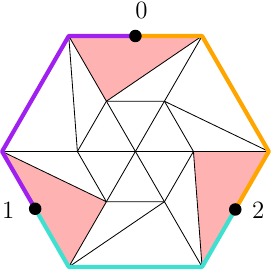}
	}
	\hfill
	\subcaptionbox{
		$\geomreal{L}\cup_{\geomreal{g}}\geomreal{B(K)}$
		\label{fig:mapping_cones_3}
		}[0.24\textwidth]{%
			\includegraphics[width=.99\linewidth]{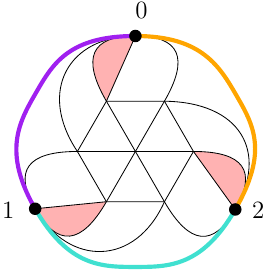}
	}
	\subcaptionbox{
		$L \cup_{g} B(K)$
		\label{fig:mapping_cones_4}
		}[0.24\textwidth]{%
			\includegraphics[width=.99\linewidth]{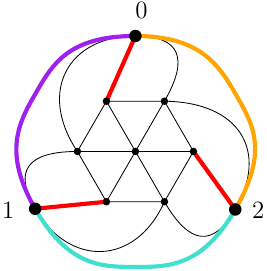}
	}
	\caption{
		The mapping cones involved in our problem. Here, $f\colon \geomreal{K}\to \geomreal{L}$ is the identity between two triangulations of $S^1$ on 6 and 3 vertices, and $g$ is the simplicial map $0,1\mapsto 0$; $2,3\mapsto 1$; $4,5\mapsto 2$.
	}
	\label{fig:mapping_cones}
\end{figure}

We still consider a simplicial approximation $g\colon K\to L$ to $f\colon S^d\to \geomreal{L}$.
In the previous section we built a simplicial ball and a homeomorphism $B^{d+1}\to\geomreal{B(K)}$.
We now face three distinct gluings, represented in \Cref{fig:mapping_cones}, which we will show are homotopy equivalent:
\[
\geomreal{L}\cup_f B^{d+1}
\longrightarrow
\geomreal{L}\cup_{\geomreal{g}} B^{d+1}
\longrightarrow
\geomreal{L\cup_g B(K)}.
\]
The first two are the (standard) mapping cones of $f$ and $\geomreal{g}$, also denoted $\cone{f}$ and $\cone{\geomreal{g}}$.
The latter is the \textit{simplicial mapping cone} of $g$, also denoted $\conesimp{g}$.
We define it as the quotient of $B(K)\sqcup L$ by the relation $v\sim g(v)$ for all vertices $v$ in the outer layer $K\subset B(K)$.

First, it is a standard fact that mapping cones built from homotopic maps are homotopy equivalent \cite[Proposition 0.18]{hatcher_algebraic_2002}.
More precisely, since our domains are balls, an explicit homotopy equivalence $\geomreal{L}\cup_{f}B^{d+1}\to \geomreal{L}\cup_{\geomreal{g}}B^{d+1}$ is given by
\begin{equation}
	x \longmapsto
	\begin{cases}
		x                                         & \text{if }~ x \in \geomreal{L},                                              \\
		x/\rho_\mathrm{hom}                                    & \text{if }~ x \in B^{d+1} ~\text{ and }~ \|x\|< \rho_\mathrm{hom}, \\
		H\big(x/\|x\|,~(\|x\|-\rho_\mathrm{hom})/(1-\rho_\mathrm{hom})\big) & \text{if }~ x \in B^{d+1} ~\text{ and }~ \|x\|\geq \rho_\mathrm{hom}.
	\end{cases}
	\label{eq:homotopy_equivalence_cones}
\end{equation}
where $H$ is a homotopy between $\geomreal{g}$ and $f$, and $\rho_\mathrm{hom}\in(0,1)$ is a parameter, chosen as $0.9$ in our implementation.
Visually, the homotopy is performed by scaling the ball $B^{d+1}$ by a factor $1/\rho_\mathrm{hom}$ and by using the outer shell $\{x \mid \|x\| \geq \rho_\mathrm{hom}\}$ to interpolate between $\geomreal{g}$ and~$f$.

Second, to compare the remaining two gluings, we can simply use the quotient map
\[
q\colon \geomreal{L}\cup_{\geomreal{g}}\geomreal{B(K)} \to \geomreal{L\cup_g B(K)}.
\]
It has the effect of collapsing simplices on which $g$ is not injective; compare \Cref{fig:mapping_cones_3,fig:mapping_cones_4}.

\fullonly{
\begin{proposition}[%
	name={proof p.~\pageref{proof:equivalencemappingcones}},%
	restate=equivalencemappingcones%
	]
	\label{prop:equivalencemappingcones}
	The quotient map $q$ is a homotopy equivalence.
\end{proposition}
}

\fullonly{
	Equipped with this explicit homotopy equivalence, the simplicial mapping cone $\geomreal{L\cup_g B(K)}$ can be endowed with a point location routine, provided it has already been implemented in $L$ and $B(K)$.
	Given a point in $\cone{f}=\geomreal{L}\cup_f B^{d+1}$, we first compute its image in $\cone{\geomreal{g}}=\geomreal{L}\cup_{\geomreal{g}}B^{d+1}$ via \Cref{eq:homotopy_equivalence_cones}.
	If it lies inside the ball, we locate it in $B(K)$, and return its carrier and barycentric coordinates, after potentially identifying vertices under $g$.
	If the image lies on the boundary $\geomreal{L}$, we apply the point location routine for $L$.
}

\shortonly{
\begin{proposition}
	\label{prop:equivalencemappingcones}
	The quotient map $q$ is a homotopy equivalence.
\end{proposition}
}

We close this section with a result that will help us navigate the mapping cone.

\fullonly{
\begin{lemma}[%
	name={proof p.~\pageref{proof:raydescendstoquotient}},%
	restate=raydescendstoquotient%
	]
	\label{lem:raydescendstoquotient}
	Let $x\in\geomreal{B(K)}$.
	Under the projection $p\colon\geomreal{B(K)}\to\geomreal{L\cup_g B(K)}$, the image of the partial ray $\{tx\mid 0\leq t\leq 1/\|x\|\}$ only depends on the image of $x$.
\end{lemma}
}

\shortonly{
\begin{lemma}
	\label{lem:raydescendstoquotient}
	Let $x\in\geomreal{B(K)}$.
	Under the projection $p\colon\geomreal{B(K)}\to\geomreal{L\cup_g B(K)}$, the image of the partial ray $\{tx\mid 0\leq t\leq 1/\|x\|\}$ only depends on the image of $x$.
\end{lemma}
}

\fullonly{
An illustration of \Cref{lem:raydescendstoquotient} is given in \Cref{fig:apothem_normalization_1}.
Suppose the vertices $0'$ and $1'$ are identified in the quotient.
In this case, all points on the dashed grey line are identified.
Accordingly, the corresponding partial ray segments through these points coincide in the quotient.
Note that this need not hold for the initial portions of the rays emanating from the origin.
Indeed, the radial rays intersect the edge $[1,0']$ at different points.
}

\section{Simplicial approximation as a list homomorphism problem}
\label{sec:approximation}

The final ingredient in our construction is a more efficient simplicial approximation. 
We consider a continuous map $f \colon S^d \to \geomreal{L}$ and a triangulation $K$ of $S^d$.
For each vertex $v$ of $K$, the point $f(v)$ lies in the geometric realization of a unique simplex of $L$, called its \emph{carrier} and denoted $\carr{f(v)}$.
We seek a simplicial map $g\colon K\to L$ such that $g(v)\in\carr{f(v)}$ for all vertices.
The existence of such a map is a purely combinatorial question that we address as a constraint satisfaction problem.
A further issue is the existence of a homotopy $H$ between $f$ and $\geomreal{g}$; we formulate the problem in the framework of locally equiconnected spaces.

\subsection{Constructing the homotopy}
\label{subsec:lec}

A practical framework for explicitly constructing a homotopy is provided by the theory of \textit{locally equiconnected spaces} (LEC) introduced by Dugundji in 1965 \cite{dugundji_locally_1965}.
Namely, $Y$ is LEC if there exists a neighborhood $U\subset Y\times Y$ of the diagonal and a continuous map $\Pi\colon U \times I \to Y$ such that $\Pi(x,y,0)=x$, $\Pi(x,y,1)=y$ and $\Pi(x,x,t)=x$ for all $(x,y,t) \in U\times I$.
The map $\Pi$ is called an \emph{equiconnecting map}, and the pair $(U,\Pi)$ is called \emph{LEC-data}.

In this section, we aim to build LEC-data for simplicial mapping cones $Y=\conesimp{g}$ on simplicial maps $g\colon K\to L$.
Dyer and Eilenberg \cite{dyer_adjunction_1972} have shown how to build LEC-data of \textit{standard} mapping cones $\cone{\geomreal{g}}$, provided that both $\geomreal{K}$ and $\geomreal{L}$ are LEC.
Their construction, however, does not descend to the quotient $\conesimp{g}$.
We present here a closely related construction, adapted to simplicial mapping cones, in the particular case where $K$ is a sphere.

In general, it is too much to expect an equiconnecting map on $Y$: we are able to build one only when $g\colon K\to L$ is \textit{2-distance injective}, i.e., injective when restricted to the closed star of each vertex.
In the language of graphs, this is equivalent to saying that $g$ is a 2-distance coloring of the 1-skeleton of $K$ \cite{kramer2008survey,brevsar2023injective}.
Without this hypothesis, we instead construct a \emph{local motion planner} \cite{farber2003topological}, i.e., a continuous map $\Pi \colon U \times I \to Y$, where $U \subset Y \times Y$ is a neighborhood of the diagonal,  satisfying $\Pi(x,y,0) = x$ and $\Pi(x,y,1) = y$ for all $(x,y) \in U$.
In contrast with equiconnecting maps, the path $t \mapsto \Pi(x,x,t)$ need not be constant. 

\fullonly{
\begin{theorem}[%
	name={proof p.~\pageref{proof:lec_data_gluing}},%
	restate=lecdatagluing%
	]
	\label{th:lec_data_gluing}
	If $g\colon K\to L$ is 2-distance injective and $\geomreal{L}$ is endowed with LEC-data, then $\geomreal{\conesimp{g}}$ also admits LEC-data $(U,\Pi)$.
	When $g$ is not 2-distance injective, the same result holds for local motion planners instead of equiconnecting maps.	
\end{theorem}
}

\shortonly{
\begin{theorem}
	\label{th:lec_data_gluing}
	If $g\colon K\to L$ is 2-distance injective and $\geomreal{L}$ is endowed with LEC-data, then $\geomreal{\conesimp{g}}$ also admits LEC-data $(U,\Pi)$.
	When $g$ is not 2-distance injective, the same result holds for local motion planners instead of equiconnecting maps.	
\end{theorem}
}

\begin{figure}[!htbp]
	\centering
	\subcaptionbox{
		Points are connected through a combination of elementary paths.
	}[0.32\textwidth]{%
		\includegraphics[width=.99\linewidth]{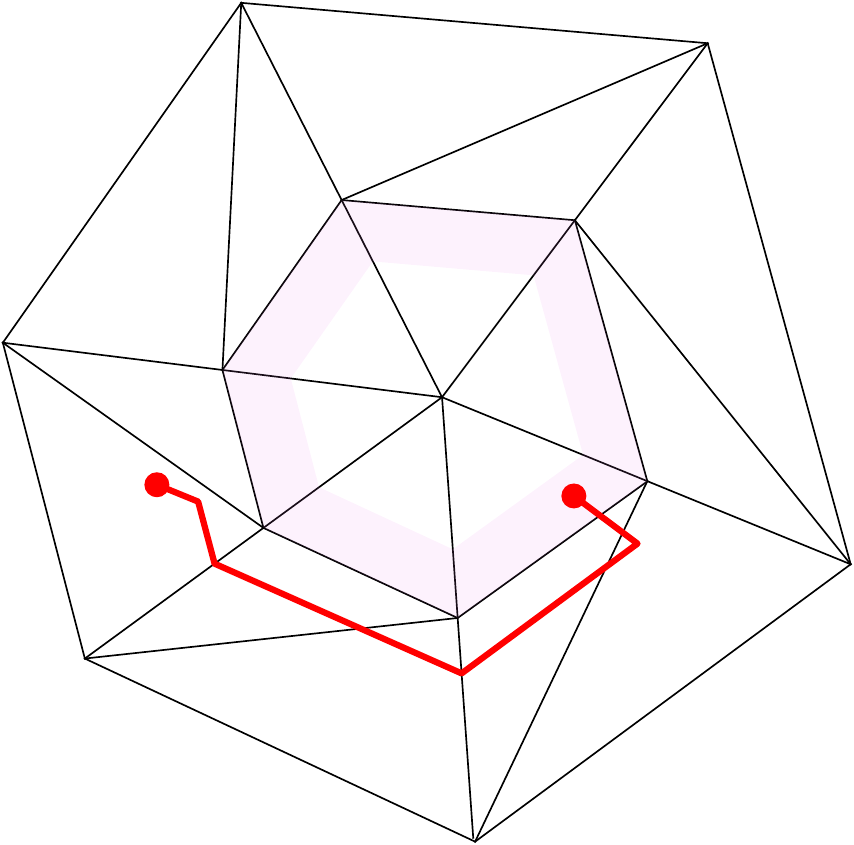}%
		\vspace{.2cm}%
	}
	\hfill
	\subcaptionbox{
		For distant points, the paths are pushed to the boundary.
	}[0.32\textwidth]{%
		\includegraphics[width=.99\linewidth]{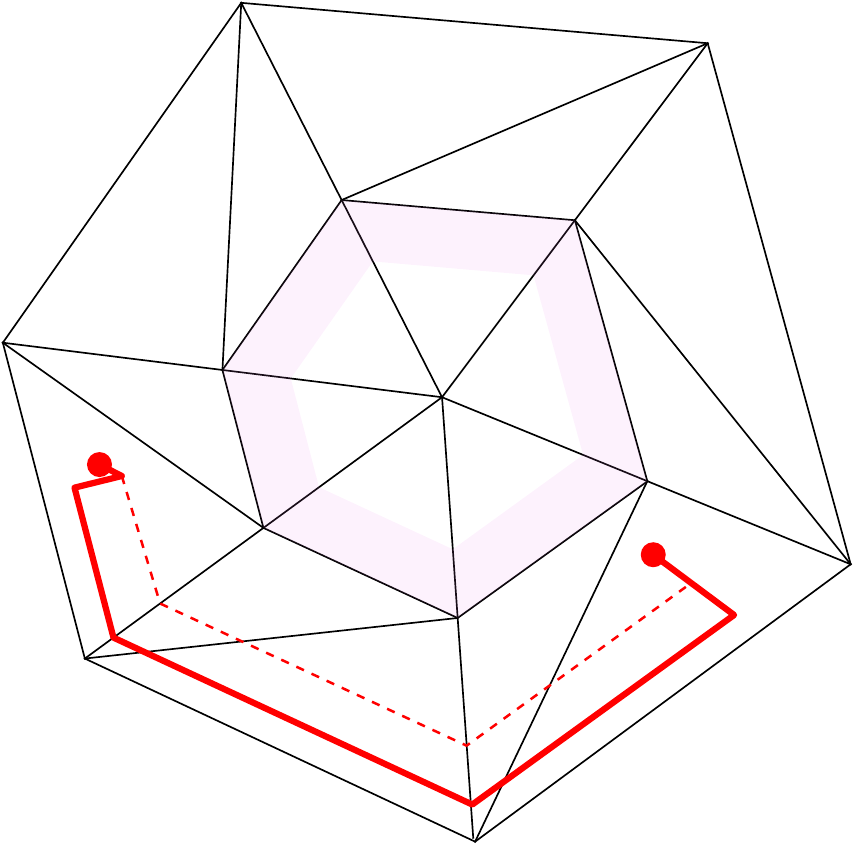}%
		\vspace{.2cm}%
	}
	\hfill
	\subcaptionbox{
		Close to the boundary, paths are interpolated with those on $\geomreal{L}$.
	}[0.32\textwidth]{%
		\includegraphics[width=.99\linewidth]{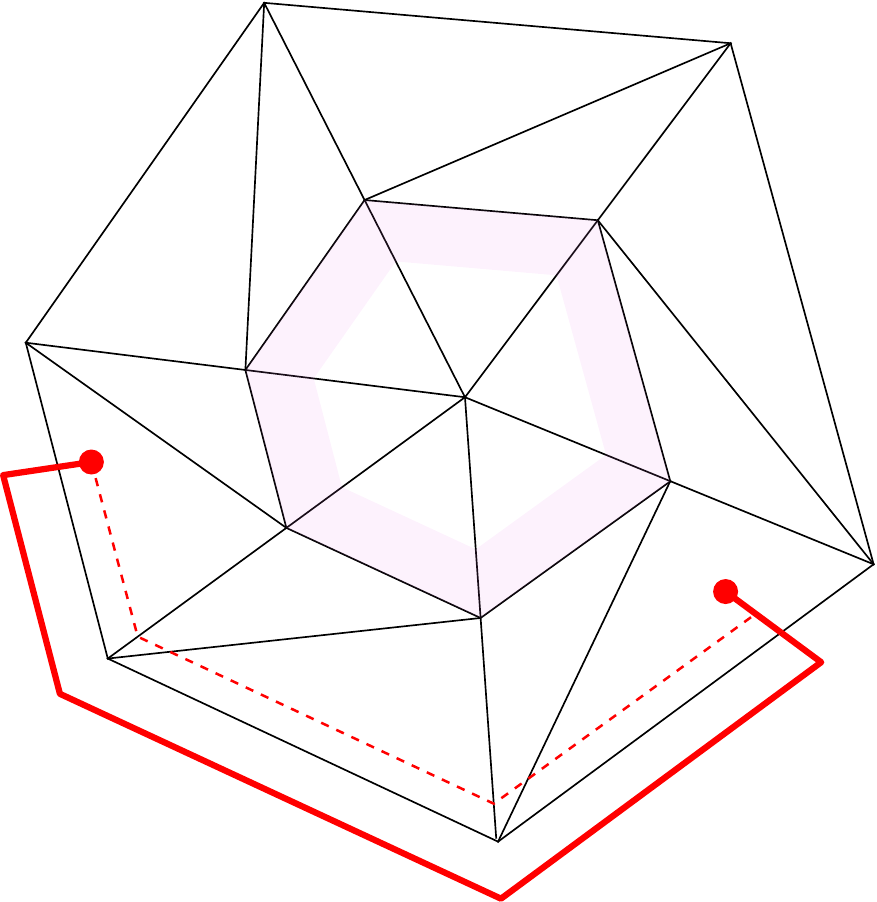}
	}
	\caption{
		We build a local motion planner on $\conesimp{g}=L\cup_g B(K)$ by first defining it on $B(K)$.
	}
	\label{fig:local_planner}
\end{figure}

As illustrated in \Cref{fig:local_planner}, we construct a planner on $\conesimp{g}$ by first defining it on $B(K)$, through a combination of ``elementary paths'' (rays, straight paths, circular arcs).
We prove that they descend along the projection map $\geomreal{B(K)}\to\geomreal{\conesimp{g}}$, yielding a well-defined local planner that can subsequently be spliced with the original planner on $\geomreal{L}$.

\begin{note}
	\label{note:lec_data_gluing}
	Equiconnecting maps or planners allow us to test the homotopy between maps.
	Indeed, two maps $a,b\colon X\to Y$ are homotopic whenever $(a(x),b(x))\in U$ for all $x\in X$; a homotopy is given by $H(x,t) = \Pi(a(x),b(x),t)$.
	We say that the maps are \textit{$U$-close}.
	During the proof of \Cref{th:lec_data_gluing}, we describe $U$ explicitly.
	In particular, for points $x,y$ in the ball $\geomreal{B(K)}$ and their images in the mapping cone via $\geomreal{B(K)}\to\geomreal{\conesimp{g}}$, the corresponding pair lies in $U$ when $x$ and $y$ are sufficiently close to the origin or when they are not antipodal.
	
	\fullonly{
		We recall that, in the proof of the simplicial approximation theorem \cite[Theorem~2C.1]{hatcher_algebraic_2002}, a map $a$ and a simplicial approximation $b$ are shown to be homotopic via the straight-line homotopy
		$H(x,t) = (1-t)a(x) + t\,\geomreal{b}(x)$.
		This is precisely the homotopy induced by the \textit{linear} motion planner $\Pi(x,y,t)=(1-t)x+ty$.
		The caveat is that $\Pi$ applies only for pairs of points that lie in a common simplex. 
		By contrast, the planner of \Cref{th:lec_data_gluing} also connects points in simplices that are far apart, allowing fewer subdivisions and thus smaller complexes.
	}
\end{note}

\subsection{Simplicial approximation routine}
\label{subsec:approximation_routine}

\paragraph*{Planner condition}

We return to the problem of simplicial approximation for $f \colon S^d \to \geomreal{L}$ where $L$ is a mapping cone $L = L_0 \cup_{g_0} B(K_0)$ now endowed with a local motion planner $(U,\Pi)$. 
Given a triangulation $K$ of $S^d$, we ask whether it is fine enough so that the planner can be applied on each facet.

More precisely, for each facet $\sigma = [v_0,\dots,v_d] \in K$ we look for a simplicial assignment
\[
v_i \mapsto g(v_i) \in \carr{f(v_i)}
\]
such that, on $\geomreal{\sigma}$, the continuous map $f$ and the linear map $\geomreal{g}$ are $U$-close, i.e., $(f(x), \geomreal{g}(x)) \in U$ for all $x \in \geomreal{\sigma}$. 
Let us assume for simplicity that $f(\geomreal{\sigma})$ is contained in the last cell $B(K_0)$ of $L = L_0 \cup_{g_0} B(K_0)$; the general case is treated recursively along the filtration.

As explained in \Cref{note:lec_data_gluing}, a pair $(x,y) \in B(K_0) \times B(K_0)$ belongs to $U$ provided the points are sufficiently close to the origin or are not antipodal ($x/\|x\|\neq -y/\|y\|$).
To guarantee this condition uniformly on $f(\geomreal{\sigma})$, we estimate its size using the edge lengths of $\geomreal{\sigma}$ and a Lipschitz bound $\lambda$ for $f$.
This Lipschitz constant is computed explicitly for each $f$.

\fullonly{
\begin{lemma}[%
	name={proof p.~\pageref{proof:lipschitz_deformation_simplices}},%
	restate=lipschitzdeformationsimplices%
	]
	\label{lem:lipschitz_deformation_simplices}
	Let $\sigma = [v_0,\dots,v_d]$ be a geometric simplex in $\R^n$ and $f\colon\geomreal{\sigma} \to \R^m$ a $\lambda$-Lipschitz continuous map.
	For $I\subset\{0,\dots,d \}$ nonempty, define the barycenter and length
	\[
	c_I = \frac{1}{|I|}\sum_{i\in I} f(v_i)
	\quad
	\text{and}
	\quad
	r_I = \max_{0\leq k \leq d } \frac{1}{|I|}\sum_{i\in I}\|v_k-v_i\|.
	\]
	Then $f(\geomreal{\sigma})$ is included in the closed ball $B(c_I, \lambda r_I)$.
\end{lemma}
}

\shortonly{
\begin{lemma}
	\label{lem:lipschitz_deformation_simplices}
	Let $\sigma = [v_0,\dots,v_d]$ be a geometric simplex in $\R^n$ and $f\colon\geomreal{\sigma} \to \R^m$ a $\lambda$-Lipschitz continuous map.
	For $I\subset\{0,\dots,d \}$ nonempty, define the barycenter and length
	\[
	c_I = \frac{1}{|I|}\sum_{i\in I} f(v_i)
	\quad
	\text{and}
	\quad
	r_I = \max_{0\leq k \leq d } \frac{1}{|I|}\sum_{i\in I}\|v_k-v_i\|.
	\]
	Then $f(\geomreal{\sigma})$ is included in the closed ball $B(c_I, \lambda r_I)$.
\end{lemma}
}

In our implementation we use the case $I = \{0,\dots,d \}$ only, corresponding to the barycenter of all vertices $f(v_i)$, although the estimate can be sharpened by considering other subsets $I$.

\begin{figure}[!htbp]
	\centering
	\subcaptionbox{
		Image $f(\geomreal{\sigma})$ close to the origin.
	}[0.49\textwidth]{%
		\includegraphics[width=.7\linewidth]{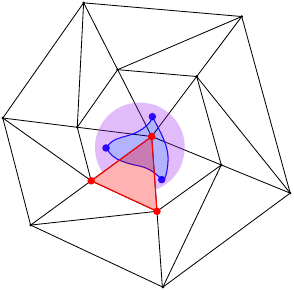}
		}
	\hfill
	\subcaptionbox{
		Image $f(\geomreal{\sigma})$ away from the origin.
	}[0.49\textwidth]{%
		\includegraphics[width=.7\linewidth]{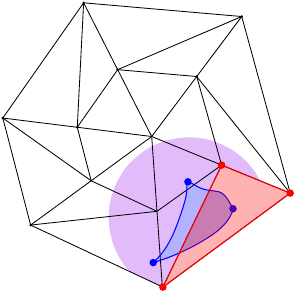}}
	\caption{
		For every facet $\sigma\in K$, the \textit{planner condition} checks whether the image $f(\geomreal{\sigma})$ (in blue) is $U$-close to a simplex $\tau$ of $L$ (in red).
		\Cref{lem:lipschitz_deformation_simplices} allows us to enclose $f(\geomreal{\sigma})$ in a ball (in purple).
		}
	\label{fig:planner_condition}
\end{figure}

If $\sigma$ is small enough, \Cref{lem:lipschitz_deformation_simplices} and the description of $U$ ensure that the required assignment $g$ exists: $\sigma$ is mapped to inner vertices if $f(\geomreal{\sigma})$ is close to the origin, or $g(\sigma)$ avoids the origin if $f(\geomreal{\sigma})$ is close to the boundary; see \Cref{fig:planner_condition}.
We refer to this requirement on $\sigma$ as the \emph{planner condition}. 
If the condition fails on some facet, we subdivide $K$ until it is satisfied.
The termination of this procedure is established in \Cref{prop:approximationroutine}.

\paragraph*{List homomorphism problem}

Once the planner condition is satisfied on $K$, we obtain, for each facet $\sigma\in K$, a collection of admissible simplices ${\tau_1,\tau_2,\dots}$ of $L$.
From these, we derive for each vertex $v \in V(K)$ a set $\mathcal{L}(v) \subset V(L)$ of admissible images, thus defining a map to the power set
\[
\mathcal{L}\colon V(K)\to \mathcal{P}(V(L))
\]
Given these lists of candidates, the remaining task is to find a simplicial map $g$ with $g(v) \in \mathcal{L}(v)$ for every vertex.
The motion planner ensures that such a map is homotopic to $f$.

In this form, the problem is essentially an instance of the \emph{list homomorphism problem}, denoted \texttt{LHom}.
Conventionally, it is formulated for graphs rather than simplicial complexes, and one typically forbids collapsing an edge to a vertex.
It is known that \texttt{LHom} can be solved in polynomial time when $L$ is a bi-arc graph and is NP-complete otherwise \cite{feder2003bi,egri2012complexity}.

In practice, we treat our instance as a constraint satisfaction problem and solve it using the \texttt{CP-SAT} solver from the \texttt{OR-Tools} suite \cite{cpsatlp,ortools}. If the resulting problem is unsatisfiable, we refine the triangulation $K$ and repeat the procedure, as formalized in \Cref{alg:algorithm_approx}.

\paragraph*{Delaunay simplifications}

Once a simplicial map $g \colon K \to L$ has been found, we apply a postprocessing step that we call \emph{Delaunay simplification}. 
Since the simplicial sphere $K$ arises as a Delaunay complex $K = \Del{X}$ on a finite set $X \subset S^d$, we seek a subset $X' \subset X$ such that the induced vertex map $g' \colon \Del{X'} \to L$ is still simplicial and homotopic to $f$.

Concretely, we inspect the vertices of $X$ one by one, remove a candidate vertex, recompute the Delaunay complex, and then check whether the new facets still satisfy the planner condition and whether the induced map remains simplicial. 
This procedure is efficient, since deleting a point $v \in X$ only affects the open star of $v$ in $\Del{X}$. 
Besides, to promote larger simplices, we preferentially remove vertices that belong to facets with the smallest inradius.

\paragraph*{Local Delaunay refinements}

A drawback of our method is that it refines the entire complex, even when the obstruction is localized on a few facets. 
To address this, we introduce a \textit{local refinement} strategy. 
If the solver reports that the instance is infeasible, we instead consider a weaker problem that we call \texttt{LHomDrop}: determine a minimal set of facets $\sigma_0,\dots,\sigma_k$ of $K$ whose removal makes the instance feasible. 
We then perform a local refinement by inserting Steiner points (see \Cref{subsec:global_delaunay_refinements}) only on the facets blamed by the solver.
In practice, we impose a time limit and use the best solution found by the solver so far (30 seconds in our implementation).

To accelerate the procedure, we apply \texttt{LHom} iteratively on the $i$-skeleta and use the solution in dimension $i$ as a hint for dimension $i+1$. 
If the instance is infeasible in dimension~$i$, we call \texttt{LHomDrop} to identify which $i$-simplices are to blame, and refine their maximal cofaces.

In the same spirit, when enforcing the planner condition we may refine only those facets on which the condition fails.
We write down our procedure, both for global and local refinements, in  \Cref{alg:algorithm_approx} below.
We are currently able to prove termination only for the variant with global refinement. However, in all our experiments the local refinement strategy also terminated and produced significantly smaller triangulations.

\begin{algorithm}[!htbp]
	\caption{Simplicial approximation with global or local refinement and simplification}
	\label{alg:algorithm_approx}
	\KwIn{map $f \colon S^d \to \geomreal{L}$, triangulation $K$ of $S^d$, planner $(U,\Pi)$ on $L$, scalar $\lambda>0$}
	\KwOut{refinement of $K$ and simplicial map $g \colon K \to L$ homotopic to $f$}
	
	\Repeat{a solution $g$ is found}{
		compute admissible lists $\mathcal{L}(v) \subset V(L)$ via the planner condition\;
		solve \texttt{LHom} with a SAT solver to obtain $g$\;
		\If{no solution}{
			apply Delaunay refinement on $K$
			(global, or only facets blamed by \texttt{LHomDrop})\;
		}
	}
	apply Delaunay simplification to $(K,g)$\;
	\Return{$(K,g)$}\;
\end{algorithm}

\fullonly{
\begin{proposition}[%
	name={proof p.~\pageref{proof:approximationroutine}},%
	restate=approximationroutine%
	]
	\label{prop:approximationroutine}
	With global refinement, \Cref{alg:algorithm_approx} terminates and produces a simplicial map homotopic to the input continuous map.
	With local refinement, if it terminates, then the output is also homotopic to the given map.
\end{proposition}
}

\shortonly{
\begin{proposition}
	\label{prop:approximationroutine}
	With global refinement, \Cref{alg:algorithm_approx} terminates and produces a simplicial map homotopic to the input continuous map.
	With local refinement, if it terminates, then the output is also homotopic to the given map.
\end{proposition}
}

\subsection{Full algorithm}\label{subsec:full_algorithm}

We now assemble the ingredients of \Cref{sec:delaunay,sec:simplicial_mapping_cones,sec:approximation} into the full procedure, summarized in \Cref{alg:algorithm_full}. The input is a finite CW complex $X$ together with explicit attaching maps for its cells and Lipschitz constants.
The output is a finite simplicial complex $L$ together with a homotopy equivalence $X\to\geomreal{L}$ encoded as a point location routine on $\geomreal{L}$.

\begin{algorithm}[!htbp]
	\caption{Simplicial approximation to CW complexes with global or local refinement}
	\label{alg:algorithm_full}
	\KwIn{CW complex $X$ of dimension $n$ with cells $e_i^d$, attaching maps $\phi_i^d$, scalars $\lambda_i$}
	\KwOut{simplicial complex $L$ endowed with a homotopy equivalence $X\to\geomreal{L}$}
	
	$L \gets$ discrete simplicial complex on the $0$-cells of $X$\;
	
	\For{$d = 1,\dots,n$}{
		\ForEach{$d$-cell $e_i^d$ with attaching map $\phi_i^d \colon S^{d-1} \to X^{d-1}$}{
			compose $\phi_i^d$ with $X^{d-1} \to \geomreal{L}$ to obtain
			$f_i^d \colon S^{d-1} \to \geomreal{L}$\;
			
			use \Cref{alg:algorithm_approx} to obtain
			$g_i^d \colon K_i \to L$ homotopic to $f_i^d$
			(with global or local refinement, and parameter $\lambda_i$)\;
			
			construct a simplicial ball $B(K_i)$ and glue it to $L$ along $K_i$ via $g_i^d$, i.e., let $L \gets L\cup_{g_i^d}B(K_i)$\;
			
			endow $\geomreal{L\cup_{g_i^d}B(K_i)}$ with an equiconnecting map or a local motion planner\;
		}
	}
	\Return{$L$}\;
\end{algorithm}

\fullonly{
\begin{theorem}[%
	name={proof p.~\pageref{proof:algorithmcorrect}},%
	restate=algorithmcorrect%
	]
	\label{th:algorithmcorrect}
	With global refinement, \Cref{alg:algorithm_full} terminates and produces a simplicial complex homotopy equivalent to the given CW complex. With local refinement, if it terminates, then the output is also homotopy equivalent to the input CW complex.
\end{theorem}
}

\shortonly{
\begin{theorem}
	\label{th:algorithmcorrect}
	With global refinement, \Cref{alg:algorithm_full} terminates and produces a simplicial complex homotopy equivalent to the given CW complex. With local refinement, if it terminates, then the output is also homotopy equivalent to the input CW complex.
\end{theorem}
}



\bibliography{cw2simp}

\appendix
	
\iffullversion
	
\section{Notation}
\label{sec:background}

\begin{longtable}{p{0.26\linewidth}p{0.72\linewidth}}
	\hline
	\textbf{Symbol}  & \textbf{Meaning}	
	\\\hline	
	\multicolumn{2}{l}{\textbf{Standard notation} } \\
	
	$\langle x, y \rangle$, $x\cdot y$ & Inner product between $x,y\in\R^n$ \\

	$\R P^n$, $\C P^n$, $\mathrm{SO}(n)$, $\mathrm{SU}(n)$, $\mathrm{U}(n)$, $\VV(d,n)$, $\GG(d,n)$ & Classical manifolds defined in \Cref{tab:known_triangulations}\\
	
	$X$, $S^{d}$, $B^{d}$, $e^d$ & Topological space, unit sphere of $\R^{d+1}$, unit ball of $\R^{d}$, $d$-cell \\
	
	$X\cong Y$, $X\simeq Y$ & Homeomorphic and homotopy equivalent topological spaces \\
	
	$f\colon X\to Y$ & Continuous map between topological spaces \\
	
	$K$, $\geomreal{K}$, $\sigma$ & Simplicial complex, its geometric realization, simplex\\
	
	$\Delta^d$ & Standard $d$-simplex\\
	
	$g\colon K\to L$, $\geomreal{g}\colon\geomreal{K}\to\geomreal{L}$ & Simplicial map between simplicial complexes, its geometric realization \\
	
	$\Phi_i^d$, $\phi_i^d$ & Characteristic map and gluing map in a CW structure (\Cref{subsec:overview}) \\
	
	\multicolumn{2}{l}{\textbf{Notation introduced in \Cref{sec:delaunay}} } \\
	
	$\Del{X}$ & Delaunay complex on a finite subset $X\subset S^d$ (\Cref{subsec:delaunay_triangulations})\\
	
	$\rho_\mathrm{circ}(\Del{X})$ & Maximal (spherical) circumradius of simplices of $\Del{X}$ (\Cref{subsec:delaunay_triangulations})\\
	
	$\rho_\mathrm{cov}(X)$ & Covering (spherical) radius of a finite subset $X\subset S^d$ (\Cref{subsec:global_delaunay_refinements})\\
	
	\multicolumn{2}{l}{\textbf{Notation introduced in \Cref{sec:simplicial_mapping_cones}} } \\
	
	$B(K)$ & Simplicial ball built from a triangulation $K$ of the sphere (\Cref{subsec:triangulation_ball}) \\
	
	$\rho_{\mathrm{inner}}$ & Common norm of inner vertices of $\geomreal{B(K)}$ (\Cref{subsec:triangulation_ball}) \\
	
	$\nu \colon \geomreal{B(K)} \to B^{d+1}$ & Radial normalization map (\Cref{subsec:triangulation_ball}) \\
	
	$\cone{f}$ & Mapping cone of a continuous map $f\colon S^{d}\to Y$, equal to the adjunction space $Y\cup_f B^{d+1}$ (\Cref{subsec:equivalence_mapping_cones}) \\
	
	$\conesimp{g}$ & Simplicial mapping cone of a simplicial map $g\colon K\to L$, equal to the quotient simplicial complex $L\cup_g B(K)$ (\Cref{subsec:equivalence_mapping_cones}) \\
	
	\multicolumn{2}{l}{\textbf{Notation introduced in \Cref{sec:approximation}} } \\
	
	$(U,\Pi)$ & LEC-data or local motion planner on a topological space $Y$, where $U\subset Y\times Y$ and $\Pi\colon U\times[0,1]\to Y$ (\Cref{subsec:lec}) \\
	
	$\carr{x}$ & Carrier simplex of a point $x$ in a simplicial complex $\geomreal{K}$ (\Cref{subsec:approximation_routine}) \\
	
	$\lambda$ & Lipschitz constant for a continuous map (\Cref{subsec:approximation_routine}) \\
	
	\texttt{LHom}, \texttt{LHomDrop} & List homomorphism problem, without or with dropping (\Cref{subsec:approximation_routine}) \\
	
	\multicolumn{2}{l}{\textbf{Notation used in the proof of \Cref{th:lec_data_gluing}} (page~\pageref{proof:lec_data_gluing})} \\
	
	$\epsilon_\mathrm{inner}$ & Offset parameter, in $(0,\rho_\mathrm{inner})$ \\

	$B(K)/g$ & Quotient of $B(K)$ along its boundary with respect to a simplicial map $g$ \\

	$\geomreal{\mathring{B}(K)}$, $\geomreal{\mathring{B}(K)/g}$ & Interior of $\geomreal{B(K)}$, its image in the quotient $\geomreal{B(K)/g}$ \\

	$R_{\mathrm{inner}}$, $R_{\mathrm{equal}}$ & Inner representative map, equal-norm representative map \\

	$\gamma^x_\mathrm{ray}$ or $x\xleadsto[\text{ray}]y$
	&
	Radial path from the origin through $x$
	\\
	
	$\gamma^x_\mathrm{climb}$ or $x\xleadsto[\text{climb}]y$
	&
	Climb path towards the origin from $x$
	\\
	
	$\gamma^{x\leadsto y}_\mathrm{straight}$ or $x\xleadsto[\text{straight}]y$
	&
	Straight path from $x$ to $y$
	\\
	
	$\gamma^{x\leadsto y}_\mathrm{arc}$ or $x\xleadsto[\text{arc}]y$
	&
	Polygonal circular arc from $x$ to $y$ (two different definitions, depending on whether $g$ is 2-distance injective or not)
	\\
	
	$\gamma_\mathrm{middle}$ or $x\xleadsto[\text{middle}] y$
	&
	Interpolation between straight path and circular arc
	\\
		
	$\d_{/g}(\bar u,\bar v)$ & Quotient distance between $\bar u,\bar v$ in a common simplex of $B(K)/g$. \\
	\hline
\end{longtable}

\section{Applications}
\label{sec:applications}

The two applications below are fully implemented and can be found in our repository.\footnote{Haldane \url{https://github.com/raphaeltinarrage/cw2simp/blob/master/demos/demo_haldane.ipynb}
Hopfion \url{https://github.com/raphaeltinarrage/cw2simp/blob/master/demos/demo_hopfion.ipynb}}

\begin{example}
	In machine learning, \textit{vector bundles} provide a rigorous framework to endow data with extra information at each point, such as features, orientations, or tangent directions.
	They let models work in local coordinates and ``glue'' them together, supporting coordinate-independent architectures \cite{cohen2019gauge,aronsson2022homogeneous} and dimensionality reduction methods \cite{singer2012vector,scoccola_et_al:LIPIcs.SoCG.2023.56}.
	On the other hand, \textit{characteristic classes} allow one to explore the global structure of vector bundles, and their computation has recently been the focus of several works \cite{tinarrage2022computing,scoccola2023approximate,gang2025persistentstiefelwhitneyclassestangent}.
	The tools developed in this work facilitate practical computations with vector bundles, as we illustrate.
	
	The \textit{Haldane Bundles} dataset consists of synthetically generated complex line bundles over the torus $T^2$, designed as a benchmark for machine learning \cite{pmlr-v228-tipton24a}.
	Concretely, they can be described as maps $T^2\to \C P^1$, where the projective space $\C P^1$ encodes the (complex) 1-dimensional subspaces of $\C^2$; see \Cref{fig:haldane}.
	These bundles are classified by their \textit{Chern class}, an integer $c_1$ in the cohomology group $H^2(T^2; \Z)\cong\Z$.
	The authors propose approximating $c_1$ via an integral formula. 
	We can compute it directly using simplicial approximation.
	
	We obtained, via \Cref{alg:algorithm_full}, a simplicial complex $L$ homotopy equivalent to $ \C P^1$. 
	For a sufficiently refined triangulation $K$ of $T^2$, a bundle $f\colon T^2\to \C P^1$ will induce a simplicial map $g\colon K\to L$; here we use \Cref{alg:algorithm_approx}.
	Now, the induced homomorphism in cohomology,
	\[
	g^*\colon H^2(T^2; \Z) \leftarrow H^2(\C P^1; \Z),
	\]
	sends the generator of $H^2(\C P^1; \Z)\cong\Z$ to the Chern class of the bundle.
	Our notebook presents a concrete computation for a bundle with Chern class 3.
	The procedure above returns $c_1=3$, as desired.
	Homology computations were carried out with \texttt{HAP} \cite{Ellis2025}.
	
	Another interesting viewpoint is to treat the complex line bundle as a \textit{real plane bundle}.
	This simulates a situation where we did not know that the bundle had a complex structure.
	In this context, one computes the second \textit{Stiefel-Whitney} class of the bundle, an integer modulo 2 in the cohomology group $H^2(T^2; \Z/2\Z)\cong\Z/2\Z$.
	It is denoted $w_2$ and is equal to the reduction of the Chern class $c_1$ modulo 2.
	For the bundle above, one has $w_2=1$.
	
	In the same vein as the complex case, this real bundle is encoded by a classifying map $f_\R\colon T^2\to \GG(2,4)$ to the Grassmannian of planes in $\R^4$.
	We obtained a triangulation of $\GG(2,4)$ through \Cref{alg:algorithm_full}.
	Moreover, by \Cref{alg:algorithm_approx}, we obtain a simplicial approximation $g_\R\colon K\to L$ homotopic to $f_\R$.
	As before, the induced homomorphism
	\[
	g^*_\R\colon H^2(T^2; \Z/2\Z) \leftarrow H^2(\GG(2,4); \Z/2\Z)
	\]
	sends the generator of $H^2(\GG(2,4); \Z/2\Z)\cong\Z/2\Z$ to the Stiefel-Whitney class $w_2$.
	We computed that it is equal to 1, as expected.
	
	\begin{figure}[!htbp]
		\centering
		\includegraphics[width=1\linewidth]{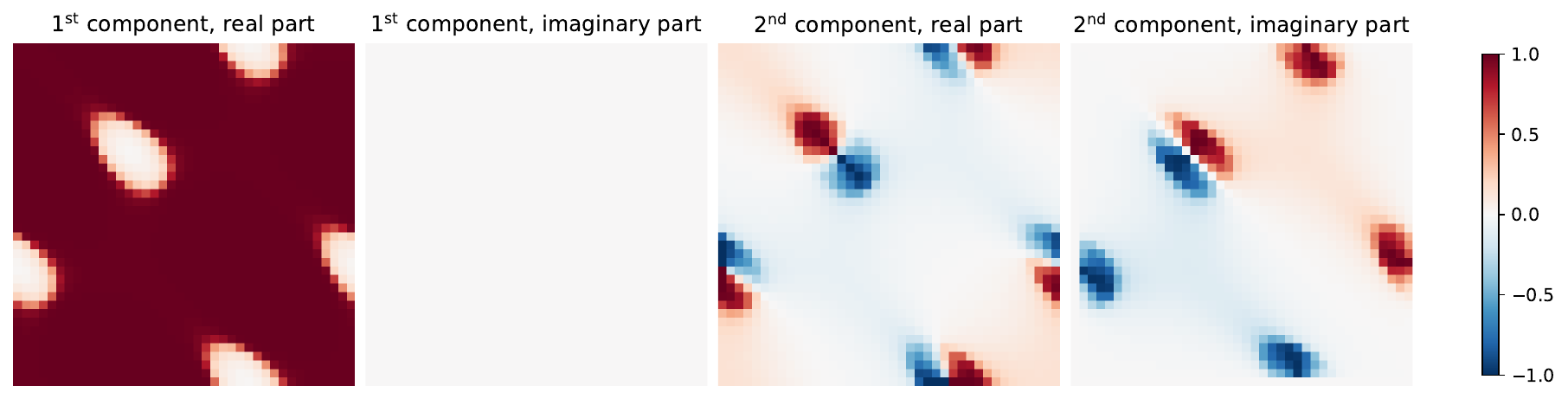}
		\caption{
			Visualization of a complex line bundle on the torus, from the Haldane dataset \cite{pmlr-v228-tipton24a}.
			A section $T^2\to\C^2$ is represented; the four plots represent the real and imaginary parts of the first and second components of the section, respectively.
			Note that this section admits zeros.
		}
		\label{fig:haldane}
	\end{figure}
\end{example}

\begin{example}
	In condensed matter physics, \textit{hopfions} are a class of three-dimensional topological solitons; they are modeled as smooth unit vector fields $\R^3\to S^2\subset\R^3$ with a ``knotted'' structure.
	They were first proposed in the Skyrme--Faddeev model and are classified by their \textit{Hopf invariant}, an integer in the homotopy group $\pi_3(S^2)\cong\Z$ \cite{faddeev1976some,faddeev1997stable}. 
	Since then, hopfions have been predicted and experimentally realized in chiral magnets \cite{sutcliffe2018hopfions,zheng2023hopfion} and nematic liquid crystals \cite{chen2013generating,tai2022geometric}.
	Their Hopf invariant is usually estimated numerically, e.g., via Whitehead's integral formula \cite{guslienko2023emergent,knapman2025numerical} or linking numbers \cite{ackerman2017diversity,wu2022hopfions}.
	Through the methods developed here, we can compute this invariant \textit{exactly}, at the cohomological level.
	
	As a concrete example, we use the public dataset \cite{HerguedasGomezSanchez2025} provided with the \texttt{MARTApp} software \cite{herguedas2025martapp,herguedas2023fast,hierro20183d}.
	It consists of a three-channel array of shape $250\times 440\times 440$ representing the reconstructed magnetization of a simulated magnetic hopfion, originally used to benchmark a tomography workflow; see \Cref{fig:hopfion}.
	After normalization, we view this as a unit-length vector field $\Omega\to S^2$ defined on a rectangular box $\Omega\subset\R^3$.
	It is nearly constant near the boundary of $\Omega$.
	By collapsing this boundary to a point, we obtain a continuous map $f\colon S^3\to S^2$.
	
	We model $S^2$ as $L=\partial \Delta^3$, the boundary of the $3$-simplex.
	Using successive centroid Delaunay refinement (see \Cref{sec:delaunay}), we obtain a triangulation $K$ of $S^3$ (with 3736 vertices) and a simplicial map $g\colon K\to L$ homotopic to $f$.
	To study its homotopy class, we build its simplicial mapping cone (see \Cref{sec:simplicial_mapping_cones}).
	This is a simplicial complex with integral cohomology
	\[
	H^0\cong\Z,\quad
	H^1=0,\quad
	H^2\cong\Z,\quad
	H^3=0,\quad
	H^4\cong\Z.
	\]
	Using \texttt{HAP}, we compute the cup product and obtain $\alpha_2 \smile \alpha_2 = \alpha_4$, where $\alpha_2$ and $\alpha_4$ are generators of $H^2$ and $H^4$, respectively.
	By the classical cohomological definition of the Hopf invariant, this shows that $f$ has Hopf invariant $1$, as expected for a hopfion.
	\begin{figure}[!htbp]
		\centering
		\includegraphics[width=1\linewidth]{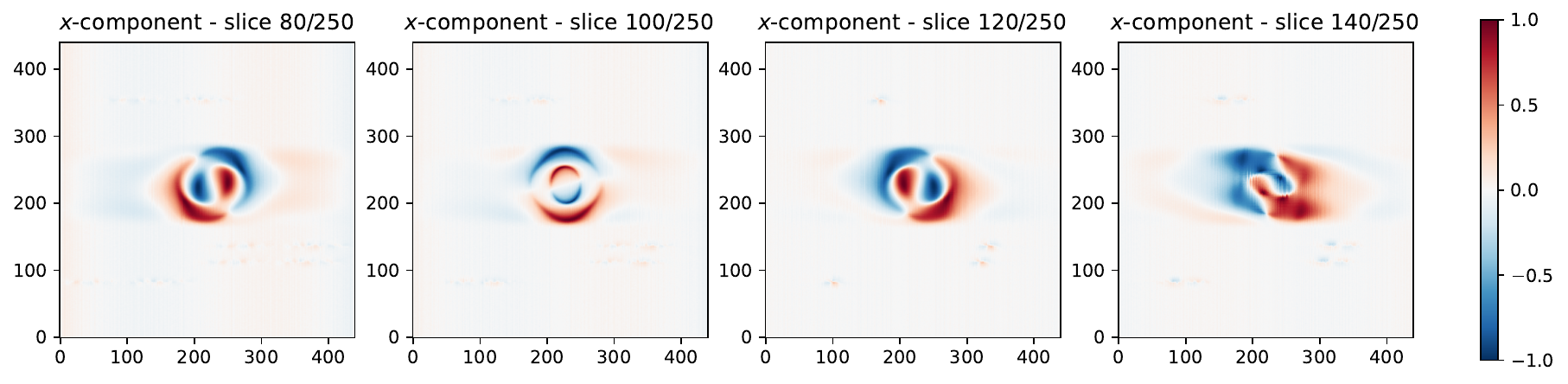}
		\caption{
			Visualization of the 3D vector field from \cite{HerguedasGomezSanchez2025}. The data are a three-channel array of shape $250\times440\times440$ encoding a hopfion $\Omega\subset\R^3\to S^2$; we show the first component on a few slices.
		}
		\label{fig:hopfion}
	\end{figure}
\end{example}

\section{Examples of execution of the algorithm}
\label{sec:execution}

We applied \Cref{alg:algorithm_full} to a selection of manifolds\footnote{Notebook triangulations: \url{https://github.com/raphaeltinarrage/cw2simp/blob/master/demos/demo_triangulations.ipynb}} whose CW structures are well understood.
The triangulations were obtained via local Delaunay refinement, using centroids as Steiner points.
In each case, we applied the Delaunay simplification procedure, which substantially reduced the number of vertices.
In \Cref{tab:outputs}, we report the $f$-vectors of the resulting complexes (numbers of vertices, edges, and higher-dimensional simplices).
All computations were completed within a few minutes on a personal laptop.
Pushing these experiments to higher dimensions will require either more memory or more efficient approximation schemes.

\begin{table}[ht]
	\centering
	\begin{tabular}{p{.3\linewidth}llllll}
		\hline
		\textbf{Space}                    
		& \textbf{$f_0$}
		& \textbf{$f_1$}
		& \textbf{$f_2$}
		& \textbf{$f_3$}
		& \textbf{$f_4$}
		& \textbf{$f_5$}
		\\
		\hline
		\multicolumn{7}{l}{\textbf{Real projective space} }
		\\
		$\R P^2$
		& 10
		& 27
		& 18
		& 
		& 
		& 
		\\
		$\R P^3$
		& 112
		& 658
		& 1093
		& 547
		& 
		& 
		\\
		$\R P^4$
		& 462
		& 5478
		& 17355
		& 20582
		& 8244
		& 
		\\
		\multicolumn{7}{l}{\textbf{Special orthogonal group} } 
		\\
		$\mathrm{SO}(3)$
		& 79
		& 473
		& 788
		& 394
		& 
		& 
		\\
		$\mathrm{SO}(4)$ (only 4-skeleton)
		& 975
		& 12006
		& 38395
		& 45645
		& 18281
		& 
		\\
		\multicolumn{7}{l}{\textbf{Complex projective space} } 
		\\
		$\C P^2$
		& 1363 
		& 14112
		& 42992
		& 50458
		& 20218
		& 
		\\
		\multicolumn{7}{l}{\textbf{Real Grassmannian} } 
		\\
		$\GG(3,4)$
		& 81
		& 479
		& 796
		& 398
		& 
		& 
		\\
		$\GG(2,4)$
		& 991
		& 12226
		& 39135
		& 46540
		& 18642
		& 
		\\
		\multicolumn{7}{l}{\textbf{Real Stiefel manifold} } 
		\\
		$\VV(2,3)$
		& 81
		& 486
		& 810
		& 405
		& 
		& 
		\\
		$\VV(2,4)$
		& 1961
		& 44230
		& 239828
		& 510411
		& 470395
		& 157543
		\\
		\hline
	\end{tabular}
	\caption{
		$f$-vectors of the complexes produced by \Cref{alg:algorithm_full} on a selection of manifolds.
	}
	\label{tab:outputs}
\end{table}

In \Cref{lst:execution} we also report the console output of a full run of the algorithm on the Grassmannian $\GG(2,4)$, given as the input CW complex. 
We use the standard CW structure in Schubert cells, which has one $0$-cell, one $1$-cell, two $2$-cells, one $3$-cell, and one $4$-cell; accordingly, the output is organized by cell.
For each cell---except the $0$- and $1$-cells, whose gluings are trivial---the algorithm prints information regarding the following steps:

\begin{description}
	\item[Spherical Delaunay]
	Generate a set of well-spaced sample points on the sphere, as in \Cref{subsec:delaunay_triangulations}.
	The number of sampled points is a hyperparameter for each cell.
	
	\item[Locate in triangulation]
	For each sampled point, compute its image under the gluing map and locate it in the current complex (via its carrier simplex and barycentric coordinates).
	
	\item[Check planner condition]
	Using these data, verify that all facets satisfy the planner condition from \Cref{subsec:approximation_routine}. 
	If the condition fails, refine the offending facets as in \Cref{subsec:global_delaunay_refinements}.
	
	\item[Solve LHom problem]
	When the planner condition holds, search for a simplicial approximation via the \texttt{LHom} problem. 
	If \texttt{LHom} is infeasible, call \texttt{LHomDrop} instead, refine the offending facets, and go to the previous step. 
	The solver is applied iteratively over skeleta. 
	Together with the previous step, this forms the simplicial approximation routine (\Cref{alg:algorithm_approx}).
	
	\item[Delaunay simplification]
	Once a solution to \texttt{LHom} is found, simplify the domain triangulation by iteratively removing vertices while preserving the homotopy class; see \Cref{subsec:approximation_routine}.
	
	\item[Triangulation of sphere, ball, and mapping cone]
	Given the resulting simplicial map $g\colon K\to L$, we construct three ``Triangulation'' objects: the sphere (realizing $K$), the ball (realizing $B(K)$), and the simplicial mapping cone $\conesimp{g}$; see \Cref{sec:simplicial_mapping_cones}.
	
	\item[Homology groups]
	Finally, for illustration, we compute the integral homology of the resulting complex, via \texttt{GAP} (more specifically, \texttt{HAP}). 
	An output of the form 
	\begin{center}
		\verb|[ [ 0 ], [ 2 ], [ 2 ], [  ], [ 0 ] ]|
	\end{center}
	indicates the homology groups
	\[
	H_0=\Z,\quad
	H_1=\Z/2\Z,\quad
	H_2=\Z/2\Z,\quad
	H_3=0,\quad
	H_4=\Z.
	\]
\end{description} 
The total running time for this session was approximately 5 minutes.
The resulting complex is, by \Cref{th:algorithmcorrect}, homotopy equivalent to $\GG(2,4)$.
Its $f$-vector has been given in \Cref{tab:outputs}.

\lstdefinestyle{consolestyle}{
	basicstyle=\scriptsize\ttfamily,
	columns=fullflexible,
	keepspaces=true,
	literate=*%
	{Condition\ satisfied}{{\textcolor{blue}{Condition satisfied}}}{19}%
	{Condition\ not\ satisfied}{{\textcolor{red}{Condition not satisfied}}}{23}%
	{Problem\ feasible}{{\textcolor{blue}{Problem feasible}}}{16}%
	{Problem\ not\ feasible}{{\textcolor{red}{Problem not feasible}}}{20}%
}

\begin{lstlisting}[style=consolestyle,caption={Console output of \Cref{alg:algorithm_full} on the Grassmannian $\GG(2,4)$},label={lst:execution}]
------- Triangulate G(2,4) -------
---- Init cell of dimension 0 ---- 
| Triangulation of 0-cell        | Dim/Verts/Facets/Splx = 0/1/1/1. Cell #0 initialized.
| Homology groups with <gap>     | [ [ 0 ] ]. Duration 0:02.201.
---- Glue cell of dimension 1 ---- 
| Triangulation of sphere        | Dim/Verts/Facets/Splx = 0/2/2/2. Mesh ratio 1.0e+00.
| Triangulation of ball          | Dim/Verts/Facets/Splx = 1/4/3/7. Min dist 6.7e-01.
| Triangulation of mapping cone  | Dim/Verts/Facets/Splx = 1/3/3/6. Cell #1 glued.
| Homology groups with <gap>     | [ [ 0 ], [ 0 ] ]. Duration 0:02.382.
---- Glue cell of dimension 2 ---- 
| Spherical Delaunay             | Generate 10 points. Duration 0:00.000. Build hull. Duration 0:00.000.
| Locate in triangulation        | Vertex 10/10. Duration 0:00.004.
| Check planner condition        | Facet 10/10... Duration 0:00.002. Condition satisfied.
| Solve LHom problem             | On 1-skeleton... Problem feasible. Duration 0:00.004.
| Delaunay simplification        | Initialize. Duration 0:00.003. Vertex 5/10. Duration 0:00.004.
| Triangulation of sphere        | Dim/Verts/Facets/Splx = 1/6/6/12. Mesh ratio 5.0e-01.
| Triangulation of ball          | Dim/Verts/Facets/Splx = 2/13/18/61. Min dist 3.1e-01.
| Triangulation of mapping cone  | Dim/Verts/Facets/Splx = 2/10/18/55. Cell #2 glued.
| Homology groups with <gap>     | [ [ 0 ], [ 2 ], [  ] ]. Duration 0:02.176.
---- Glue cell of dimension 2 ---- 
| Spherical Delaunay             | Generate 10 points. Duration 0:00.000. Build hull. Duration 0:00.000.
| Locate in triangulation        | Vertex 10/10. Duration 0:00.006.
| Check planner condition        | Facet 10/10... Duration 0:00.004. Condition not satisfied for 20.0% 
				   of vertices (2/10).
| Delaunay refinement            | Blame 2 0-simplices. Add 4 centroids. Duration 0:00.000.
| Locate in triangulation        | Vertex 4/4. Duration 0:00.002.
| Check planner condition        | Facet 14/14... Duration 0:00.001. Condition satisfied.
| Solve LHom problem             | On 1-skeleton... Problem feasible. Duration 0:00.003.
| Delaunay simplification        | Initialize. Duration 0:00.007. Vertex 9/14. Duration 0:00.008.
| Triangulation of sphere        | Dim/Verts/Facets/Splx = 1/6/6/12. Mesh ratio 1.7e-01.
| Triangulation of ball          | Dim/Verts/Facets/Splx = 2/13/18/61. Min dist 1.6e-01.
| Triangulation of mapping cone  | Dim/Verts/Facets/Splx = 2/17/36/104. Cell #3 glued.
| Homology groups with <gap>     | [ [ 0 ], [ 2 ], [ 0 ] ]. Duration 0:02.246.
---- Glue cell of dimension 3 ---- 
| Spherical Delaunay             | Generate 100 points. Duration 0:00.000. Build hull. Duration 0:00.001.
| Locate in triangulation        | Vertex 100/100. Duration 0:00.071.
| Check planner condition        | Facet 196/196... Duration 0:00.046. Condition satisfied.
| Solve LHom problem             | On 2-skeleton... Problem feasible. Duration 0:00.052.
| Delaunay simplification        | Initialize. Duration 0:00.124. Vertex 35/100. Duration 0:00.215.
| Triangulation of sphere        | Dim/Verts/Facets/Splx = 2/66/128/386. Mesh ratio 3.3e-01.
| Triangulation of ball          | Dim/Verts/Facets/Splx = 3/133/512/2441. Min dist 1.7e-01.
| Triangulation of mapping cone  | Dim/Verts/Facets/Splx = 3/84/436/1911. Cell #4 glued.
| Homology groups with <gap>     | [ [ 0 ], [ 2 ], [ 2 ], [  ] ]. Duration 0:02.304.
---- Glue cell of dimension 4 ---- 
| Spherical Delaunay             | Generate 1000 points. Duration 0:05.734. Build hull. Duration 0:00.009.
| Locate in triangulation        | Vertex 1000/1000. Duration 0:00.695.
| Check planner condition        | Facet 5892/5892... Duration 0:01.778. Condition satisfied.
| Solve LHom problem             | On 1-skeleton... Problem not feasible. Blame 2.597% of 1-simplices 
				   (179/6892). Duration 0:20.582.
| Delaunay refinement            | Blame 179 1-simplices. Add 596 centroids. Duration 0:00.295. 
| Locate in triangulation        | Vertex 596/596. Duration 0:00.593.
| Check planner condition        | Facet 9834/9834... Duration 0:01.899. Condition satisfied.
| Solve LHom problem             | On 1-skeleton... Problem not feasible. Blame 2.301% of 1-simplices 
				   (263/11430). Duration 0:21.046.
| Delaunay refinement            | Blame 263 1-simplices. Add 987 centroids. Duration 0:00.411.
| Locate in triangulation        | Vertex 987/987. Duration 0:00.810. 
| Check planner condition        | Facet 16092/16092... Duration 0:02.861. Condition satisfied.
| Solve LHom problem             | On 1-skeleton... Problem not feasible. Blame 0.005% of 1-simplices 
				   (1/18675). Duration 0:16.032.
| Delaunay refinement            | Blame 1 1-simplices. Add 4 centroids. Duration 0:00.318.
| Locate in triangulation        | Vertex 4/4. Duration 0:00.004.
| Check planner condition        | Facet 16120/16120... Duration 0:00.170. Condition satisfied.
| Solve LHom problem             | On 3-skeleton... Problem feasible. Duration 0:30.731.
| Delaunay simplification        | Initialize. Duration 0:16.279. Vertex 1682/2587. Duration 2:12.183.
| Triangulation of sphere        | Dim/Verts/Facets/Splx = 3/906/5587/24160. Mesh ratio 6.1e-02.
| Triangulation of ball          | Dim/Verts/Facets/Splx = 4/1813/27935/187845. Min dist 2.5e-02.
| Triangulation of mapping cone  | Dim/Verts/Facets/Splx = 4/991/18644/117534. Cell #5 glued.
| Homology groups with <gap>     | [ [ 0 ], [ 2 ], [ 2 ], [  ], [ 0 ] ]. Duration 0:15.327.
\end{lstlisting}

\section{Background on real Grassmannians}
\label{sec:background_grassmannian}

\paragraph*{Definition}

The real Grassmannians $\GG(d,n)$ form a family of smooth manifolds indexed by integers
$d$ and $n$ with $1 \le d \le n$.
For fixed $d$ and $n$, the space $\GG(d,n)$ is the set of $d$-dimensional linear subspaces of $\R^n$.
In particular, $\GG(1,n)$ identifies with the real projective space $\R P^{n-1}$.

More generally, $\GG(d,n)$ carries the structure of a smooth manifold of dimension $d(n-d)$.
It admits a concrete realization as a submanifold of $\mathcal M(\R^n)$, the space of $n\times n$ real matrices.
Namely, $\GG(d,n)$ can be identified with the set of orthogonal projection matrices of rank $d$.

\paragraph*{Cohomology}

The cohomology ring of the real Grassmannians $\GG(d,n)$ over $\Z/2\Z$ is well understood; see \cite[Problem 7-B, p.~95]{Milnor_1974} or \cite[p.~190]{Borel1953}.
It is generated by the Stiefel--Whitney classes $w_1,\dots,w_d$ and the dual classes $\bar w_1,\dots,\bar w_{n-d}$, subject to certain relations: 
\[
H^*(\GG(d,n);\Z/2\Z)
\cong
\frac{\Z/2\Z[w_1,\dots,w_d,\bar w_1,\dots,\bar w_{n-d}]}
{(1+w_1+\cdots+w_d)(1+\bar w_1+\cdots+\bar w_{n-d})=1}.
\]

With integral coefficients, the structure is more subtle: the cohomology generally contains torsion, but all torsion is of order $2$ \cite{JMPA_1937_9_16_1-4_69_0,hiller1980cohomology}.
As an example, $\GG(2,4)$ has cohomology groups
\[
H^k(\GG(2,4);\Z)\cong
\begin{cases}
	\Z & k=0,4,\\
	\Z/2\Z & k=2,3,\\
	0 & k=1 \text{ or } k\ge 5,
\end{cases}
\]
and all the cup products are trivial.

\paragraph*{Cell structure for $\GG(2,4)$}

The classical CW structure on $\GG(d,n)$ is given by the Schubert cell decomposition; see \cite[Section~6]{Milnor_1974}.
We describe this construction here in the special case of $\GG(2,4)$, which already illustrates the general picture.

We begin by parametrizing the planes in $\R^4$ by their Schubert symbols.
Let $(e_1,e_2,e_3,e_4)$ be the standard basis of $\R^4$, and for each $k\in\{1,\dots,4\}$, let $\R^k\subset \R^4$ denote the subspace spanned by the first $k$ basis vectors.
For any plane $T\in \GG(2,4)$, the sequence
\[
\dim(T\cap \R^1),\quad \dim(T\cap \R^2),\quad \dim(T\cap \R^3),\quad \dim(T\cap \R^4)
\]
is nondecreasing, and each successive increment is at most $1$.
It follows that there exists a unique pair $\sigma=(\sigma_1,\sigma_2)$ with
$1\le \sigma_1 < \sigma_2 \le 4$ such that
\[
\dim(T\cap \R^{\sigma_1-1})=0,\qquad \dim(T\cap \R^{\sigma_1})=1,
\]
and
\[
\dim(T\cap \R^{\sigma_2-1})=1,\qquad \dim(T\cap \R^{\sigma_2})=2.
\]
The pair $\sigma$ is called the \emph{Schubert symbol} of $T$.

In the case of planes in $\R^4$, the admissible Schubert symbols are
\[
(1,2),\ (1,3),\ (1,4),\ (2,3),\ (2,4),\ (3,4).
\]
If $e_\sigma$ denotes the set of planes with Schubert symbol $\sigma$, then we have a partition of $\GG(2,4)$:
\[
\GG(2,4)
=
e_{1,2}\sqcup e_{1,3}\sqcup e_{1,4}\sqcup e_{2,3}\sqcup e_{2,4}\sqcup e_{3,4}.
\]

Each set $e_\sigma$ is in fact homeomorphic to an open ball.
To describe this parametrization, define, for each $k\in\{1,\dots,4\}$, the subset
$H^k\subset \R^4$ consisting of vectors whose $k$-th coordinate is positive and whose last $4-k$ coordinates vanish.
Then every plane in $e_\sigma$ admits a unique orthonormal basis $(v_1,v_2)$ such that
\[
v_1\in H^{\sigma_1}\quad \text{and}\quad v_2\in H^{\sigma_2}.
\]
We refer to this basis as the \emph{reduced echelon form} of the plane.
In matrix form, the reduced echelon representatives of the six cells are of the following type
(the stars indicate arbitrary real entries, the symbol $+$ indicates a positive entry, and the two rows are orthonormal):
\[
e_{1,2}:\begin{pmatrix}
	1&0&0&0\\
	0&1&0&0
\end{pmatrix},
~~~~~~~~
e_{1,3}:\begin{pmatrix}
	1&0&0&0\\
	0&*&+&0
\end{pmatrix},
\qquad
e_{1,4}:\begin{pmatrix}
	1&0&0&0\\
	0&*&*&+
\end{pmatrix},
\]
\[
e_{2,3}:\begin{pmatrix}
	*&+&0&0\\
	*&*&+&0
\end{pmatrix},
\qquad
e_{2,4}:\begin{pmatrix}
	*&+&0&0\\
	*&*&*&+
\end{pmatrix},
\qquad
e_{3,4}:\begin{pmatrix}
	*&*&+&0\\
	*&*&*&+
\end{pmatrix}.
\]

Let $s_{\sigma_1}\in \R^4$ be the vector whose entries are all zero except for a $1$ in the $\sigma_1$-th position.
Then one obtains a homeomorphism $e_\sigma \to \mathring{B}^{\sigma_1-1}\times \mathring{B}^{\sigma_2-2}$ from the cell to a product of open balls by the composition
\[
(v_1,v_2)\mapsto (v_1,R(v_1,v_2))
\mapsto (v_1',R(v_1,v_2)'),
\]
where
\begin{itemize}
	\item 
		$
		R(v_1,v_2)
		=
		v_2-[((s_{\sigma_1}+v_1)\cdot v_2)/(1+s_{\sigma_1}\cdot v_1)](s_{\sigma_1}+v_1)
		+2(s_{\sigma_1}\cdot v_2)v_1,
		$
	\item 
		$v_1'$ denotes the vector formed by the first $\sigma_1-1$ coordinates of $v_1$,
	\item 
		$R(v_1,v_2)'$ denotes the vector formed by the first $\sigma_2-1$ coordinates of $R(v_1,v_2)$, with the $\sigma_1$-th coordinate removed.
\end{itemize}

The inverse homeomorphism $\mathring{B}^{\sigma_1-1}\times \mathring{B}^{\sigma_2-2}\to e_\sigma$
is given by
\[
(x,y)\mapsto (x',y')\mapsto (x',R^{-1}(x',y')),
\]
where $x'$ is obtained from $x$ by appending the coordinate $\sqrt{1-\|x\|^2}$, and
$y'$ is obtained from $y$ by appending the coordinate $\sqrt{1-\|y\|^2}$ and inserting a zero in the $\sigma_1$-th position.

Last, this homeomorphism extends continuously to a map $B^{\sigma_1-1}\times B^{\sigma_2-2}\to \overline{e_\sigma}$,
where $\overline{e_\sigma}$ denotes the closure in $\GG(2,4)$.
This proves that the Schubert cells form a CW structure.

\paragraph*{An exceptional homeomorphism}

A special feature of $\GG(2,4)$ is that it admits an explicit description as a quotient of a product of spheres; see Lin's notes \cite{Lin_RealGrassmannian}. 
More precisely, the Grassmannian $\GG^+(2,4)$ of \textit{oriented} $2$-planes in $\R^4$ is homeomorphic to $S^2\times S^2$, and the projection $\GG^+(2,4)\to \GG(2,4)$ is a double covering. 
The deck transformation is the involution $(x,y)\mapsto (-x,-y)$, so that
\[
\GG(2,4)\cong (S^2\times S^2)/((x,y) \sim (-x,-y)).
\]

This description suggests an alternative approach to the construction of a triangulation of $\GG(2,4)$: starting from a triangulation of $S^2\times S^2$ compatible with the involution and then passing to the quotient. 
We did not pursue this point of view further here.

\section{Proofs}
\label{sec:proofs}

\subsection{Proofs for Section~\ref*{sec:delaunay}}

\originstereographicprojection*

\begin{proof}[Proof of \Cref{lem:originstereographicprojection}]
	\label{proof:originstereographicprojection}
	Seeing the tangent space $T_x S^d$ as a linear hyperspace of $\R^{d+1}$, the stereographic projection of a point $y\in S^d$, with $y\neq -x$, can be written as
	\[
	p(y) = \frac{y-\langle y,x\rangle x}{1+\langle y,x\rangle}.
	\]
	In particular, $p(x) = 0$.
	Let $(b_0,\dots,b_d)$ be the barycentric coordinates of $r(x)$ in $[v_0,\dots,v_d]$.
	We craft barycentric coordinates for $p(x)$ in $[p(v_0),\dots,p(v_d)]$.
	For every $i\in\{0,\dots,d\}$, define
	\[
	b'_i = b_i\frac{1+\langle v_i,x\rangle}{1+\langle r(x), x\rangle}.
	\]
	These coefficients are non-negative and sum to 1:
	\begin{align*}
		\sum_{i=0}^{d} b_i' 
		&= \frac{1}{1+\langle r(x), x\rangle}\bigg(\sum_{i=0}^{d} b_i + \big\langle\sum_{i=0}^{d} b_i v_i, x\big\rangle\bigg)\\
		&= \frac{1}{1+\langle r(x), x\rangle}\big(1+\langle r(x),x\rangle\big)\\
		&= 1.
	\end{align*}
	On the other hand, one has
	\begin{align*}
		\sum_{i=0}^{d} b_i' p(v_i)
		&= \sum_{i=0}^{d} b_i \frac{1+\langle v_i,x\rangle}{1+\langle r(x), x\rangle}\cdot \frac{v_i-\langle v_i,x\rangle x}{1+\langle v_i,x\rangle}\\
		&= \frac{1}{1+\langle r(x), x\rangle}\sum_{i=0}^{d} b_i \big(v_i-\langle v_i,x\rangle x\big)\\
		&= \frac{1}{1+\langle r(x), x\rangle}\big(r(x) - \langle r(x),x\rangle x\big)\\
		&= 0.
	\end{align*}
	In other words, $p(x) = 0$ belongs to the convex hull of $p(v_0),\dots,p(v_d)$, as desired.
\end{proof}

\decreasecoveringradius*

\begin{proof}[Proof of \Cref{lem:decreasecoveringradius}]
	\label{proof:decreasecoveringradius}
	The proof is based on the three auxiliary lemmas stated below.
	To prove that $X\cup Y$ is sufficiently dense, we proceed facet by facet in $\Del{X}$.
	We begin by estimating the covering radius in the Euclidean facets (i.e., in the convex hulls of the vertices) using \Cref{lem:steiner_covering}.
	We then transfer this estimate to the sphere by means of \Cref{lem:spherical_distorsion_distances}.
	Finally, \Cref{lem:spherical_convexity} allows us to express the bound in terms of geodesic distance on the sphere.
	
	Given a facet $X_\sigma\in\Del{X}$, let $Y_\sigma$ denote the Euclidean Steiner points associated with $X_\sigma$, before projection on the sphere.
	According to \Cref{lem:steiner_covering}, for every point $y$ in the (Euclidean) simplex $\geomreal{\sigma}\subset\R^{d+1}$, there exists a vertex $x\in X_\sigma\cup Y_\sigma$ such that
	\[
	\|x-y\| \leq \alpha' \delta,
	\]
	where $\delta$ is the Euclidean circumradius of $X_\sigma$.
	It is linked to the spherical circumradius $\Delta$ via
	\[
	\delta = \sin(\Delta).
	\]
	Consequently, \Cref{lem:spherical_distorsion_distances} yields
	\[
	\left\|\frac{x}{\|x\|}-\frac{y}{\|y\|} \right\|
	\leq
	\frac{1}{\cos(\Delta)} \|x-y\|
	\leq
	\frac{1}{\cos(\Delta)} \alpha'\sin(\Delta).
	\]
	By \Cref{lem:equality_covering_circumradius}, the circumradius $\Delta$ is at most the covering radius $\rho_\mathrm{cov}(X)$.
	Taking the union over all facets, we conclude that for all $y\in S^d$, there exists $x\in X\cup Y\subset S^d$ such that
	\[
	\left\| x-y\right\|
	\leq
	\frac{\alpha'}{\cos(\rho_\mathrm{cov}(X))} \sin(\rho_\mathrm{cov}(X)).
	\]
	To conclude, we apply \Cref{lem:spherical_convexity} with $c=\alpha'/\cos(\rho_\mathrm{cov}(X))$ and $x=\sin(\rho_\mathrm{cov}(X))$: the geodesic distance is upper bounded by
	\[
	\d(x,y)
	\leq
	\frac{\alpha'}{\cos(\rho_\mathrm{cov}(X))} \rho_\mathrm{cov}(X).
	\]
	The lemma holds provided two conditions: that 
	\[
	\frac{\alpha'}{\cos(\rho_\mathrm{cov}(X))}\leq2,
	\] 
	which is true since $\alpha'\leq 1$ and $\rho_\mathrm{cov}(X)\leq \pi/3$ by assumption; and that 
	\[
	\sin(\rho_\mathrm{cov}(X))\leq1,
	\]
	which obviously holds.
\end{proof}

\begin{lemma}\label{lem:steiner_covering}
	Let $X=\{v_0,\dots,v_{d}\}\subset\R^{n}$, $\delta$ be their (Euclidean) circumradius, and $\geomreal{\sigma}\subset \R^{n}$ denote the (linear) simplex they span.
	Let $Y\subset\R^{n}$ be the Euclidean Steiner point(s) associated with $X$ (edge midpoints, the minicenter, or the centroid) before projection to the sphere.
	Then for every $y \in \geomreal{\sigma}$, there exists $x\in X\cup Y$ such that
	\[
	\|x-y\| \leq \alpha' \delta,
	\]
	where $\alpha'$ is defined in \Cref{lem:decreasecoveringradius}.	
\end{lemma}

\begin{proof}[Proof of \Cref{lem:steiner_covering}]
	We treat each refinement separately.
	
	\proofsubparagraph*{Edgewise refinement}
	The Steiner points $Y$ are the midpoints $(v_i+v_j)/2$ of pairs $(v_i,v_j)$ of distinct vertices in $X$.
	On the other hand, by Maurey's empirical method with two samples \cite[Theorem 0.0.2]{VershyninHDP2}, for any $y \in \geomreal{\sigma}$, there exists $x_1,x_2\in X$, possibly equal, such that
	\[
	\left\|y-\frac{x_1+x_2}{2} \right\|
	\leq
	\frac{\delta}{\sqrt{2}}.
	\]
	The case $x_1=x_2$ corresponds to a point in $X$, and the case $x_1\neq x_2$ to a point in $Y$.
	
	\proofsubparagraph*{Minicenter refinement}
	Let $w$ be $X$'s minicenter, and denote $r_*=\|y-w\|$.
	Moreover, write $y = \sum_{i=0}^{d}\lambda_i v_i$ in barycentric coordinates, and denote $r_i = \|y-v_i\|$ for all $i\in\{0,\dots,d\}$.
	We aim to show that at least one value among $r_*,r_0,\dots,r_d$ is at most $\delta/\sqrt{2}$.
	
	Since every vertex $v_i$ satisfies $\|w-v_i\|\leq \delta$, a direct computation shows that	
	\[
	\sum_{i=0}^{d}\lambda_i \|y-v_i\|^2 
	= \sum_{i=0}^{d}\lambda_i \left(\|y-w\|^2+\|w-v_i\|^2-2\langle y-w,w-v_i\rangle\right) 
	\leq \delta^2 - \|y-w\|^2. 
	\]  
	In other words, 
	\[
	r_*^2+	\sum_{i=0}^{d}\lambda_ir_i^2 \leq \delta^2.
	\]
	Let $t$ be the smallest value among $r_*,r_0,\dots,r_d$.
	From the previous equation, we deduce that $2 t^2 \leq \delta^2$, as desired.
	
	\proofsubparagraph*{Centroid refinement}
	This last refinement requires a finer geometric analysis.
	We start with the 2-dimensional case, i.e., $X = \{v_0,v_1,v_2\}$ is a triangle.
	We denote by $c$ its centroid.
	We must show that the distance from any point $y\in\geomreal{\sigma}$ to $\{c,v_0,v_1,v_2\}$ is at most $(2/3)\delta$.
	
	We define a closed cover of the triangle $\geomreal{\sigma}$ by four sets $\mathcal{C}(c)$, $\mathcal{C}(v_0)$, $\mathcal{C}(v_1)$, and $\mathcal{C}(v_2)$.
	For $i=0,1,2$, we denote by $v_i^1$ and $v_i^2$ the intersection points between the boundary of $\geomreal{\sigma}$ and the perpendicular bisector of the segment $[v_i,c]$, and we let $\mathcal{C}(v_i)$ be the triangle $[v_i, v_i^1, v_i^2]$.
	In addition, we define $\mathcal{C}(c)$ as the convex hull of $\{v_0^1, v_0^2, v_1^1, v_1^2, v_2^1, v_2^2\}$.
	As visualized in \Cref{subfig:proof_shrinking_Delaunay_centroid_voronoi}, this covering of $\geomreal{\sigma}$ coincides with the Voronoi tessellation of $\{c,v_0,v_1,v_2\}$ restricted to $\geomreal{\sigma}$, provided the triangle is sufficiently regular.
	
	\begin{figure}[!htbp]
		\centering
		\begin{subfigure}[t]{0.32\linewidth}
			\centering
			\includegraphics[width=.8\textwidth]{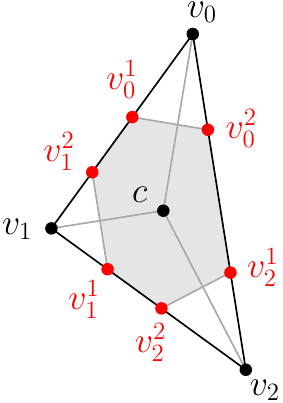}
			\subcaption{The triangle is covered by four polygonal cells.}
			\label{subfig:proof_shrinking_Delaunay_centroid_voronoi}
		\end{subfigure}
		\hfill
		\begin{subfigure}[t]{0.32\linewidth}
			\centering
			\includegraphics[width=.99\textwidth]{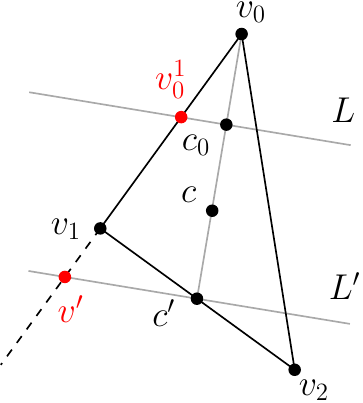}
			\subcaption{Configuration for the intercept theorem.}
			\label{subfig:proof_shrinking_Delaunay_centroid_2D}
		\end{subfigure}
		\hfill
		\begin{subfigure}[t]{0.32\textwidth}
			\centering
			\includegraphics[width=.85\textwidth]{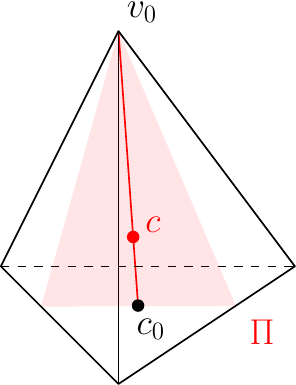}
			\subcaption{The section of a simplex by a plane through $v_0$ is a triangle.}
			\label{subfig:proof_shrinking_Delaunay_centroid_nD}
		\end{subfigure}
		\caption{Configurations involved in the proof of \Cref{lem:steiner_covering} for centroid refinement.}
	\end{figure}
	
	To prove the result, we must show that, for $i=0,1,2$, the points $v_i^1$ and $v_i^2$ lie at distance at most $(2/3)\delta$ from $c$ and $v_i$.
	Without loss of generality, we will prove it for $v_0^1$ only.	
	
	Let $c_0$ be the midpoint of $[c,v_0]$, and $L$ be the perpendicular bisector of the segment $[c,v_0]$.	
	Let $c'$ be the midpoint of the opposite edge $[v_1,v_2]$, and let $L'$ be the line parallel to $L$ and passing through $c'$.
	The ray $[v_0,v_1)$ intersects $L'$; we denote their intersection by $v'$.
	The two rays $[v_0,c')$ and $[v_0,v')$ emanate from $v_0$ and intersect the parallel segments $[v_0^1,c_0]$ and $[v',c']$; see \Cref{subfig:proof_shrinking_Delaunay_centroid_2D}.
	We can therefore apply the intercept theorem, which yields
	\begin{equation}
		\label{eq:proof_shrinking_Delaunay_centroid_2D}
		\frac{\|v_0-v_0^1\|}{\|v_0-v'\|} = \frac{\|v_0-c_0\|}{\|v_0-c'\|}.
	\end{equation}
	On the other hand, since the centroid $c$ divides the median $[v_0,c']$ in the ratio $2:1$, and since $c_0$ is the midpoint of $[v_0,c]$, we deduce that
	\begin{equation}
		\label{eq:proof_shrinking_Delaunay_centroid_2D_median}
		\frac{\|v_0-c_0\|}{\|v_0-c'\|} = \frac{1}{2}\cdot\frac{2}{3} = \frac{1}{3}.
	\end{equation}
	Together with \Cref{eq:proof_shrinking_Delaunay_centroid_2D}, we obtain
	\[
	\|v_0-v_0^1\| = \frac{1}{3}\|v_0-v'\|.
	\]
	To conclude, we observe that $\|v_0-v'\|$ is at most the diameter of the triangle, which is, in turn, at most twice the circumradius $\delta$.	
	We conclude that
	\[
	\|v_0-v_0^1\| \leq \frac{1}{3}\cdot2\cdot\delta.
	\]
	
	We now suppose that $\sigma$ has dimension $d$, with vertices $v_0,\dots,v_d$.
	Its centroid is still denoted by $c$.
	Let us choose $v_0$ as a base vertex, and denote by $c'$ the centroid of the opposite face $[v_1,\dots,v_d]$.
	Given any affine plane $P$ passing through $v_0$ and $c$, the section $P\cap\geomreal{\sigma}$ is a triangle $[v_0,v_1',v_2']$; see \Cref{subfig:proof_shrinking_Delaunay_centroid_nD}.
	In this triangle, we can repeat the argument above: the triangle is covered by cells, and \Cref{eq:proof_shrinking_Delaunay_centroid_2D} is still valid.
	\Cref{eq:proof_shrinking_Delaunay_centroid_2D_median}, however, has to be modified: the centroid $c$ now divides the median $[v_0,c']$ in the ratio $d:1$, yielding
	\[
	\frac{\|v_0-c_0\|}{\|v_0-c'\|} = \frac{1}{2}\frac{d}{d+1}.
	\]
	Together with $\|v_0-v'\|\leq 2\delta$, we deduce the result:	
	\[
	\|v_0-v_0^1\| \leq \frac{1}{2}\frac{d}{d+1} \cdot 2\delta = \frac{d}{d+1}\delta.
	\]
\end{proof}

\begin{lemma}\label{lem:spherical_distorsion_distances}
	Let $\sigma=\{v_0,\dots,v_{d}\}\subset S^d$, $\Delta$ be their (spherical) circumradius, and $\geomreal{\sigma}\subset \R^{d+1}$ denote the (Euclidean) simplex they span.
	Then for all $x,y \in \geomreal{\sigma}$, one has
	\[
	\left\|\frac{x}{\|x\|}-\frac{y}{\|y\|} \right\|
	\leq
	\frac{1}{\cos(\Delta)} \|x-y\|.
	\]
\end{lemma}

\begin{proof}[Proof of \Cref{lem:spherical_distorsion_distances}]
	Let $u\in S^d$ be the circumcenter of $\sigma$.
	Its great-circle distance to every vertex $v_i$ is $\Delta$.
	In terms of angles, this means that 
	\[
	\langle u, v_i \rangle = \cos(\Delta).
	\]
	In particular, the vertices all lie in the hyperplane
	\[
	H = \{v \in \R^{d+1} \mid \langle u, v \rangle = \cos(\Delta)\}.
	\]
	Consequently, $\geomreal{\sigma}\subset H$.
	By Cauchy–Schwarz, it follows that every $p\in \geomreal{\sigma}$ satisfies
	\[
	\|p\|\geq \cos(\Delta).
	\]	
	Next, let $\pi\colon \R^{d+1}\setminus\{0\} \to S^d$ be the projection onto the sphere.
	Its operator norm at $p$ is
	\[
	\|D \pi_p\|_\mathrm{op} = \frac{1}{\|p\|}.
	\]
	We deduce that $\pi$ is $1/\cos(\Delta)$-Lipschitz on $H$, and the result follows.
\end{proof}

\begin{lemma}\label{lem:spherical_convexity}
	Let $h\colon t \mapsto 2\arcsin(t/2)$ be the map that converts Euclidean into spherical distance.
	For all $c \in [0,2]$ and $x\in[0,1]$, it holds that $h(cx)\leq c\arcsin(x)$.
\end{lemma}

\begin{proof}[Proof of \Cref{lem:spherical_convexity}]
	This follows directly from the convexity of $h$ on the segment $[0, 2x]$:
	\begin{align*}
		h(cx) &= h\left(\left(1-\frac{c}{2}\right)\cdot0 + \left(\frac{c}{2}\right)\cdot 2x\right)\\ 
		&\leq \left(1-\frac{c}{2}\right)\cdot h(0) + \left(\frac{c}{2}\right)\cdot h\left(2x\right),
	\end{align*}
	which is equal to $c\arcsin(x)$.
\end{proof}

\subsection{Proofs for Section~\ref*{sec:simplicial_mapping_cones}}

\apothemnormalization*

\begin{proof}[Proof of \Cref{lem:apothemnormalization}]
	\label{proof:apothemnormalization}
	The first statement is obvious from the definition of $\nu$.
	As for the second statement, we consider a circular arc
	\[
	C = S(0,r) \cap P,
	\]
	where $S(0,r)$ is the sphere of radius $r$ and $P$ is a linear plane.
	Every simplex $\sigma\in K$ spans an affine hyperplane
	\[
	H = \{x\in \R^{d+1} \mid \langle x, n\rangle = h\}
	\]
	where $n$ is a normal vector and $h$ a scalar.
	Consider a point $r u \in C$ on the arc, where $u$ is a unit vector, and such that the ray $\{tu\mid t\geq 0\}$ intersects $\geomreal{\sigma}$.
	In particular, $J^{-1}(u)u$ is the intersection point of the line segment $[0,u]$ and the hyperplane $H$.
	Consequently,
	\[
	\langle J^{-1}(u)u, n\rangle = h.
	\]
	Similarly, the point $\nu^{-1}(ru)=J^{-1}(u)ru$ satisfies
	\[
	\langle \nu^{-1}(ru), n\rangle = rh.
	\]
	In other words, it belongs to the hyperplane
	\[
	H' = \{x\in \R^{d+1} \mid \langle x, n\rangle = rh\},
	\]
	which is parallel to $H$.
	On the other hand, $\nu^{-1}(ru)$ still lies in the plane $P$.
	This shows that the section of $C$ in the sector defined by $\sigma$ is included in $H'\cap P$, as claimed.
\end{proof}

\equivalencemappingcones*

\begin{proof}[Proof of \Cref{prop:equivalencemappingcones}]
	\label{proof:equivalencemappingcones}
	We first give the topological mapping cone $\geomreal{L}\cup_{\geomreal{g}} \geomreal{B(K)}$ a CW structure.
	Since $B(K)$ and $L$ are simplicial complexes, hence CW complexes, the quotient map $\geomreal{L} \sqcup \geomreal{B(K)} \rightarrow \geomreal{L}\cup_{\geomreal{g}} \geomreal{B(K)}$ restricted to each simplex can be seen as a characteristic map. 
	Their collection forms a CW structure (more particularly, a $\Delta$-complex structure).
	
	Now, we inspect the fibers of $q$ and show they are contractible.
	Both the codomain $\geomreal{L}$ and the inner ball of $\geomreal{B(K)}$ are mapped injectively into $\geomreal{L\cup_g B(K)}$ via $q$, hence the fibers over these points are singletons.
	
	Next, let $\bar{x}$ be a point in the outer shell of $\geomreal{B(K)}$, viewed in $\geomreal{L\cup_g B(K)}$.
	Let $\bar{\sigma}$ be its carrier simplex.
	Its vertices decompose into inner vertices $v_k,\dots,v_d$ and vertices $w_0,\dots,w_l$ of $L$.
	We label the inner vertices according to the local order used for staircase triangulation (see \Cref{subsec:triangulation_ball}).
	Let $(\bar{b}_k,\dots,\bar{b}_d,\bar{c}_0,\dots,\bar{c}_l)$ denote the barycentric coordinates of $\bar{x}$ in $\bar{\sigma}$.

	A point $y\in\geomreal{B(K)}$ maps to $\bar{x}$ in $\geomreal{L\cup_g B(K)}$ if it lies in an outer-shell simplex $\sigma_k = [v_k,\dots,v_d,v_0',\dots,v_k']$ whose inner vertices coincide with those of $\bar{\sigma}$, and whose outer vertices map to those of $\bar{\sigma}$.
	In other words, the following sets are equal:
	\[
	\{g(v_k'),\dots,g(v_d')\}  = \{w_0,\dots,w_l\}.
	\]
	For all $i\in\{0,\dots,l\}$, define the nonempty set
	\[
	V(i) = \{j\in\{0,\dots,k\} \mid  g(v_j') = w_i \}.
	\]
	
	Let $(b_k,\dots,b_d,b'_0,\dots,b_k')$ be the barycentric coordinates of $y$ in $\sigma_k = [v_k,\dots,v_d,v_0',\dots,v_k']$.
	If $q(y)=\bar{x}$, then the coordinates of the inner vertices must agree:
	\[
	\bar{b}_i = b_i  \text{ for all } i\in\{k,\dots,d\},
	\]
	and the coordinates of the outer vertices must match after passing to the quotient:
	\[
	\bar{c}_i = \sum_{j \in V(i) } b'_j \text{ for all } i\in\{0,\dots,l\}.
	\]
	
	These linear constraints define convex polytopes inside the cells (each is the intersection of a simplex with an affine subspace), and the fiber $q^{-1}(\bar{x})$ is a finite union of such polytopes.	
	Moreover, their union is contractible in $\geomreal{L}\cup_{\geomreal{g}} \geomreal{B(K)}$.
	Indeed, each cell deformation retracts linearly onto an arbitrary common point $x_0 \in q^{-1}(\bar{x})$.
	
	Finally, by Smale's Vietoris mapping theorem (under the hypotheses in our setting), a surjective map between finite CW complexes with contractible fibers is a weak homotopy equivalence \cite{smale1957vietoris}.
	Since both spaces are CW complexes, Whitehead's theorem upgrades this to a homotopy equivalence, which proves the claim.
	We note that the same argument can also be phrased in the language of \emph{cell-like} maps; see \cite[Theorem~1.3]{lacher1969cell}.
\end{proof}

\begin{remark}
	Although $K$ can be ``filled'' in many ways to obtain a triangulation $B(K)$ of $B^{d+1}$ (even without introducing additional vertices), our problem requires introducing new vertices.
	This is illustrated by the following example.
	If a graph $K$ is a cycle with $3p$ vertices and $g\colon K\to \partial\Delta^2$ is a simplicial map of degree $p$, then the simplicial gluing $\partial\Delta^2\cup_g B(K)$ has exactly the three vertices of $\partial\Delta^2$ plus the new vertices added in $B(K)$.
	In particular, for the simplicial gluing to faithfully model the mapping cone of $g$ as $p\to\infty$ (which has infinitely many homotopy types), the number of new vertices in $B(K)$ must grow without bound.
\end{remark}

\raydescendstoquotient*

\begin{proof}[Proof of \Cref{lem:raydescendstoquotient}]
	\label{proof:raydescendstoquotient}
	Let $p\colon\geomreal{B(K)}\to\geomreal{L\cup_g B(K)}$ denote the projection map, and consider two points $x,y\in\geomreal{B(K)}$ such that $p(x)=p(y)$.
	If $x$ belongs to the inner ball, then its equivalence class in the quotient is a singleton, hence $x=y$ and the claim is immediate.
		
	Henceforth assume that $x$ (and therefore $y$) lies in the outer shell.
	Given $s\in[0,1]$, consider the following points of $\geomreal{B(K)}$:
	\begin{align*}
		x(s) &= s x + (1-s) \frac{x}{\|x\|},\\
		y(s) &= s y + (1-s) \frac{y}{\|y\|}.
	\end{align*}
	We show that $p(x(s)) = p(y(s))$ holds for all $s\in[0,1]$, which proves the result.
	Our strategy consists in computing explicitly the barycentric coordinates of $x(s)$ in simplices of $B(K)$, and showing that their images in the quotient coincide.
	
	\proofsubparagraph*{Quotient barycentric coordinates}

	To start, let $\sigma^x$ be the carrier of $x$ in $B(K)$ (the unique simplex in which it admits positive barycentric coordinates).
	The vertices of $\sigma^x$ can be divided into \textit{inner vertices} $V_\mathrm{inner}(x) = \{v_0^x,\dots,v_k^x\}$ and \textit{outer vertices} $V_\mathrm{outer}(x) = \{w_0^x,\dots,w_m^x\}$:
	\[
	\sigma^x = V_\mathrm{inner}(x)\sqcup V_\mathrm{outer}(x).
	\]
	They are associated with positive barycentric coordinates $\{b_0^x,\dots,b_k^x\}$ and $\{c_0^x,\dots,c_m^x\}$, respectively.
	Similarly, if $\sigma^y$ denotes the carrier of $y$, then we can decompose
	\[
	\sigma^y = V_\mathrm{inner}(y)\sqcup V_\mathrm{outer}(y),
	\]
	with vertices $\{v_0^y,\dots,v_l^y\}$ and $\{w_0^y,\dots,w_n^y\}$, and barycentric coordinates $\{b_0^y,\dots,b_l^y\}$ and $\{c_0^y,\dots,c_n^y\}$, respectively.
	
	The inner vertices are untouched by the quotient. In particular, one has
	\[
	V_\mathrm{inner}(x) = V_\mathrm{inner}(y),
	\]
	and their barycentric coordinates coincide: $k=l$ and $b_i^x=b_i^y$ for $i\in\{0,\dots,k\}$; we denote it by $b_i$.
	On the other hand, the outer vertices coincide in the quotient:
	\[
	p(V_\mathrm{outer}(x)) = p(V_\mathrm{outer}(y)),
	\]
	and their barycentric coordinates agree, after taking the sum over the preimages.
	More precisely, given a vertex $\bar z \in p(V_\mathrm{outer}(x))$, define the indices 
	\begin{align*}
		F^{\bar z}(x) &= \{i \in \{0,\dots,m\}\mid p(w_i^x) = \bar z \},\\
		F^{\bar z}(y) &= \{i \in \{0,\dots,n\}\mid p(w_i^y) = \bar z \}.
	\end{align*}
	Then it holds that
	\[
	\sum_{i \in F^{\bar z}(x)} c_i^x = \sum_{i \in F^{\bar z}(y)} c_i^y.
	\]
	Let us denote by $c(\bar z)$ this quantity.
	We show that the paths $s\mapsto p(x(s))$ and $s\mapsto p(y(s))$ only depend on the \textit{inner coordinates} $b_i, i\in\{0,\dots,k\}$ and \textit{quotient coordinates} $c(\bar z), \bar z \in p(V_\mathrm{outer}(x))$.

	\proofsubparagraph*{Barycentric coordinates at ridges}
	
	From now on, we change notation and work in the staircase triangulation.	
	Let $\sigma=[v_0',\dots,v_d']$ be a facet of the outer layer $K\subset B(K)$ such that $x$ belongs to $\sect{\sigma}$ and let $\sigma_k = [v_k,\dots,v_d,v_0',\dots,v_k']$ be a simplex in $\sect{\sigma}$ that contains $x$.
	The inner and outer vertices are respectively $\{v_k,\dots,v_d\}$ and $\{v_0',\dots,v_k'\}$.
	By the staircase construction, the ray from the origin through $x$ crosses the facets of the outer shell in consecutive order:
	\[
	\sigma_0<\sigma_1<\dots<\sigma_d.
	\]
	This has been observed in \Cref{subsec:triangulation_ball} (see  \Cref{fig:triangulation_prism_2}).
	The outer vertices are added one by one and do not leave the simplices: if $v'$ is an outer vertex in $\sigma_l$, then $v'\in\sigma_m$ for all $m\geq l$.
	
	For $l\in\{k,\dots,d-1\}$, define the \textit{ridge} $\tau_l=\sigma_l\cap\sigma_{l+1}$.
	The radial ray from $x$ intersects $\geomreal{\tau_l}$ in a unique point, denoted $x_l$, whose barycentric coordinates can be explicitly described.
	We compute $x_k$ in \Cref{claim:coordinates_quotient_first}, and the other $x_{k+1},\dots,x_d$ in \Cref{claim:coordinates_quotient_rest}.
	
	We start with the first intersection point, $x_k$.
	To simplify the notation, let us order the vertices of $\sigma_k$ as $ [v_0',\dots,v_k',v_k,\dots,v_d]$, and the vertices of $\tau_k$ as $ [v_0',\dots,v_k',v_{k+1},\dots,v_d]$ (only $v_k$ is dropped).
	From the barycentric coordinates $(b_0',\dots,b_k',b_k,\dots,b_d)$ of $x$ in $\sigma_k$, one obtains the coordinates of $x_k$ in $\tau_k$ by replacing the pair $(b_k',b_k)$ with $b_k'+\rho_{\mathrm{inner}}b_k$, where $\rho_{\mathrm{inner}}$ is the norm of the inner vertices, and normalizing all coordinates to sum to 1.
	
	\begin{claim}\label{claim:coordinates_quotient_first}
		The coordinates of $x_k$ in $\tau_k=[v_0',\dots,v_{k-1}',v_k',v_{k+1},\dots,v_d]$ are 
		\[
		\frac{1}{1-(1-\rho_{\mathrm{inner}})b_k}\big(b_0',\dots,b_{k-1}',b_k'+\rho_{\mathrm{inner}}b_k,b_{k+1},\dots,b_d\big).
		\]
	\end{claim}
	
	\begin{claimproof}
		Using that $v_i = \rho_{\mathrm{inner}}v_i'$, the barycentric coordinate decomposition reads
		\begin{align*}
			x &= b_0'v_0'+\dots+\underline{b_k'v_k'+b_kv_k}+\dots+b_d v_d\\
			&= b_0'v_0'+\dots+\underline{(b_k'+\rho_{\mathrm{inner}}b_k)v_k'}+\dots+b_d v_d.
		\end{align*}
		Now, the point $tx$, with $t>0$, belongs to $\geomreal{\tau_k}$ whenever the vector
		\[
		t(b_0',\dots,b_k'+\rho_{\mathrm{inner}}b_k,\dots,b_d)
		\]
		sums to 1, that is, when
		\[
		t = 1/(b_0'+\dots+b_k'+\rho_{\mathrm{inner}}b_k+\dots+b_d).
		\]
		Using that 
		\[
		b_0'+\dots+b_k'+\rho_{\mathrm{inner}}b_k+\dots+b_d = (1-b_k) + \rho_{\mathrm{inner}}b_k,
		\]
		we deduce
		\[
		t = 1/\big(1-(1-\rho_{\mathrm{inner}})b_k).\claimqedhere{}
		\]
	\end{claimproof}

	We already see that these coordinates descend to the quotient. 	
	The behavior for the next intersection points is similar.
	More precisely, the barycentric coordinates of $x_{l}$ in $\tau_{l}$ can be obtained from those of $x_{k}$ in $\tau_{k}$ by multiplying all corresponding inner-vertex coordinates by $\rho_{\mathrm{inner}}$, and normalizing accordingly.
	
	\begin{claim}\label{claim:coordinates_quotient_rest}
		For $l\in\{k,\dots,d-1\}$, the coordinates of $x_l$ in $\tau_l=[v_0',\dots,v_l',v_{l+1},\dots,v_d]$ are 
		\[
		\frac{1}{1-(1-\rho_{\mathrm{inner}})(b_k+\dots+b_l)} 
		\big(\widetilde b_0, \dots, \widetilde b_d\big)
		\]
		where the unnormalized coordinates are
		\[
		\widetilde b_i =
		\begin{cases}
			b_i' & \text{if}\ i<k 
				\quad~~~~~(\text{at}\ v_i'),\\
			b_i'+ \rho_{\mathrm{inner}}b_i& \text{if}\ i=k 
				\quad~~~~~(\text{at}\ v_k'), \\
			\rho_{\mathrm{inner}}b_i & \text{if}\ k<i\leq l 
				\quad(\text{at}\ v_i'), \\
			b_i & \text{if}\ l<i\leq d 
				\quad(\text{at}\ v_i).
		\end{cases}
		\]
	\end{claim}
	
	\begin{claimproof}
		We prove the claim by induction.
		The base case $l=k$ is the content of \Cref{claim:coordinates_quotient_first}.
		Next, assume the statement holds for $x_{l-1}$.
		The barycentric coordinates of $x_{l-1}$ in $\tau_{l-1}=[v_0',\dots v_k',\dots,v_{l-1}',v_{l},\dots,v_d]$ are:
		\[
		r\big(
		\underset{ v_0'}{\underline{\vphantom{\rho_{\mathrm{inner}} b_k}\,
				b_0'}},
		\dots,
		\underset{ v_k'}{\underline{\vphantom{\rho_{\mathrm{inner}} b_k}\,
				b_k' + \rho_{\mathrm{inner}} b_k}},
		\dots,
		\underset{ v_{l-1}'}{\underline{\vphantom{\rho_{\mathrm{inner}} b_k}\,
				\rho_{\mathrm{inner}} b_{l-1}}},
		\underset{ v_l}{\underline{\vphantom{\rho_{\mathrm{inner}} b_k}\,
				b_l}},
		\dots,
		\underset{ v_d}{\underline{\vphantom{\rho_{\mathrm{inner}} b_k}\,
				b_d}}
		\big)
		\]
		where $r =1 /\big(1-(1 - \rho_{\mathrm{inner}})(b_k+\dots+b_{l-1})\big)$.
		Indeed, the coordinates sum to
		\[
		b_0'+\dots+b_k'+\rho_{\mathrm{inner}}(b_k+\dots+b_{l-1})+b_l\dots+b_d = 1-(1 - \rho_{\mathrm{inner}})(b_k+\dots+b_{l-1}).
		\]
		We obtain the barycentric coordinates of $x_{l-1}$ in the coface $\sigma_l = [v_0',\dots,v_k',\dots,v_{l-1}',v_l',v_l,\dots,v_d]$ by inserting a zero at the coordinate $v_l'$:
		\[
		r\big(
		\underset{ v_0'}{\underline{\vphantom{\rho_{\mathrm{inner}} b_k}\,
				b_0'}},
		\dots,
		\underset{ v_k'}{\underline{\vphantom{\rho_{\mathrm{inner}} b_k}\,
				b_k' + \rho_{\mathrm{inner}} b_k}},
		\dots,
		\underset{ v_{l-1}'}{\underline{\vphantom{\rho_{\mathrm{inner}} b_k}\,
				\rho_{\mathrm{inner}} b_{l-1}}},
		\underset{ v_l'}{\underline{\vphantom{\rho_{\mathrm{inner}} b_k}\,
		 		0}},
		\underset{ v_l}{\underline{\vphantom{\rho_{\mathrm{inner}} b_k}\,
				b_l}},
		\dots,
		\underset{ v_d}{\underline{\vphantom{\rho_{\mathrm{inner}} b_k}\,
				b_d}}
		\big)
		\]
		As in the proof of \Cref{claim:coordinates_quotient_first}, we use $v_{l} = \rho_{\mathrm{inner}}v_l'$ to write $x_{l-1}$ as		
		\begin{align*}
			x_{l-1}/r &= 
				b_0'v_0'
				+\dots
				+(b_k'+\rho_{\mathrm{inner}}b_k)v_k'
				+\dots
				+\rho_{\mathrm{inner}}b_{l-1}v_{l-1}'
				+\underline{0\cdot v_l'+b_l v_l}
				+\dots
				+b_d v_d\\
					&= 
				b_0'v_0'
				+\dots
				+(b_k'+\rho_{\mathrm{inner}}b_k)v_k'
				+\dots
				+\rho_{\mathrm{inner}}b_{l-1}v_{l-1}'
				+\underline{ \rho_{\mathrm{inner}} b_l v_l'}
				+\dots
				+b_d v_d.
		\end{align*}
		The point $x_{l}$ is obtained as the scaling $tx_{l-1}$, $t>0$, for which the vector 
		\[
		t\cdot r\big(
		\underset{ v_0'}{\underline{\vphantom{\rho_{\mathrm{inner}} b_k}\,
				b_0'}},
		\dots,
		\underset{ v_k'}{\underline{\vphantom{\rho_{\mathrm{inner}} b_k}\,
				b_k' + \rho_{\mathrm{inner}} b_k}},
		\dots,
		\underset{ v_{l-1}'}{\underline{\vphantom{\rho_{\mathrm{inner}} b_k}\,
				\rho_{\mathrm{inner}} b_{l-1}}},
		\underset{ v_l'}{\underline{\vphantom{\rho_{\mathrm{inner}} b_k}\,
				\rho_{\mathrm{inner}}b_l}},
		\dots,
		\underset{ v_d}{\underline{\vphantom{\rho_{\mathrm{inner}} b_k}\,
				b_d}}
		\big)
		\]
		sums to $1$. 
		The claim follows.
	\end{claimproof}
	
	\proofsubparagraph*{Barycentric coordinates along the ray}
		
	We can now prove the lemma.
	\Cref{claim:coordinates_quotient_rest} shows that, while following the radial ray through $x$, inner coordinates $b_i$ associated to $v_i$ are transferred to the corresponding outer vertex $v_i'$, with a scale factor $\rho_\mathrm{inner}$.
	More precisely, for each ridge $\tau_l=[v_0',\dots,v_l',v_{l+1},\dots,v_d]$, the barycentric coordinates of the intersection point $x_l$ are
	\[
	\lambda_{v_i}(x_l)=\frac{b_i}{D_l},
	\]
	for inner vertices $v_i$, $i\in\{l+1,\dots,d\}$, and where $D_l=1-(1-\rho_{\mathrm{inner}})(b_k+\cdots+b_l)$ is the normalizing factor from
	\Cref{claim:coordinates_quotient_rest}.
	On the other hand, given an outer vertex $\bar z$ in the quotient, the corresponding barycentric coordinate is
	\[
	\lambda_{\bar z}\big(p(x_l)\big)
	=\frac{1}{D_l}\bigg(
	\sum_{i\in F^{\bar z}_\mathrm{outer}} b_i' 
	+
	\rho_{\mathrm{inner}} \sum_{i\in F^{\bar z}_\mathrm{inner}}  b_i
	\bigg),
	\]
	where we define the fiber index sets
	\begin{align*}
		F^{\bar z}_\mathrm{outer} &= \{i \in \{0,\dots,k\}\mid p(v_i') = \bar z \},\\
		F^{\bar z}_\mathrm{inner} &= \{i \in \{k,\dots,l\}\mid p(v_i') = \bar z \}.
	\end{align*}
	In particular, $\lambda_{\bar z}\big(p(x_l)\big)$ depends only on the inner coordinates $\{b_i\mid k\leq i\leq l\}$ and on the total mass
	$\sum_{i\in F^{\bar z}_\mathrm{outer}} b_i'$ and $\sum_{i\in F^{\bar z}_\mathrm{inner}} b_i$ in the fiber.
		
	Since $p(x)=p(y)$, these inner coordinates and fiber-sums agree; hence $p(x_l)=p(y_l)$ for every $l$.
	On each simplex $\sigma_l$, the projection map $p$ is affine, so it preserves barycentric
	interpolation along the segment of the ray contained in $\sigma_l$.
	It follows that $p(x(s))=p(y(s))$ for all $s\in[0,1]$, as claimed.
\end{proof}

\subsection{Proofs for Section~\ref*{sec:approximation}}

\lecdatagluing*

\begin{proof}[Proof of \Cref{th:lec_data_gluing}]
	\label{proof:lec_data_gluing}
	Let $(U^L,\Pi^L)$ denote the prescribed LEC-data or local motion planner on $\geomreal{L}$.
	We start by defining a planner on the simplicial ball $\geomreal{B(K)}$.
	To ensure it survives the quotient $\geomreal{L\cup_g B(K)}$, we restrict our construction to a few elementary paths, and prove in \Cref{claim:representatives} and \Cref{claim:elementary_paths} that they descend to the quotient.
	This allows us to define a local motion planner on the simplicial ball; see \Cref{claim:planner_ball}.
	By taking the quotient distance, we obtain a planner on the \textit{interior} $\geomreal{\mathring{B}(K)/g}$ of the quotient ball, as in \Cref{claim:planner_quotient}.
	This will be spliced with the planner already defined on $L$, yielding in \Cref{claim:planner_boundary} a genuine planner on $\geomreal{B(K)/g}$.
	Last, we show in \Cref{claim:planner_full} how to extend it to the simplicial mapping cone.
	
	Extending a planner from a space $X$ to a gluing $Y\cup_f X$ encounters two difficulties, illustrated in \Cref{fig:extending_LEC_difficulties}.
	First, when $f\colon A\subset X\to Y$ is not injective, a point $y\in Y$ may have several preimages in $X$. 
	In this case, a path from a given $x\in X$ to $y$ should first advance towards the boundary of $X$, independently from $y$, and then connect to $y$.
	Second, a given path in $X$ between $x,y\in X$, when projected onto $Y$, may not coincide with the path in $Y$ between their projections.
	This is remedied by introducing a ``buffer band'' to interpolate between them, in the spirit of Dyer and Eilenberg's construction \cite{dyer_adjunction_1972}.
		
	\begin{figure}[!htbp]
		\centering
		\subcaptionbox{
			When $y\in Y$ admits several preimages, connecting $x\in X$ to $y$ requires pushing $x$ towards the boundary, followed by a path in $Y$.
		}[0.49\textwidth]{%
			\includegraphics[width=.8\linewidth]{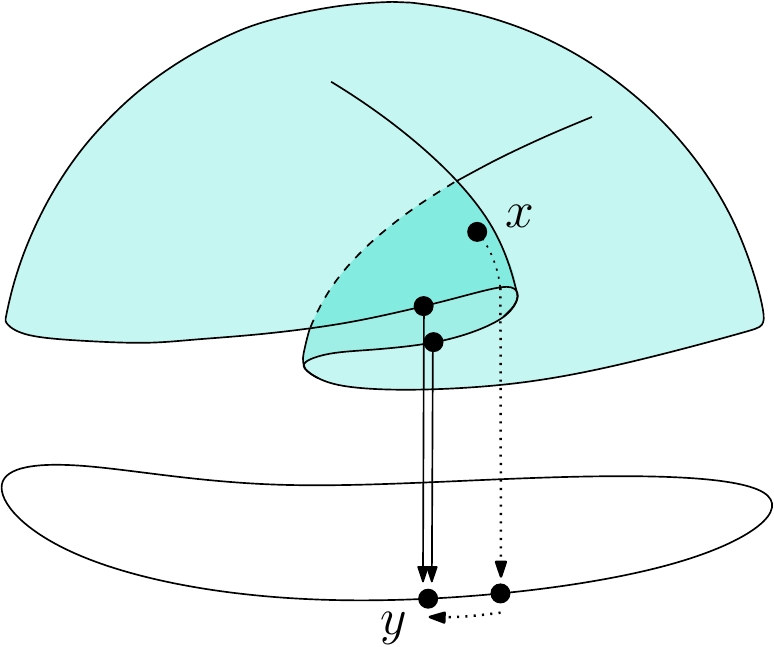}
			\vspace{.8cm}}
		\hfill
		\subcaptionbox{
			If $x,y \in X$ are close to the boundary, their path in $X$ must be interpolated with the path between their projections in $Y$.
		}[0.49\textwidth]{%
			\includegraphics[width=.8\linewidth]{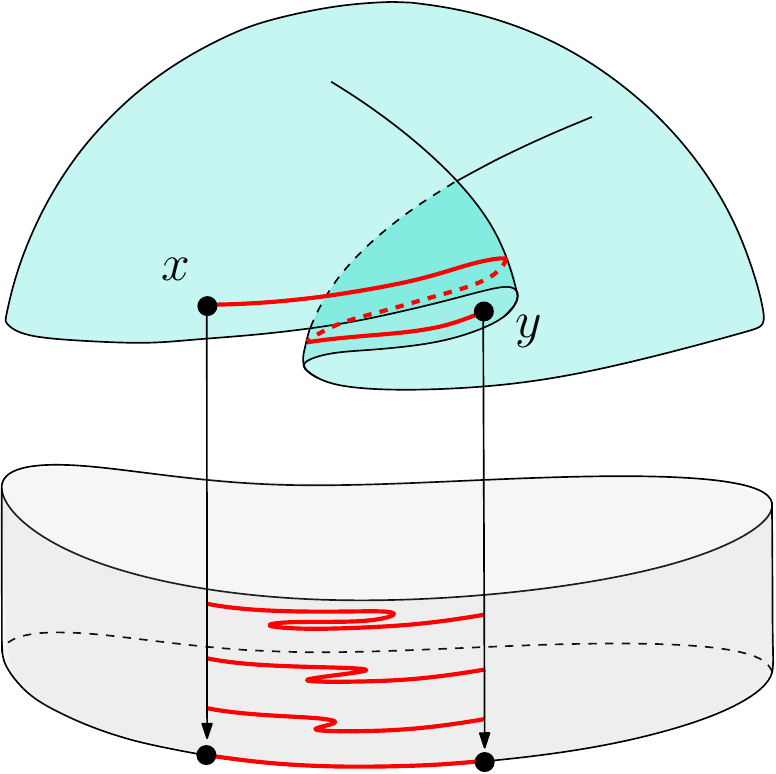}}
		\caption{
			To join two equiconnecting maps along a gluing $Y\cup_f X$, interpolation is required. This is already the case when $X$ has dimension 1 and the gluing ``folds''.}
		\label{fig:extending_LEC_difficulties}
	\end{figure}
	
	\proofsubparagraph*{Elementary paths}
	
	To define paths $\Pi(x,y,t)$ on the simplicial mapping cone, our strategy is to consider paths $\gamma \colon I\to \geomreal{B(K)}$ in the simplicial ball and take their image under $\geomreal{B(K)}\to \geomreal{L\cup_g B(K)}$.
	However, since simplices of $B(K)$ may degenerate, $\Pi$ may not descend to the quotient; this has been observed at the end of \Cref{subsec:equivalence_mapping_cones}.	
	We are compelled to consider only paths on $\geomreal{B(K)}$ that are well-defined after arbitrary identification of outer vertices.
	More precisely, given a simplex $\sigma\in B(K)$ and base point $x\in\geomreal{\sigma}$, we will build two kinds of paths, $\gamma^x_\mathrm{ray}$ and $\gamma^x_\mathrm{climb}\colon I\to \geomreal{B(K)}$, which satisfy the following property: for any other $x'\in\geomreal{B(K)}$ that is equal to $x$ after quotient, the corresponding paths $\gamma_\mathrm{ray}^x$ and $\gamma_\mathrm{ray}^{x'}$ (and similarly for $\gamma^x_\mathrm{climb}$ and $\gamma^{x'}_\mathrm{climb}$) have the same image in $\geomreal{L\cup_g B(K)}$.
	Moreover, given another $y\in\geomreal{B(K)}$ at equal distance to the origin, we will define $\gamma^{x\leadsto y}_\mathrm{arc},\gamma^{x\leadsto y}_\mathrm{straight}\colon I\to \geomreal{B(K)}$ that satisfy the analogous property with respect to $(x,y)$.
	We do not require the paths after quotient to be equally parametrized; choosing a parametrization is handled in \Cref{claim:planner_quotient}.
	
	We note that identifying boundary vertices of $B(K)$ does not affect the inner ball.
	Thus, every path in the inner ball descends to the quotient;
	the difficulty lies solely in the outer shell.
	Accordingly, certain paths will be defined by distinguishing between these cases.
	
	On the other hand, when $g$ is 2-distance injective, then the interior $\geomreal{\mathring{B}(K)}$ of $B(K)$ is mapped injectively into the quotient $\geomreal{L\cup_g B(K)}$; our construction in this case is simpler.

	We will make use of the \textit{inner representative} map 
	\[
	R_{\mathrm{inner}} \colon \geomreal{\mathring{B}(K)}\to \geomreal{B(K)},
	\]
	defined on the interior of $B(K)$ and taking values in the inner ball of $B(K)$.
	It is obtained from any $x\in\geomreal{\mathring{B}(K)}$ by identifying the simplex $\sigma\in B(K)$ to which it belongs, and zeroing out the barycentric coordinates corresponding to outer vertices.
	More precisely, if $x$ belongs to an inner simplex, then $R_{\mathrm{inner}}(x)=x$.
	Otherwise, $x$ is in a simplex $\sigma_k=[v_k,\dots,v_d,v_0',\dots,v_k']$ of the outer shell (see the staircase construction in \Cref{subsec:triangulation_ball}).
	If $(b_k,\dots,b_d,b_0',\dots,b_k')$ are its barycentric coordinates, we define $R_{\mathrm{inner}}(x)$ as the point obtained by zeroing out the last $k+1$ coordinates (corresponding to outer vertices). Accordingly, $R_{\mathrm{inner}}(x)$ belongs to the inner layer $K\subset B(K)$.
	Using the radial normalization $\nu\colon\geomreal{B(K)}\to  B^{d+1}$ defined in \Cref{subsec:triangulation_ball}, this means that $\|\nu(R_{\mathrm{inner}}(x))\| = \rho_{\mathrm{inner}}$ whenever $\|\nu(x)\|\geq \rho_{\mathrm{inner}}$.
	
	Similarly, we define the \textit{equal-norm representative} map 
	\[
	R_{\mathrm{equal}} \colon \geomreal{\mathring{B}(K)}\to \geomreal{B(K)}
	\]
	by first applying $R_{\mathrm{inner}}$ and then radially rescaling to obtain a point of the same norm as $x$ (after radial normalization).
	More precisely, given $x\in\geomreal{\mathring{B}(K)}$, $R_{\mathrm{equal}}(x)$ is the point on the ray $[0, R_{\mathrm{inner}}(x)/\|R_{\mathrm{inner}}(x)\|)$ such that $\|\nu(R_{\mathrm{equal}}(x))\| = \|\nu(x)\|$.
	We note that $R_{\mathrm{equal}}(x)=x$ when $x$ belongs to the inner ball.
	Both the inner and equal-norm representative maps descend to the quotient, meaning that they induce well-defined maps $\geomreal{\mathring{B}(K)}\to\geomreal{\conesimp{g}}$.
	They are visualized in \Cref{fig:representative_maps}.
		
	\begin{claim}\label{claim:representatives}
		If $x,y\in\geomreal{\mathring{B}(K)}$ coincide in the quotient $\geomreal{L\cup_g B(K)}$, then $R_{\mathrm{inner}}(x)=R_{\mathrm{inner}}(y)$ and $R_{\mathrm{equal}}(x)=R_{\mathrm{equal}}(y)$ in $\geomreal{B(K)}$.
	\end{claim}
	
	\begin{claimproof}
		We can suppose that $x$ and $y$ lie in the outer shell, otherwise their equivalence classes are singletons and the result is trivial.
		Consider a sector $\sect{\sigma}$ that contains $x$, and denote the vertices of $\sigma$, seen in the outer layer, as $v_0',\dots,v_d'$.
		It is associated to a simplex $[v_0,\dots,v_d]$ in the inner layer.
		The point $x$ belongs to a simplex $\sigma_k = [v_k,\dots,v_d,v_0',\dots,v_k']$ of the staircase triangulation of the prism on $\sigma$.
		Denote the barycentric coordinates of $x$ in $\sigma_k$ as $(b_k,\dots,b_d,b_0',\dots,b_k')$.
		Its representative $R_{\mathrm{inner}}(x)$ is the point of $B(K)$ in the inner simplex $[v_k,\dots,v_d]$ with barycentric coordinates $(b_k,\dots,b_d)$.
		Let us consider the other point $y$.
		It may belong to $\sect{\sigma}$ or to another sector $\sect{\tau}$.
		As before, let $w_0',\dots,w_d'$ be the vertices of $\tau$, $\tau_l = [w_l,\dots,w_d,w_0',\dots,w_l']$ a facet that contains $y$, and $(c_l,\dots,c_d,c_0',\dots,c_l')$ its barycentric coordinates.
		We consider the inner vertices of $\sigma_k$ and $\tau_l$ associated to nonzero barycentric coordinates:
		\begin{align*}
			V(x) &= \{v_i \mid i \in \{k,\dots,d\}, b_i > 0 \},\\
			V(y) &= \{w_i \mid i \in \{l,\dots,d\}, c_i > 0 \}.
		\end{align*}
		Since $x$ and $y$ coincide after quotient, we have $V(x)=V(y)$.
		Moreover, the associated barycentric coordinates must coincide.
		Therefore, $R_{\mathrm{inner}}(x)=R_{\mathrm{inner}}(y)$.
		
		The result for equal-norm representatives directly follows since  $R_{\mathrm{equal}}(x)$ and $R_{\mathrm{equal}}(y)$ are, by definition, built from $R_{\mathrm{inner}}(x)$ and $R_{\mathrm{inner}}(y)$.
	\end{claimproof}
	
	\begin{figure}[!htbp]
		\centering
		\subcaptionbox{
			$R_{\mathrm{inner}}(x)$ is obtained by zeroing out the barycentric coordinates of $x$ associated to outer vertices. 
		}[0.49\textwidth]{%
			\includegraphics[width=.65\linewidth]{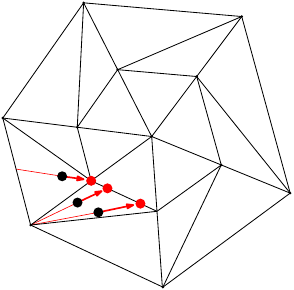}
		}
		\hfill
		\subcaptionbox{
			$R_{\mathrm{equal}}(x)$ is obtained by scaling $R_{\mathrm{inner}}(x)$ so it has the same norm as $x$.
		}[0.49\textwidth]{%
			\includegraphics[width=.65\linewidth]{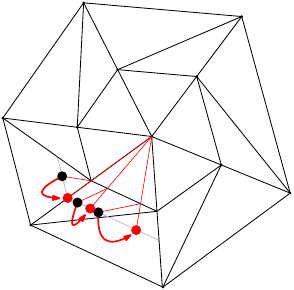}}
		\caption{
			Inner representative $R_{\mathrm{inner}}(x)$ and equal-norm representative $R_{\mathrm{equal}}(x)$ in $\geomreal{\mathring{B}(K)}$.
		}
		\label{fig:representative_maps}
	\end{figure}

	We now define the elementary paths that will be used in our construction.
	
	\begin{description}
		\item[Ray away from the origin:]
		Starting at $x\in\geomreal{B(K)}$, follow the ray emanating from the origin and passing through $x$ (see \Cref{fig:admissible_moves_1}).
		This path is denoted $\gamma^x_\mathrm{ray}$ and is defined for all $x\in\geomreal{B(K)}\setminus\{0\}$, either in the inner or outer part.
		Its image is the line segment $[x,x']$, where $\|\nu(x')\|=1$.
		If only the subpath ending at $y\in[x,x']$ is needed, we  write 
		\[
		x\xleadsto[\text{ray}]y.
		\]
		
		\item[Climb towards the origin:]
		If $x\in\geomreal{B(K)}$ belongs to the inner ball, follow the ray up to the origin.
		Otherwise, $x$ belongs to the outer shell (i.e., $\|\nu(x)\|\geq \rho_{\mathrm{inner}}$), and the path is defined in two pieces: $x$ first follows the straight path towards its representative $R_{\mathrm{inner}}(x)$ in the inner layer, then the ray to the origin; see \Cref{fig:admissible_moves_2}.
		It is denoted $\gamma^x_\mathrm{climb}$ and is defined for all $x\in\geomreal{\mathring{B}(K)}$.
		Its image is the union of line segments $[x,R_{\mathrm{inner}}(x)]\cup[R_{\mathrm{inner}}(x),0]$.
		As before, subpaths are denoted 
		\[
		x\xleadsto[\text{climb}]y.
		\]
		
		\item[Straight path:] 
		The path $\gamma^{x\leadsto y}_\mathrm{straight}$, also denoted
		\[
		x\xleadsto[\text{straight}]y
		\] 
		is the line segment from $x$ to $y$ in $\geomreal{B(K)}$ (see \Cref{fig:admissible_moves_3}).
		We will use it only to connect points at equal distance to the origin, i.e., $\|\nu(x)\|=\|\nu(y)\|$.
		The path is well-defined when $x,y$ belong to the inner ball---i.e., $\|\nu(x)\|,\|\nu(y)\|\leq\rho_\mathrm{inner}$---or to the same sector.
		In this latter case, its radial normalization $\nu\circ\gamma^{x\leadsto y}_\mathrm{straight}$ draws a circular arc.
		In general, however, this path does not descend to the quotient; thus we define another path.
		
		\item[Circular arc in the outer shell, for 2-distance injective mappings:]
		Let $x,y\in\geomreal{B(K)}$ such that $\|\nu(x)\|=\|\nu(y)\|$.
		This path is more conveniently defined in the Euclidean ball: $\nu\circ\gamma^{x\leadsto y}_\mathrm{arc}$ is the arc connecting $\nu(x)$ to $\nu(y)$ in $B^{d+1} = \nu(\geomreal{B(K)})$.
		In other words, it is the length-minimizing path in the circle 
		\[
		C = S(0,\|\nu(x)\|) \cap P(\nu(x),\nu(y))
		\]
		obtained as the intersection of the sphere of radius $\|\nu(x)\|$ centered at the origin and the linear plane passing through $\nu(x)$ and $\nu(y)$.
		Taken back to the polyhedron, the path $\gamma^{x\leadsto y}_\mathrm{arc}$ is linear on each sector, as visualized in \Cref{fig:apothem_normalization,fig:admissible_moves_3}.
		It is also denoted 
		\[
		x\xleadsto[\text{arc}]y.
		\] 
		Note that it is not defined for antipodal points ($\nu(x)=-\nu(y)$) since two length-minimizing arcs exist in $C$.
		In particular, it is not defined on a neighborhood of the origin $(0,0)\in\geomreal{B(K)}\times\geomreal{B(K)}$.
		This is remedied by the use of the straight path defined above.
				
		\item[Circular arc in the outer shell, for general mappings:]
		When the mapping $g$ is not 2-distance injective, the path above may not descend to the quotient.
		Accordingly, we introduce the following modification: before connecting $x$ and $y$ by a circular arc, we connect $x$ to its equal-norm representative $R_{\mathrm{equal}}(x)$, as well as $y$ to $R_{\mathrm{equal}}(y)$, via a straight path.
		In other words, this modified path can be written as
		\[
		x
		\xleadsto[\text{straight}]
		R_{\mathrm{equal}}(x)
		\xleadsto[\text{arc}]
		R_{\mathrm{equal}}(y)
		\xleadsto[\text{straight}]
		y.
		\] 
		When working with general mappings instead of 2-distance injective mappings, we will denote this path as $\gamma^{x\leadsto y}_\mathrm{arc}$ and $x\xleadsto[\text{arc}]y$.
		It is well-defined provided $\|\nu(x)\|,\|\nu(y)\|<1$ (for $R_{\mathrm{equal}}$ to be defined) and $R_{\mathrm{equal}}(x),R_{\mathrm{equal}}(y)$ are not antipodal (for the circular arc).
	\end{description}
	
	\begin{claim}\label{claim:elementary_paths}
		The images $\gamma^{x}_\mathrm{ray}(I)$, $\gamma^{x}_\mathrm{climb}(I)$, $\gamma^{x\leadsto y}_\mathrm{straight}(I)$, and $\gamma^{x\leadsto y}_\mathrm{arc}(I)$ in $\geomreal{L\cup_g B(K)}$ only depend on the equivalence classes of $x$ and $y$.
	\end{claim}
	
	\begin{claimproof}
	For the ray, the result has already been stated in \Cref{lem:raydescendstoquotient};
	for the climb, straight path, and circular arc, the result is a direct consequence of \Cref{claim:representatives}.
	\end{claimproof}
	
		\begin{figure}[!htbp]
		\centering
		\subcaptionbox{
			\textbf{Ray} away from the origin
			\label{fig:admissible_moves_1}
		}[0.315\textwidth]{%
			\includegraphics[width=\linewidth]{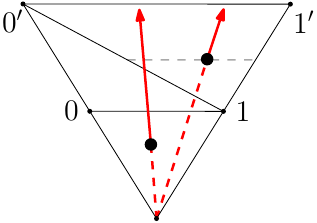}}
		\hfill
		\subcaptionbox{
			\textbf{Climb} towards the origin
			\label{fig:admissible_moves_2}
		}[0.29\textwidth]{%
			\includegraphics[width=.95\linewidth]{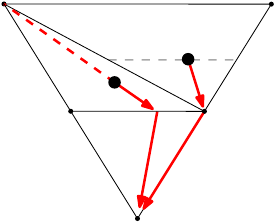}}
		\hfill
		\subcaptionbox{
			\textbf{Straight} path and circular \textbf{arc}.
			\label{fig:admissible_moves_3}
		}[0.375\textwidth]{%
			\includegraphics[width=\linewidth]{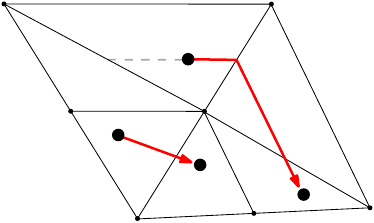}}
		\caption{
			To define paths on the triangulated ball that descend to the quotient, we shall only use a combination of four elementary moves: ray, climb, straight line and arc.
			Our definition of the arc depends on whether $g$ is 2-distance injective; we illustrate here the case where it is.
		}
		\label{fig:admissible_moves}
	\end{figure}
	
	\proofsubparagraph*{Planner on the ball}
	
	To define an equiconnecting map on the triangulated ball $B(K)$ that descends to the quotient $L\cup_g B(K)$, we shall only use the elementary paths defined above.
	Let $x,y\in \geomreal{B(K)}$.
	We measure distances in the Euclidean $B^{d+1}$ through the radial normalization
	\[
	\nu\colon \geomreal{B(K)}\to B^{d+1}.
	\]
	By ``the distance from $x$ (resp.\ $y$) to the origin'' we mean $\|\nu(x)\|$ (resp.\ $\|\nu(y)\|$).
	Without loss of generality, we may assume $\|\nu(x)\|\leq\|\nu(y)\|$, i.e., $x$ is closer to the origin.
	Our construction of a path $x\leadsto y$ uses two intermediate points $x',y'$ of equal norm $\delta$ (defined below),
	\[
	\|\nu(x)\| \leq \|\nu(x')\| = \delta = \|\nu(y')\| \leq \|\nu(y)\|,
	\]
	where $x'$ is the point of norm $\delta$ on the radial path $x\xleadsto[\text{ray}]x'$---i.e., $\nu(x') = \delta \nu(x)/\|\nu(x)\|$---and $y'$ is found by climbing towards the origin via $y\xleadsto[\text{climb}]y'$.
	Connecting $x'\leadsto y'$ is the challenging part: as pointed out previously, circular paths are not well defined in neighborhoods of the origin (for they contain antipodal points), nor are straight lines in the outer shell (for they may not descend to the quotient).
	Thus we interpolate between them.
	
	We introduce an offset parameter $\epsilon_\mathrm{inner}\in (0,\rho_\mathrm{inner})$ that will be used to thicken the inner layer; it is equal to $0.1$ in our implementation.
	Define the quantities
	\begin{align*}
		q(x,y) &= \mathrm{clamp}\left(
		\frac{\|\nu(y)\|-\rho_\mathrm{inner}}{1-\rho_\mathrm{inner}}
		\right)
		~~~~~~~~~~~~~~~~\text{(normalized distance to inner ball)},\\
		\delta(x,y) &= (1-q(x,y))\|\nu(x)\| + q(x,y)\|\nu(y)\| 
		~~~~\text{(intermediate norm)},\\
		s(x,y) &= \mathrm{clamp}\left(
		\frac{\delta(x,y)-(\rho_\mathrm{inner}-\epsilon_\mathrm{inner})}{\epsilon_\mathrm{inner}}\right) 
		~~~~\text{(interpolation parameter)},
	\end{align*}
	where $\mathrm{clamp}(a)=\min(1,\max(0,a))$ is the projection of a real number on the interval $[0,1]$. 
	
	We define the middle connector $x'\leadsto y'$ of $x \leadsto x' \leadsto y' \leadsto y$ as the $s(x,y)$-interpolation between straight and circular paths:
	\[
	\gamma_\mathrm{middle} = (1-s)\gamma_\mathrm{straight}^{x'\leadsto y'} + s\gamma_\mathrm{arc}^{x'\leadsto y'},
	\]
	where the sum is understood in the polyhedron $\geomreal{B(K)}$.
	
	Note that $q(x,y)$ is $0$ when both points lie in the inner ball, in which case $\delta(x,y)=\|\nu(x)\|$ and $x'=x$.
	If, in addition, $x$ is at distance at least $\epsilon_\mathrm{inner}$ from its boundary---i.e., $\|\nu(x)\|\leq \rho_\mathrm{inner}-\epsilon_\mathrm{inner}$---then $s(x,y)$ is zero and $x'\xleadsto[\text{middle}] y'$ is a straight path.
	On the other hand, $q(x,y)$ is $1$ when $\|\nu(y)\|= 1$ (i.e., $y$ is on the boundary), in which case $\delta(x,y)=1$ and $y'=y$.
	Our construction avoids radial paths at the origin and climbing walks from the boundary.
	More precisely, we have the following three regimes, illustrated in \Cref{fig:equiconnecting_map_ball}.
	
	\begin{description}
		\item[Inner ball, away from offset:]
		If $\delta<\rho_\mathrm{inner}- \epsilon_\mathrm{inner}$, we use a straight segment 
		\[
		\gamma_\mathrm{middle} = \gamma_\mathrm{straight}^{x'\leadsto y'}.
		\]
		
		\item[Outer shell:]
		If $\delta>\rho_\mathrm{inner}$, we use a circular arc
		\[
		\gamma_\mathrm{middle} = \gamma_\mathrm{arc}^{x'\leadsto y'}.
		\]
		
		\item[Inner ball, inside offset:]
		If $\rho_\mathrm{inner}-\epsilon_\mathrm{inner}\leq\delta\leq\rho_\mathrm{inner}$, 
		we interpolate 
		\[
		\gamma_\mathrm{middle} = (1-s)\gamma_\mathrm{straight}^{x'\leadsto y'} + s\gamma_\mathrm{arc}^{x'\leadsto y'}.
		\]
	\end{description}
	
	In each of these cases, the final path from $x$ to $y$ decomposes as 
	\[
	x
	\xleadsto[\text{ray}]
	x'
	\xleadsto[\text{middle}] 
	y' 
	\xleadsto[\text{climb}]
	y.
	\]
	We associate to each of these pieces a duration proportional to its length, seen in the polyhedron $\geomreal{B(K)}$.
	For the ray and the straight line, this is respectively $\|x'\|-\|x\|$ and $\|x'-y'\|$;
	for the climbing walk and the polygonal arc, we compute it piecewise in the polyhedron; for the interpolated middle path (between straight line and arc), we take, for implementation simplicity, the corresponding convex combination of the lengths.
	
	We denote this path as $t\to\Pi^{\geomreal{\mathring{B}(K)}}(x,y,t)$, where $t\in[0,1]$.
	It is well-defined on a neighborhood of the diagonal $U^{\geomreal{\mathring{B}(K)}}\subset\geomreal{\mathring{B}(K)}\times \geomreal{\mathring{B}(K)}$.	
	A more precise description of $U^{\geomreal{\mathring{B}(K)}}$ is given in the claim below.
	When the base points $x,y$ satisfy the reverse inequality $\|\nu(x)\|>\|\nu(y)\|$, we define $t\to\Pi^{\geomreal{\mathring{B}(K)}}(x,y,t)$ to be equal to $t\to\Pi^{\geomreal{\mathring{B}(K)}}(y,x,1-t)$.
	In particular, $U^{\geomreal{\mathring{B}(K)}}$ is symmetric: 
	\[
	(x,y)\in U^{\geomreal{\mathring{B}(K)}}\implies(y,x)\in U^{\geomreal{\mathring{B}(K)}}.
	\]
	
	We point out that the choice of $(\rho_\mathrm{inner},\epsilon_\mathrm{inner})$ controls where the construction transitions between chord and circular arc and regulates behavior near the origin.
	
	\begin{claim}\label{claim:planner_ball}
		Paths $t\mapsto\Pi^{\geomreal{\mathring{B}(K)}}(x,y,t)$ are well-defined provided one of the conditions holds:
		\begin{itemize}
			\item Both $x$ and $y$ lie in the ball of radius $\rho_\mathrm{inner}-\epsilon_\mathrm{inner}$, that is, $\|\nu(x)\|,\|\nu(y)\|<\rho_\mathrm{inner}-\epsilon_\mathrm{inner}$,
			\item 
			\begin{itemize}
				\item \textit{For 2-distance injective mappings:} the intermediate points $x',y'$ are not antipodal,
				\item \textit{For general mappings:} the intermediate points $R_{\mathrm{equal}}(x'),R_{\mathrm{equal}}(y')$ are not antipodal.
			\end{itemize}
		\end{itemize}
		Consequently, for 2-distance injective mappings, $\Pi^{\geomreal{\mathring{B}(K)}}$ endows $\geomreal{\mathring{B}(K)}$ with LEC-data; and for general mappings, it endows $\geomreal{\mathring{B}(K)}$ with a local motion planner. 
	\end{claim}
	
	\begin{claimproof}
		The fact that $\Pi^{\geomreal{\mathring{B}(K)}}$ is well-defined directly comes from the domain of definition of the ray, the climb, the straight path, and the circular arc, defined above \Cref{claim:elementary_paths}.

		In general, $\Pi^{\geomreal{\mathring{B}(K)}}$ is a local motion planner: by construction, it is continuous, and it satisfies $\Pi^{\geomreal{\mathring{B}(K)}}(x,y,0) = x$ and $\Pi^{\geomreal{\mathring{B}(K)}}(x,y,1) = y$.
		Moreover, in the 2-distance injective case, we also have that $\Pi^{\geomreal{\mathring{B}(K)}}(x,x,t)=x$ for all $t\in[0,1]$.
		Indeed, in the case $x=y$, the intermediate points $x',y'$ are all equal to $x$, hence $x \xleadsto[\text{ray}] x'$ and $y' \xleadsto[\text{climb}] y$ are constant, and the middle arc $x' \xleadsto[\text{middle}] y'$ also is constant.
		Note that, in the general case, the arc $x' \xleadsto[\text{middle}]y'$ between equal points need not be constant; in the outer shell, the arc is defined as $x\xleadsto[\text{straight}] R_{\mathrm{equal}}(x)
		\xleadsto[\text{arc}] R_{\mathrm{equal}}(y)
		\xleadsto[\text{straight}] y$.
	\end{claimproof}
	
	\begin{figure}[htbp]
		\centering
		\subcaptionbox{
			Case I: $\delta<\rho_\mathrm{inner}-\epsilon_\mathrm{inner}$.
			The joining path is a segment.
		}[0.32\textwidth]{%
			\includegraphics[width=\linewidth]{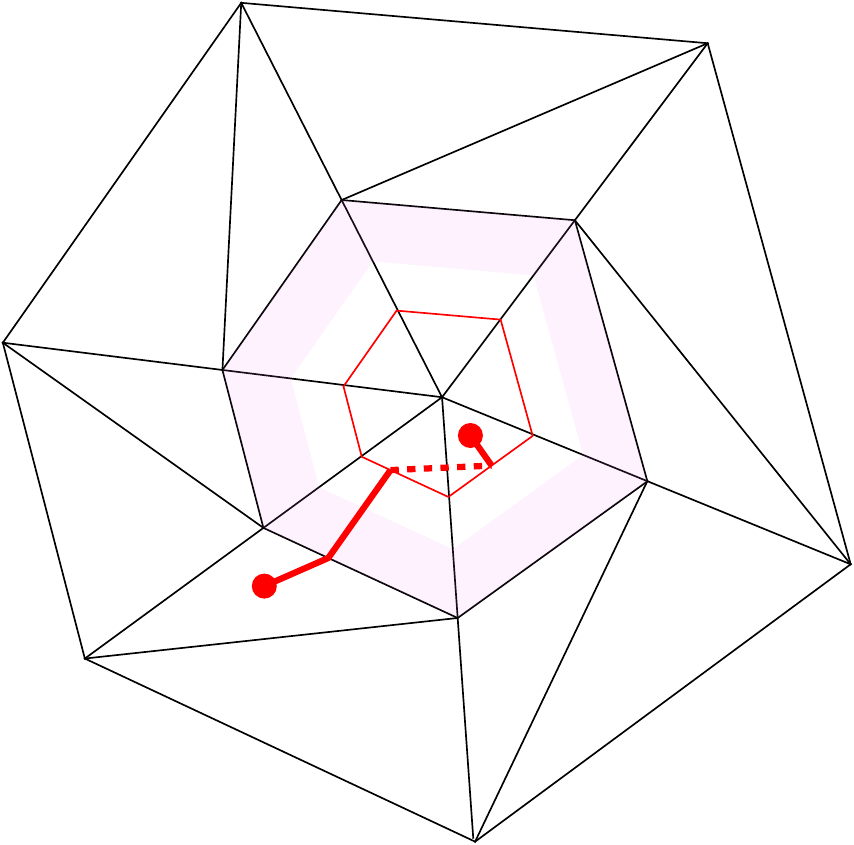}}
		\hfill
		\subcaptionbox{
			Case II: $\delta>\rho_\mathrm{inner}$.
			The joining path is a piecewise-linear arc.
		}[0.32\textwidth]{%
			\includegraphics[width=\linewidth]{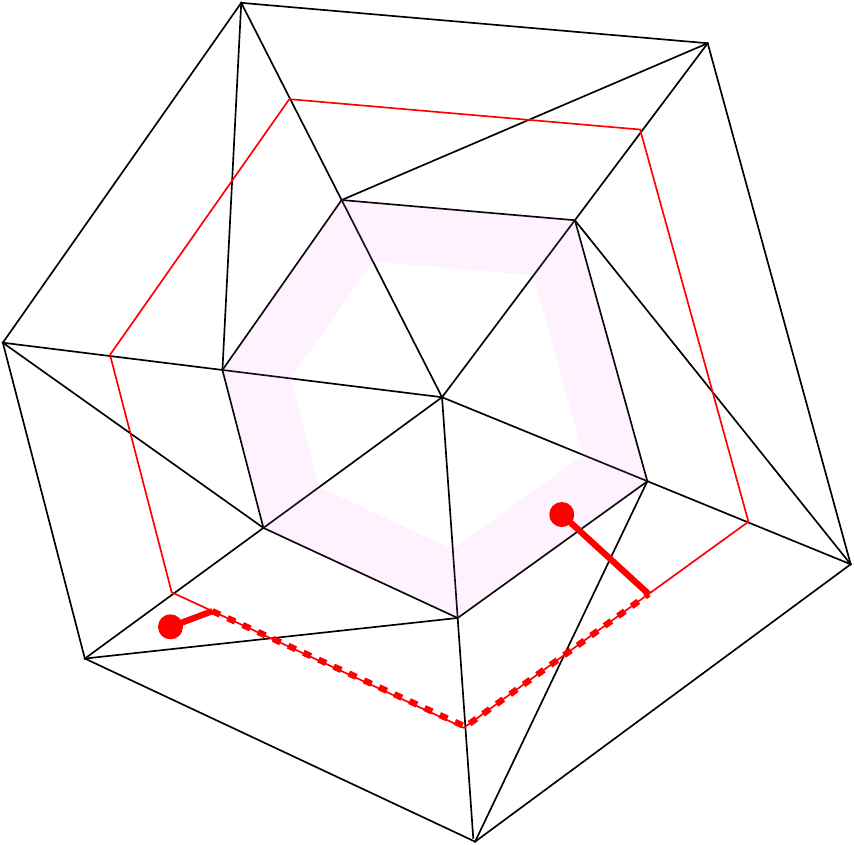}}
		\hfill
		\subcaptionbox{
			Case III: $\rho_\mathrm{inner}-\epsilon_\mathrm{inner}<\delta<\rho_\mathrm{inner}$.
			Interpolation.
		}[0.32\textwidth]{%
			\includegraphics[width=\linewidth]{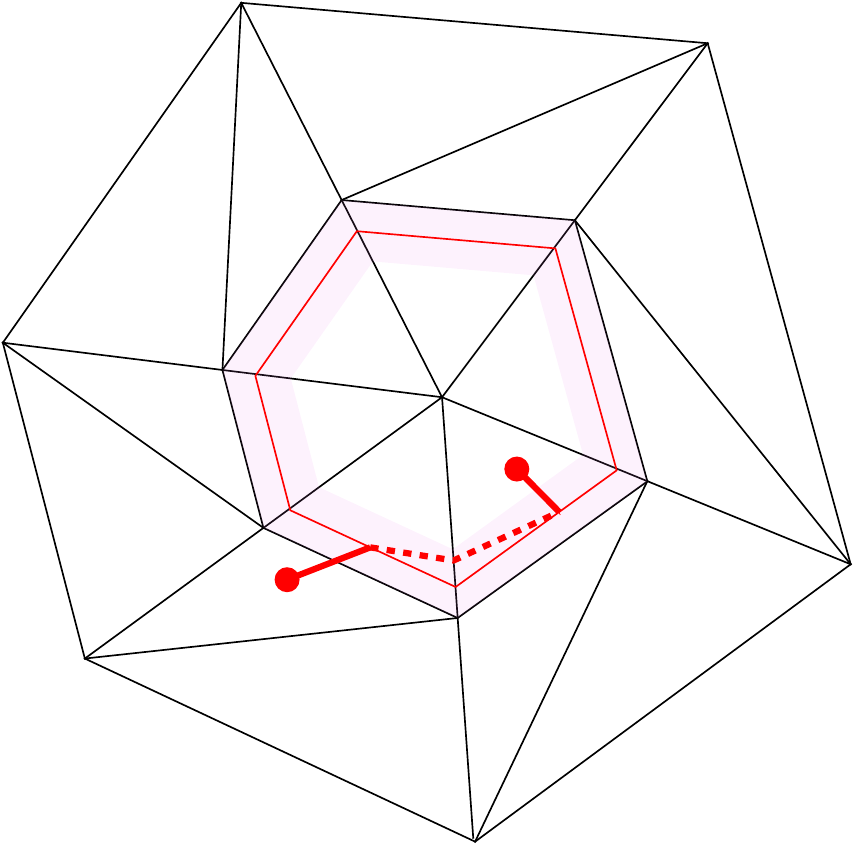}}
		\caption{
			Paths $x\leadsto y$ on the simplicial ball $B(K)$ are obtained as compositions $x\leadsto x'\leadsto y'\leadsto y$, where $x',y'$ are intermediate points of equal norm $\delta=\delta(x,y)$.
			The piece $x'\leadsto y'$ depends on $\delta$: it is a segment close to the origin, a piecewise-linear circular arc in the outer shell, and an interpolation between these paths otherwise.
			When both points belong to the inner ball, $\delta$ is equal to the norm of $x$, and $x'=x$. 
			That is, $x$ does not follow a ray; this ensures the paths are well-defined in a neighborhood of the origin.
		}
		\label{fig:equiconnecting_map_ball}
	\end{figure}
	
	\proofsubparagraph*{Planner on the ball after quotient}
	Let $B(K)/g$ be the quotient of the simplicial ball $B(K)$ obtained by identifying outer vertices
	$v\sim w$ whenever $g(v)=g(w)$. It is a simplicial subcomplex of the simplicial mapping cone
	$\conesimp{g}=L\cup_g B(K)$. 
	Write
	\[
	\geomreal{\mathring{B}(K)/g}:=\geomreal{B(K)/g}\setminus \geomreal{K/g}
	\]
	for its interior.
	
	A key property of the planner $\Pi^{\geomreal{\mathring{B}(K)}}$ from the previous paragraph is that each of its elementary pieces (ray, climb, straight segment, circular arc, and straight--arc interpolation) descends to the quotient; this is exactly the content of \Cref{claim:elementary_paths}. 
	Thus, for every pair of points $x,y\in\geomreal{\mathring{B}(K)}$, the \emph{image} of the path $t\mapsto \Pi^{\geomreal{\mathring{B}(K)}}(x,y,t)$ depends only on the
	quotient classes $\bar x,\bar y\in \geomreal{\mathring{B}(K)/g}$.
	To obtain a genuine planner on the quotient, it remains to choose a parametrization that uses length measured in the quotient.
	
	On the inner ball (which is unaffected by the quotient), we keep Euclidean lengths.
	On the outer shell, each elementary piece is linear on simplices, and we use the induced
	\emph{quotient distance}: if $\bar u,\bar v\in|\bar\sigma|$ lie in a common simplex
	$\bar\sigma\in B(K)/g$, define
	\[
	\d_{/g}(\bar u,\bar v)
	=\min\{\|u-v\|\mid (u,v) \text{ maps to }(\bar u,\bar v)\}.
	\]
	Polygonal arcs are split into finitely many linear segments
	$x=x_0\leadsto\cdots \leadsto x_m=y$, and their length is computed by summing $\d_{/g}(\bar x_{i-1},\bar x_i)$.
	See \Cref{note:cvxpy,note:arc_distance,note:splitting_arc} for practical computation.
	
	This gives a continuous time-rescaling of the same image paths, hence a well-defined map
	\[
	\Pi^{\geomreal{\mathring{B}(K)/g}}:U^{\geomreal{\mathring{B}(K)/g}}\times[0,1]\to
	\geomreal{\mathring{B}(K)/g}
	\]
	on the image of $U^{\geomreal{\mathring{B}(K)}}$ under the quotient
	$\geomreal{\mathring{B}(K)}\to \geomreal{\mathring{B}(K)/g}$.
	
	\begin{claim}\label{claim:planner_quotient}
		The map $\Pi^{\geomreal{\mathring{B}(K)/g}}$ endows $\geomreal{\mathring{B}(K)/g}$ with LEC-data
		(for $2$-distance injective mappings) or with a local motion planner (for general mappings).
		Its domain of definition is the image in the quotient of the domain from
		\Cref{claim:planner_ball}.
	\end{claim}
	
	\proofsubparagraph*{Joining with boundary planner}
	Assume an equiconnecting map or local motion planner $(U^{\geomreal{L}},\Pi^{\geomreal{L}})$ is fixed
	on $\geomreal{L}$.
	We now modify $\Pi^{\geomreal{\mathring{B}(K)/g}}$ so that it agrees with $\Pi^{\geomreal{L}}$ near the
	boundary $\geomreal{K/g}=\partial \geomreal{B(K)/g}$.
	The result will be denoted $\Pi^{\geomreal{B(K)/g}}$.
	
	We start with \textit{mixed pairs} $(\bar x, \bar y)\in \geomreal{\mathring{B}(K)/g}\times \geomreal{K/g}$ where $\bar x$ is inside the ball and $\bar y$ is on the boundary.
	If $\bar x$ is not the origin, let $P(\bar x)\in \geomreal{K/g}\subset\geomreal{L}$
	be the endpoint on the boundary of the (projected) ray through $\bar x$.
	Equivalently, if $x\in \geomreal{B(K)}$ lifts $\bar x$, then $P(\bar x)$ is the image in the quotient of the point $\nu(x)/\|\nu(x)\|$.
	This is well-defined by \Cref{claim:elementary_paths}.
	Provided $(P(\bar x),\bar y)\in U^{\geomreal{L}}$, we have a well-defined path  $\Pi^{\geomreal{L}}(P(\bar x),\bar y,\cdot)$.
	We define the path $\Pi^{\geomreal{B(K)/g}}(\bar x,\bar y,\cdot)$ to be the concatenation
	\[
	\bar x \xleadsto[\text{ray}] P(\bar x) \xleadsto[\Pi^{\geomreal{L}}(P(\bar x),\bar y,\cdot)] \bar y,
	\]
	with respective durations $1\!-\!\|\nu(\bar x)\|$ and $\|\nu(\bar x)\|$.
	The reverse case $\Pi(\bar y,\bar x,\cdot)$ is defined similarly.
	
	Second, we consider pairs  $(\bar x,\bar y)\in \geomreal{\mathring{B}(K)/g}\times\geomreal{\mathring{B}(K)/g}$ \textit{inside the ball} and assume $\|\nu(\bar x)\|\le \|\nu(\bar y)\|$ (their distance to the origin is well-defined).
	Write the decomposition from \Cref{claim:planner_quotient} as
	\[
	\bar x \xleadsto[\text{ray}] \bar x' \xleadsto[\text{middle}] \bar y'
	\xleadsto[\text{climb}] \bar y,
	\]
	and let
	\begin{align*}
		\delta(\bar x,\bar y) &= \|\nu(\bar x')\|=\|\nu(\bar y')\|\in (0,1]
		~~~~\text{(intermediate norm)},\\
		\Theta(\bar x,\bar y) &= \mathrm{length}_{/g}\bigl(\bar x'\leadsto \bar y'\bigr)
		~~~~~~~~~~~~~~\text{(circular length)},
	\end{align*}
	where $\Theta(\bar x,\bar y)$ is the quotient length of the middle piece.
	To simplify the notation, we define
	\[
	\Delta(\bar x,\bar y) = 1-\delta(\bar x, \bar y)
	~~~~~~~~~~~~~~~~~~~~~~~\text{(intermediate distance to boundary)}.
	\]
	
	We distinguish three regimes; they are illustrated in \Cref{fig:local_planner}.
	
	\begin{description}
		\item[Interior regime] ($\Delta\geq\Theta$):
		We leave the path unchanged: 
		\[
		\Pi^{\geomreal{B(K)/g}}(\bar x,\bar y,\cdot)=\Pi^{\geomreal{\mathring{B}(K)/g}}(\bar x,\bar y,\cdot).
		\]
		
		\item[Push regime] ($\Theta/2<\Delta<\Theta$):
		Define the push parameter
		\[
		q_{\mathrm{push}}(\bar x,\bar y)
		=\mathrm{clamp}\!\left(\frac{2\Delta(\bar x,\bar y)}{\Theta(\bar x,\bar y)}-1\right)\in(0,1),
		\]
		and the pushed intermediate norm
		\[
		\delta_{\mathrm{push}}(\bar x,\bar y)
		=(1-q_{\mathrm{push}})\cdot 1+q_{\mathrm{push}}\cdot \delta(\bar x,\bar y).
		\]
		Define $\Pi(\bar x,\bar y,\cdot)$ by re-running the construction of
		$\Pi^{\geomreal{\mathring{B}(K)/g}}(\bar x,\bar y,\cdot)$ with the change that the whole path is pushed towards the boundary, by a length $ \delta_{\mathrm{push}}(\bar x,\bar y)-\delta(\bar x,\bar y)$.
		
		More precisely, let $\bar x''$ and  $\bar y''$ be the points of norm $\delta_{\mathrm{push}}(\bar x,\bar y)$ on the corresponding rays.
		The path reads
		\[
		\bar x \xleadsto[\text{ray}] \bar x' \xleadsto[\text{ray}] \bar x'' \xleadsto[\text{pushed middle}] \bar y''  \xleadsto[\text{ray}]  \bar y'
		\xleadsto[\text{climb}] \bar y,
		\]
		We attribute to these pieces the durations $\delta-\|\nu(\bar x)\|$, $\delta_{\mathrm{push}}-\delta$, $\tau$, $\delta_{\mathrm{push}}-\delta$, and $\|\nu(\bar y)\|-\delta$, where the \textit{pushed duration} $\tau$ is
		\begin{align*}
			\tau(\bar x,\bar y) &= \big(1-q_{\mathrm{push}}(\bar x,\bar y)\big)\cdot\|\nu(\bar x)\| + q_{\mathrm{push}}(\bar x,\bar y)\cdot\Theta(\bar x, \bar y).
		\end{align*}
		
		\item[Splice regime] ($0\leq \Delta\leq \Theta/2$):
		In this regime $\delta_{\mathrm{push}}=1$, so the pushed construction reaches the boundary.
		Let $\Gamma_{\partial}(\bar x,\bar y,\cdot)$ denote the resulting boundary connector between
		$P(\bar x)$ and $P(\bar y)$ (projections on the boundary), viewed as a path in $\geomreal{K/g}\subset\geomreal{L}$.
		Assuming $(P(\bar x),P(\bar y))\in U^{\geomreal{L}}$, consider also the planner already prescribed on the boundary,
		\[
		\Gamma_{L}(\bar x,\bar y,\cdot):=\Pi^{\geomreal{L}}\bigl(P(\bar x),P(\bar y),\cdot\bigr).
		\]
		Define the splice parameter
		\[
		s_{\mathrm{splice}}(\bar x,\bar y)
		=\mathrm{clamp}\!\left(\frac{2\Delta(\bar x,\bar y)}{\Theta(\bar x,\bar y)}\right)\in[0,1],
		\]
		so that $s_{\mathrm{splice}}=1$ when $\Delta=\Theta/2$ and $s_{\mathrm{splice}}=0$ when $\Delta=0$.
		Define an interpolated boundary connector by
		\[
		t\mapsto\Gamma_{\mathrm{splice}}(\bar x,\bar y,t)
		=
		\Pi^{\geomreal{L}}\!\bigl(\Gamma_{L}(\bar x,\bar y,t),\ \Gamma_{\partial}(\bar x,\bar y,t),\
		s_{\mathrm{splice}}(\bar x,\bar y)\bigr).
		\]
		Finally, set $\Pi^{\geomreal{B(K)/g}}(\bar x,\bar y,\cdot)$ to be the concatenation
		\[
		\bar x \xleadsto[\text{ray}] P(\bar x)
		\xleadsto[\Gamma_{\mathrm{splice}}] P(\bar y)
		\xleadsto[\text{ray}] \bar y,
		\]
		parametrized with length-based durations: $\delta-\|\nu(\bar x)\|$, $2-2\delta+\tau$, and $\|\nu(\bar y)\|-\delta$.
	\end{description}
		
	\begin{claim}\label{claim:planner_boundary}
		The map $\Pi^{\geomreal{B(K)/g}}$ endows $\geomreal{B(K)/g}\cup\geomreal{L}$ with LEC-data (for
		$2$-distance injective mappings) or with a local motion planner (for general mappings). 
		Moreover, it agrees with $\Pi^{\geomreal{L}}$ on the boundary of $\geomreal{B(K)/g}$.
	\end{claim}
	
	\begin{claimproof}
		By construction, $\Pi^{\geomreal{B(K)/g}}(\bar x,\bar y,0)=\bar x$ and
		$\Pi^{\geomreal{B(K)/g}}(\bar x,\bar y,1)=\bar y$ on its domain, since each case is defined by concatenating elementary pieces with the correct endpoints.
		
		Continuity of individual paths $t\mapsto\Pi^{\geomreal{B(K)/g}}(\bar x,\bar y,t)$ follows because each elementary path (ray, climb, straight segment, circular arc) depends continuously on its endpoints, and because we apply a time-rescaling that depends continuously on the segment lengths.
		
		It remains to check continuity across the regime boundaries:
		\begin{itemize}
			\item 		
			at $\Delta=\Theta$, we have $q_{\mathrm{push}}=1$, hence $\delta_{\mathrm{push}}=\delta$, so the push construction coincides with the unmodified one;

			\item 
			at $\Delta=\Theta/2$, we have $s_{\mathrm{splice}}=1$, hence $\Gamma_{\mathrm{splice}}=\Gamma_{\partial}$, so the splice construction coincides with the $\delta_{\mathrm{push}}=1$ limit of the pushed construction;

			\item 
			at $\Delta=0$ (boundary), $s_{\mathrm{splice}}=0$, hence $\Gamma_{\mathrm{splice}}=\Gamma_{L}$ and the middle part becomes exactly $\Pi^{\geomreal{L}}(P(\bar x),P(\bar y),\cdot)$, so the planner agrees with $\Pi^{\geomreal{L}}$ on $\geomreal{L}$.
		\end{itemize}
		
		In the $2$-distance injective case, the diagonal condition $\Pi^{\geomreal{B(K)/g}}(\bar x,\bar x,t)=\bar x$ holds: 
		\begin{itemize}
			\item 
			for $\bar x\in\geomreal{\mathring{B}(K)/g}$ in the inner ball it follows from \Cref{claim:planner_quotient}; 
			
			\item 
			for $\bar x\in\geomreal{K/g}$ on the boundary it follows from the corresponding property of $\Pi^{\geomreal{L}}$; 
			
			\item 
			in the other regimes, we only interpolate between paths that are constant on the diagonal, hence the result remains constant. 
		\end{itemize}
		This proves the claim.
	\end{claimproof}
	
	\proofsubparagraph*{Planner on the full mapping cone}
	
	Last, we extend $\Pi^{\geomreal{B(K)/g}}$ to the simplicial mapping cone $\geomreal{\conesimp{g}}=\geomreal{L\cup_g B(K)}$.
	We distinguish three cases:
	
	\begin{description}
		\item[Pairs in the domain:] 
		If $u,v\in\geomreal{B(K)/g}$, set 
		\[
		\Pi^{\geomreal{\conesimp{g}}}(u,v,t)=\Pi^{\geomreal{B(K)/g}}(u,v,t).
		\]
		\item[Pairs in the codomain:] 
		If $u,v\in\geomreal{L}$, set 
		\[
		\Pi^{\geomreal{\conesimp{g}}}(u,v,t)=\Pi^{\geomreal{L}}(u,v,t).
		\]
		\item[Mixed pairs:] 
		If $u\in\geomreal{\mathring{B}(K)/g}\setminus\{0\}$ and $v\in\geomreal{L}$, let
		$P(u)\in\geomreal{K/g}\subset\geomreal{L}$ be the projection on the boundary.
		Provided $(P(u),v)\in U^{\geomreal{L}}$, we define $\Pi^{\geomreal{\conesimp{g}}}(u,v,\cdot)$ as the concatenation
		\[
		u  \xleadsto[\text{ray}] P(u) \xleadsto[\Pi^{\geomreal{L}}] v,
		\]
		with associated durations $1-\|\nu(u)\|$ and $\|\nu(u)\|$.	
		The symmetric case $(u,v)\in\geomreal{L}\times\geomreal{\mathring{B}(K)/g}\setminus\{0\}$ is defined
		analogously.
	\end{description}
	
	Let $U^{\geomreal{\conesimp{g}}}$ be the union of the corresponding domains in the three cases above; it is an open neighborhood of the diagonal.
	
	\begin{claim}\label{claim:planner_full}
		The map $\Pi^{\geomreal{\conesimp{g}}}$ endows $\geomreal{\conesimp{g}}$ with LEC-data (for $2$-distance injective mappings) or with a local motion planner (for general mappings).
	\end{claim}
	
	\begin{claimproof}
		The endpoint conditions are immediate from the definitions.
		Continuity follows because each case is built from continuous pieces, and on overlaps the definitions agree.
		In particular, the ray segment has length $0$ on the boundary, and $\Pi^{\geomreal{B(K)/g}}$ agrees with $\Pi^{\geomreal{L}}$ on $\geomreal{L}$ by \Cref{claim:planner_boundary}.
		In the $2$-distance injective case, the diagonal condition follows since each of the cases is constant on the diagonal whenever $\Pi^{\geomreal{L}}$ is.
	\end{claimproof}
	
	This last claim proves the theorem.
\end{proof}

\begin{note}
	\Cref{th:lec_data_gluing} remains valid if the assumption that ``$g$ is 2-distance injective'' is replaced by the following weaker conditions:
	\begin{enumerate}
		\item $g$ is injective on each simplex;
		\item the $1$-skeleton of $K$ admits an orientation whose restriction to every facet is acyclic; and
		\item whenever two distinct vertices both have an outgoing edge to the same vertex, their images under $g$ are distinct.
	\end{enumerate}
	Under these hypotheses, the edge orientation induces a local ordering of vertices which can be used to build staircase triangulations.
	The construction of $\Pi$ as in the 2-distance injective case yields an equiconnecting map.
	
	Note that, when $B(K)$ is a triangulated $2$-ball, conditions~2 and~3.\ are automatically satisfied as soon as~1.\ holds: one may take the cyclic order of vertices along~$S^1$.
	As a matter of fact, for the $2$-ball, one can always construct LEC-data by choosing such a cyclic order; we do not elaborate on this here.
\end{note}

\begin{note}
	\label{note:cvxpy}
	In practice, we compute the quotient distance as a convex optimization problem.
	More precisely, given two points in a common simplex of the quotient, we minimize the Euclidean distance between their lifts in the covering simplex, subject to the linear constraints encoding the quotient identifications.
	This convex problem is implemented in \texttt{CVXPY} and solved via second-order cone programming (SOCP) using the SCS solver~\cite{diamond2016cvxpy,agrawal2018rewriting}.
\end{note}

\begin{note}
	\label{note:splitting_arc}
	From a practical point of view, the intermediate points used to split a circular arc $\gamma^{x\leadsto y}_\mathrm{arc}$ can be obtained in two steps.
	First, we identify the different sectors the arc crosses; we elaborate on this idea below.
	Second, we identify, in each sector, the simplices of the staircase triangulation that the arc crosses; this can be implemented using the same predicates.
	
	We first project the endpoints $x,y$ onto the unit sphere, so that the problem reduces to detecting the intersections between the unit circular arc and the ridges of the triangulation $\nu(K)=S^d$.
	Assuming $x$ and $y$ are not antipodal, their spherical geodesic midpoint
	\[
	z \;=\; \frac{x+y}{\|x+y\|}
	\]
	is well-defined.
	Both $x$ and $y$ lie in the open hemisphere $S^+_z$ centered at $z$.
	Let
	\[
	p\colon S^+_z \longrightarrow T_z S^d
	\]
	denote the \textit{gnomonic projection} onto the tangent space at $z$.
	This projection maps geodesics to straight lines, hence the polyhedron $p\circ \nu(K)$, as well as
	the projected arc $p\circ\gamma^{x\leadsto y}_\mathrm{arc}$, are piecewise linear.
	We are thus reduced to computing the intersections between the straight segment
	$[p(x),p(y)]\subset T_z S^d$ and the triangulation $p\circ\nu(K)$.
	This is implemented by walking on the triangulation and using standard Euclidean segment--simplex
	intersection predicates.
	The resulting intersection points, mapped back by $p^{-1}$, yield the desired subdivision of
	$\gamma^{x\leadsto y}_\mathrm{arc}$.
\end{note}

\begin{note}
	\label{note:arc_distance}
	To better mimic the spherical distance, it is preferable to convert Euclidean distances in the polyhedron to geodesic distances on the sphere.
	Consider two points $x,y\in\geomreal{B(K)}$ of equal norm $r=\|\nu(x)\|=\|\nu(y)\|$ after normalization, and denote by
	\[
	d_\mathrm{chord} = \|\nu(x)-\nu(y)\|
	\]
	their chord distance on the sphere $S(0,r)$ of radius $r$.
	Their spherical (arc) distance is
	\[
	d_\mathrm{arc} = h(r,d_\mathrm{chord})
	\qquad\text{where}\qquad
	h(r,d) \;=\; 2r\,\arcsin\!\left(\frac{d}{2r}\right).
	\]
	Note that $d_\mathrm{arc}$ is the geodesic distance on $S(0,r)$ of the circular arc
	$\nu\circ\gamma^{x\leadsto y}_\mathrm{arc}$.
	Equivalently, one can decompose $\gamma^{x\leadsto y}_\mathrm{arc}$ into a piecewise linear path in the corresponding sectors,
	\[
	x = x_0 \leadsto x_1 \leadsto \dots \leadsto x_m = y,
	\]
	and compute the geodesic length by
	\[
	\sum_{i=0}^{m-1} h\bigl(r, \|\nu(x_i)-\nu(x_{i+1})\|\bigr).
	\]
	This provides a practical way to use spherical distances while working with a piecewise linear representation in the polyhedron.
\end{note}

\lipschitzdeformationsimplices*

\begin{proof}[Proof of \Cref{lem:lipschitz_deformation_simplices}]
	\label{proof:lipschitz_deformation_simplices}
	Let $x\in\geomreal{\sigma}$.
	By Lipschitz continuity, every vertex $v_i$ satisfies
	\[
		\|f(x)-f(v_i)\|\leq \lambda	\|x-v_i\|.
	\]
	Besides, the triangle inequality yields
	\[
		\|f(x)-c_I\| \leq \frac{1}{|I|}\sum_{i\in I}\|f(x)-f(v_i)\| \leq \frac{\lambda}{|I|}\sum_{i\in I} \|x-v_i\|.
	\]
	The map $x\mapsto \sum_{i\in I} \|x-v_i\|$ attains its maximum at a vertex, proving the claim.
\end{proof}

\approximationroutine*

\begin{proof}[Proof of \Cref{prop:approximationroutine}]
	\label{proof:approximationroutine}
	The correctness of \Cref{alg:algorithm_approx} is ensured by \Cref{th:lec_data_gluing} and \Cref{note:lec_data_gluing}.
	Indeed, if the algorithm terminates, then $f$ and $\geomreal{g}$ are $U$-close for some open set $U$ that is the domain of a local motion planner on $\geomreal{L}$. Therefore, $f$ and $\geomreal{g}$ are homotopic.
	
	To prove termination under global refinement, we adapt the proof of the simplicial approximation theorem \cite[Theorem~2C.1]{hatcher_algebraic_2002}.
	First, define
	\[
	\epsilon_1=\frac{\rho_\mathrm{inner}-\epsilon_\mathrm{inner}}{2\lambda},
	\]
	where $\lambda$ is the parameter of the algorithm, assumed to be a Lipschitz constant for $f$, $\rho_\mathrm{inner}$ is the inner radius parameter, defined in \Cref{subsec:triangulation_ball}, and $\epsilon_\mathrm{inner}$ is the offset parameter, defined in the proof of \Cref{th:lec_data_gluing}.
	If a facet $\sigma\in K$ has diameter less than $\epsilon_1$, then, by \Cref{lem:lipschitz_deformation_simplices}, its image in the geometric realization of some simplicial cell of $\geomreal{L}$ is contained in a ball of radius $(\rho_\mathrm{inner}-\epsilon_\mathrm{inner})/2$.

	On the other hand, as stated in \Cref{note:lec_data_gluing}, pairs of points belong to the domain $U$ of the motion planner provided they are sufficiently close to the origin---at distance at most $(\rho_\mathrm{inner}-\epsilon_\mathrm{inner})/2$, as the proof of \Cref{th:lec_data_gluing} shows---or when they are not antipodal.
	Thus, $\sigma$ satisfies the planner condition.
	
	Second, let $\mathcal{U}$ be the open cover of $\geomreal{L}$ by open stars, and consider the pullback cover
	\[
	f^{-1}(\mathcal{U})=\{\,f^{-1}(V)\mid V\in\mathcal{U}\,\}.
	\]
	Since $S^d$ is compact, $f^{-1}(\mathcal{U})$ admits a Lebesgue number; that is, there exists $\epsilon_2>0$ such that every open ball of radius $\epsilon_2$ in $S^d$ is contained in some element of $f^{-1}(\mathcal{U})$.

	Finally, let $\epsilon=\min(\epsilon_1,\epsilon_2)$.
	By \Cref{th:shrinkingrefinements}, after finitely many refinements we have
	\[
	\rho_\mathrm{diam}(K)<\epsilon,
	\]
	where $\rho_\mathrm{diam}(K)$ denotes the maximal diameter of simplices in $K$.
	Since $\rho_\mathrm{diam}(K)<\epsilon_1$, such a $K$ satisfies the planner condition. Similarly, since $\rho_\mathrm{diam}(K)<\epsilon_2$, $K$ satisfies the star condition. 
	Hence a simplicial map $K\to L$ exists, making
	\texttt{LHom} feasible. At this stage, \Cref{alg:algorithm_approx} terminates.
\end{proof}

\algorithmcorrect*

\begin{proof}[Proof of \Cref{th:algorithmcorrect}]
	\label{proof:algorithmcorrect}
	We prove by induction on the skeleta that \Cref{alg:algorithm_full} preserves the homotopy type of the CW complex.
	First, at line~1, $L$ is defined to be the discrete simplicial complex on the $0$-cells of $X$.
	Thus $|L|$ is canonically homeomorphic to $X^0$.
	
	Assume that after processing all cells of dimension lower than $d$, the current simplicial complex $L$ satisfies $|L|\simeq X^{d-1}$, and that $|L|$ is endowed with an equiconnecting map or a local motion planner (required to call \Cref{alg:algorithm_approx} on line~5).
	Let $e_i^d$ be a $d$-cell with attaching map $\phi_i^d\colon S^{d-1}\to X^{d-1}$.
	Composing $\phi_i^d$ with the current homotopy equivalence $X^{d-1}\to |L|$ yields a continuous map $f_i^d\colon S^{d-1}\to |L|$ (line~4).
	Running \Cref{alg:algorithm_approx} on $f_i^d$ produces a triangulation $K_i$ of $S^{d-1}$ and a simplicial map $g_i^d\colon K_i\to L$.
	Under global refinement, \Cref{prop:approximationroutine} guarantees both termination of \Cref{alg:algorithm_approx} and that the output $|g_i^d|$ is homotopic to $f_i^d$.
	Under local refinement, only the latter result is guaranteed.
		
	Now, attaching the cell $e_i^d$ to $X^{d-1}$ is (up to homeomorphism) the gluing $X^{d-1}\cup_{\phi_i^d} B^d$. 
	On the simplicial side, \Cref{alg:algorithm_full} constructs a simplicial ball $B(K_i)$ and glues it to $L$ along $K_i$ via $g_i^d$, producing the new complex $L\cup_{g_i^d} B(K_i)$ (line~6).
	By \Cref{prop:equivalencemappingcones}, we have a homotopy equivalence $X^{d-1}\cup_{\phi_i^d} B^d\to\geomreal{L\cup_{g_i^d} B(K_i)}$.
	After processing all $d$-cells, we obtain a simplicial complex (still denoted $L$) with $|L|\simeq X^d$.
	Finally, line~7 endows each new gluing $L\cup_{g_i^d} B(K_i)$ with a local motion planner whose existence is ensured by \Cref{th:lec_data_gluing}.
	
	By induction, at the end of the outer loop, the simplicial complex $L$ satisfies $|L|\simeq X$. 
\end{proof}

\fi

\end{document}